%% file: main.tex
\documentclass{aastex63}
\pdfoutput=1

\usepackage{graphicx}
\usepackage{enumerate}
\usepackage{amssymb, amsmath, amsfonts}
\usepackage{gensymb}
\usepackage{mathtools}
\usepackage{physics}
\usepackage{natbib}
\usepackage{hyperref}
\usepackage{color}
\usepackage{longtable}
\usepackage{ulem}
\usepackage{soul}
\usepackage{url}
\usepackage{xspace}
\usepackage{multirow}
\usepackage{subfigure}
\usepackage{float}
\usepackage{lineno}

\newcommand{\Hipparcos}{{\sl Hipparcos}\xspace}
\newcommand{\Gaia}{\textit{Gaia}\xspace}
\newcommand{\gaia}{\textit{Gaia}\xspace}

\newcommand{\HARPS}{{\sl HARPS}\xspace}
\newcommand{\HIRES}{{\sl HIRES}\xspace}

\newcommand{\Msun}{\mbox{$\mathrm{M_{\sun}}$}\xspace}

\newcommand{\Mjup}{\mbox{$\mathrm{M_{\rm Jup}}$}\xspace}

\newcommand{\orvara}{\texttt{orvara}\xspace}
\newcommand{\htof}{\texttt{htof}\xspace}
\newcommand{\htofcodename}{\texttt{htof}\xspace}

\newcommand{\totalstar}{$143$\xspace}
\newcommand{\totalcompanion}{$156$\xspace}
\newcommand{\stellarcomp}{$111$\xspace}
\newcommand{\BDcomp}{$12$\xspace}
\newcommand{\planetcomp}{$33$\xspace}
\newcommand{\startwobd}{$130$\xspace}
\newcommand{\starthreebd}{$13$\xspace}

\def\aap{A\&A}

\begin{document}

\title{Orbits and Masses for 156 Companions from Combined Astrometry and Radial Velocities, and A Validation of Gaia Non-Single Star Solutions}

\author[0000-0003-0115-547X]{Qier An}
\affiliation{Department of Physics, University of California, Santa Barbara, Santa Barbara, CA 93106, USA}
\affiliation{Department of Physics and Astronomy Johns Hopkins University, Baltimore, MD 21218, USA}

\author[0000-0003-2630-8073]{Timothy D.~Brandt}
\affiliation{Space Telescope Science Institute, 3700 San Martin Dr., Baltimore, MD 21218, USA}
\affiliation{Department of Physics, University of California, Santa Barbara, Santa Barbara, CA 93106, USA}

\author[0000-0003-0168-3010]{G.~Mirek Brandt}
\affiliation{Department of Physics, University of California, Santa Barbara, Santa Barbara, CA 93106, USA}

\author[0000-0002-8400-1646]{Alexander Venner}
\affiliation{Centre for Astrophysics, University of Southern Queensland, Toowoomba, QLD 4350, Australia}

\correspondingauthor{Qier An}
\email{qan4@jh.edu}

\begin{abstract}

We combine absolute astrometry from \Hipparcos and \Gaia with archival radial velocities from the Keck/\HIRES and ESO/\HARPS spectrographs, as well as relative astrometry (when available), to derive masses and orbits for \totalcompanion companions around main-sequence stars, including 111 stellar companions, 12 brown dwarfs, and 33 planets. Although this sample is not compiled for occurrence‐rate statistics due to systematic biases in non‐uniform target selection and varied observing strategies, we nonetheless clearly detect the Brown Dwarf desert in the distribution of companion masses (as well as in mass ratio), out to separations of $\gtrsim$10 AU. This work also enables a validation of \Gaia DR3 non-single-star solutions by predicting \Gaia's measured Right Ascension and Declination acceleration terms. For stars with Gaia astrometric acceleration solutions, we find qualitative agreement with \Gaia DR3 results. Our predicted accelerations agree with the \Gaia DR3 values overall, showing a median offset of $1.85\sigma$, with a tail extending to $\approx$10$\sigma$. These residuals suggest modestly underestimated uncertainties, broadly consistent with previous results for parallaxes and proper motions.  Three of our systems have full \Gaia orbital fits; however, their true orbital periods are long and all three \Gaia solutions are spurious. Gaia DR4 will provide individual astrometric measurements and enable more detailed and extensive investigations of accelerating and orbital fits.

\end{abstract}


\section{Introduction} \label{sec:intro}

One of the central goals in exoplanetary science is to establish how substellar companions—ranging from massive planets to brown dwarfs—form and evolve within their host systems. Determining the masses and orbits of these companions is essential to understanding their underlying formation mechanisms: whether they arise from a disk-like process analogous to planetary formation or whether they instead condense out of a collapsing molecular cloud, like stars. The field has advanced from the first confirmed exoplanet detection around a main-sequence star in 1995 \citep{Mayor1995} to a state where thousands of exoplanets have been discovered, unveiling a vast diversity in their masses, sizes, and orbital architectures \citep{2015TESS}. Yet bridging the gap between star-like and planet-like formation channels remains a key challenge.

Stars predominantly form through the gravitational collapse of molecular clouds \citep{1987Shu_StarFormation}, whereas the planets in our Solar System are thought to have formed within a protoplanetary disk \citep{NiceSolarSystem2005}. Brown dwarfs—substellar objects in the approximate mass range of 13–80~\Mjup — occupy a critical intermediate regime between these two formation pathways. Massive brown dwarfs are likely to form like stars, but their lower-mass counterparts could share formation channels with giant planets, or follow multiple distinct processes \citep{2000Chabrier}. The observed scarcity of brown dwarf companions within a few astronomical units (au) of solar-type stars, often referred to as the “brown dwarf desert” \citep{2006Grether_BDdesert_quantify}, underscores the complexity and uniqueness of their formation. This desert suggests that different formation scenarios dominate at different mass scales: star-like formation dominates above the brown-dwarf desert and planet-like formation below it.

Understanding the precise masses of brown dwarfs and massive planets is essential for distinguishing between these formation scenarios. Simple photometric or spectroscopic observations often yield model-dependent mass estimates, with significant uncertainties due to the substellar cooling models that complicate mass-luminosity relationships \citep{1997Burrows,2003Baraffe_BDcoolingModel} and/or unknown system ages. In contrast, dynamical mass measurements derived from combining radial velocity (RV) and astrometric data are independent of evolutionary models \citep{2022Dupuy}. This combination offers a powerful means to calibrate theoretical predictions and place robust constraints on companion masses and orbital configurations, thereby refining our understanding of formation and evolution of substellar objects.

RV surveys have been pivotal in identifying and characterizing substellar companions. Initially, RV measurements—sensitive to the Doppler shifts induced by unseen companions—were limited by technology and data quality. However, decades of technical improvements and the construction of high-precision spectrographs such as HIRES on the Keck Telescope \citep{Hires} and HARPS on the ESO 3.6m telescope \citep{Harps} have enabled the detection of substellar companions with unprecedented precision. Despite these achievements, RV data alone typically yield only the minimum mass ($m\sin{i}$), leaving the true companion mass uncertain without additional geometric information from astrometry.

Astrometric measurements complement RV data by tracing the projected two-dimensional motion of the host star on the sky plane. The \Hipparcos mission \citep{Hipparcos_1997,Hip2} established a foundational astrometric baseline, and the recently completed \Gaia mission \citep{Gaiamission,GaiaDR3} has extended these measurements to billions of stars with sub-milliarcsecond precision. By comparing proper motions from \Hipparcos and \Gaia, the Hipparcos-Gaia Catalog of Accelerations  \citep[HGCA,][]{2018Brandt_HGCA0,HGCA} identifies stars exhibiting changes in their proper motion over decades ($\approx$25-year baseline between the two missions), indicative of the existence of unseen companions. A number of recent papers have combined RV with the HGCA to further constrain masses and orbits \citep[e.g.][]{2021_116250b_Li,2022Feng_3Dselection_167substellar,2023_70849_mass,Unger_2023}.

The Gaia DR3 \citep{GaiaDR3} Non-Single Star (NSS) data product provides information of stars exhibiting astrometric signals indicative of one or more companions. This catalog is particularly useful for identifying and characterizing many-body systems across various separation scales. Among the key metrics of the \Gaia NSS release is the astrometric acceleration solutions, which fit astrometric acceleration terms in addition to parallax and proper motion. \Gaia NSS also includes two-body orbital solutions for stars where the data support a binary system orbit fit. These solutions provide estimates of orbital parameters including period, semi-major axis, eccentricity, and companion mass, offering critical insights into the dynamics of the systems \citep{gaia_2bd}. By incorporating both astrometric acceleration and two-body solutions, the Gaia DR3 NSS data product expands our understanding of stellar multiplicity and the distribution of stellar and substellar companions in the Milky Way.

In this paper, we conduct orbit fitting with the open-source Python package \orvara using a combination of RV and astrometry from the HGCA for 150 individual host stars, including 10 two-companion systems. We present \totalcompanion orbital solutions, including \stellarcomp stellar companions, \BDcomp Brown Dwarf companions, and \planetcomp planets. Our approach uses the open-source Python package \orvara to perform joint orbit fits that blend RV and astrometric constraints. We derive constraints on their companions' masses and orbits, and predict observable astrometric accelerations. Where the data exist, we compare our results with \Gaia DR3 astrometric acceleration results, and with \Gaia DR3 two-body solutions. By integrating multiple observational data streams and advanced modeling techniques, we move closer to disentangling the formation channels and evolutionary histories of substellar companions.
Though this sample is not intended for occurrence-rate calculations—due to systematic biases in the original target selection, heterogeneous observing strategies across data sources, and the $\chi^2$ and baseline cuts applied in our own selection process—our results nonetheless reveal the Brown Dwarf Desert in both the mass–separation and mass-ratio–separation regimes. Unbiased data would likely only deepen this desert, not diminish it.

The structure of this paper is as follows. In Section \ref{sec: Method}, we detail our methodology, including data selection and orbit-fitting procedures. In Section \ref{sec:Results}, we present orbital solutions, highlight new brown dwarf and planet~candidates, and discuss their implications. Section \ref{sec:Gaia} compares our findings with \Gaia DR3 data products, placing our results in the broader context of recent observational progress. Finally, in Sections \ref{sec:Dis} and \ref{sec:conclude}, we discuss the implications of this study for understanding substellar formation and evolution, and summarize our key conclusions.

\section{Methodology} \label{sec: Method}

The radial velocity (RV) technique is one of the earliest and most widely used methods for exoplanet~detection, applicable to both nearby and distant stellar systems. However, because it does not detect the companion directly, it has certain limitations, including the need for significant orbital phase coverage and the inability to determine orbital inclination. Consequently, RV data alone can only provide a minimum mass measurement for the companion. Astrometry, the precise measurement of a star's position and proper motion, captures the star's movement projected onto the sky plane. By combining RV data (line-of-sight motion) with astrometry (sky-plane-projected motion), we can probe the three-dimensional acceleration of the host star in response to its companion’s gravitational influence. This combined approach enables us to derive a complete orbital solution for the stellar system.

This section discusses the selection of targets from archival RV data and a catalog of astrometrically accelerating stars, followed by orbit fitting for the chosen systems. 

\subsection{Data}

We combine three types of data for our orbital fits: radial velocities (RVs), absolute astrometry of the host star, and, when available, relative astrometry between the companion and host star. In this section, we summarize each of these data sources in detail.

We use RV data from two instruments: the High-Resolution \'Echelle Spectrometer \citep[HIRES;][]{Hires} on the Keck I Telescope and the High-Accuracy Radial velocity Planetary Searcher \citep[HARPS;][]{Harps} on the ESO 3.6-meter telescope. The HARPS data were reduced by \citet{HARPSdata}, ensuring high precision and consistency in the RV measurements. The HIRES data were compiled by \citet{HIRESdata}, with many of these RVs collected as part of the long-running California Planet Search \citep[CPS;][]{Howard2010_CPS}, providing an extensive dataset of high-precision RVs for a large sample of stars. We also incorporated additional HIRES RVs from \citet{Rosenthal2021_CLS}. Where multiple iterations of HIRES RVs were available for a particular star, we used the RV data from \citet{Rosenthal2021_CLS}. The RV data usage for each system is recorded in the fifth column of Table \ref{tab:accel_all}. 

HIRES employs an iodine absorption cell for wavelength calibration and instrumental profile modeling during RV measurements. The iodine absorption lines, imprinted on the stellar spectra, provide a precise wavelength reference, enabling high-precision RV determinations \citep{Hires}. In 2004, HIRES underwent a major upgrade that included the installation of a new CCD detector and other optical improvements, significantly enhancing the instrument's precision. As a result, RV uncertainties improved from approximately 3 m s$^{-1}$ in the pre-upgrade era (1996--2004) to around 1 m s$^{-1}$ in the post-upgrade era (2004--present). \citet{HIRESdata} addresses the calibration of pre- and post-upgrade data, ensuring that both data sets can be combined for consistent RV analysis, so we treated the RV data as originating from one single instrument. To address differences in zero points introduced by the upgrade, we adopt the recalibrated data set provided by \citet{Rosenthal2021_CLS}, which explicitly accounts for separate RV zero points ($\gamma$) for pre- and post-upgrade data in their time-series modeling. We treated the pre- and post-upgrade data as taken from two separate instruments. This ensures consistency in RV measurements across both eras.

HARPS, in contrast, does not use an iodine cell but relies on a highly stable spectrograph design and a dual-fiber system for simultaneous wavelength calibration. One fiber observes the star, while the other records a calibration spectrum from a Th-Ar lamp or a Fabry-P\'erot etalon \citep{Harps}. In 2015, HARPS underwent an instrumental upgrade that affected its RV zero-point and overall instrumental behavior \citep{2015HARPS_upgrade}. To account for this change, we labeled the RV data collected before and after the upgrade as originating from two different instruments, with independent RV zero-point offsets included in the orbit fitting. This treatment ensures that the derived orbital parameters are not biased by the instrumental change.

We use absolute astrometry---precise stellar positions and proper motions---from the Hipparcos-Gaia Catalog of Accelerations \citep[HGCA;][]{HGCA}. The HGCA cross-calibrates \Hipparcos and \Gaia astrometry onto a common reference frame and applies an error budget to achieve statistically well-behaved errors for each proper motion measurement. The catalog provides three distinct proper motions: one from \Hipparcos, one from \Gaia, and a long-term proper motion derived from the difference between the \Hipparcos and \Gaia positions. For most stars, the \Hipparcos proper motion is by far the least precise; the \Gaia proper motion and the long-term proper motion often have comparable uncertainties.  Differences among these three proper motions in the HGCA enable the calculation of stellar acceleration in an inertial reference frame. Additionally, the catalog provides a $\chi^2$ value that quantifies the significance of any proper motion anomalies for each star, serving as an indicator of its acceleration.

Relative astrometry provides direct positional measurements of a companion with respect to its host star at specific epochs. The additional information is crucial for widely separated systems in which RVs cover only a small fraction of the orbit. When only an RV trend and astrometric acceleration are available, we can determine the mass-to-separation-squared ratio  ($\mathrm{M_2/R^2}$), but not the individual values for mass or separation. A very widely separated, massive companion can induce the same astrometric acceleration as a closer, less massive companion, leading to an inherent degeneracy. Incorporating relative astrometry introduces the extra constraint needed to independently derive both the companion’s mass and its separation from the host star, thereby breaking this degeneracy \citep{Brandt2019}. 

We use three principal sources of relative astrometric measurements. First, the Washington Double Star Catalog \citep[WDS;][]{WDS} provides relative positions and separations for visual double stars over multiple epochs. This catalog covers both historical and contemporary observations, enabling the detection and characterization of bound stellar companions from their measured position angles and separations. Second, we incorporate \Gaia astrometry when a companion is individually resolved by \Gaia, using the Gaia\,DR3 measurements at its fiducial epoch of J2016.0 \citep{GaiaDR3}. Finally, for previously imaged substellar companions not cataloged in WDS or resolved by \Gaia, we draw on literature astrometry, citing each data source in Table~\ref{tab:accel_all}. The usage of relative astrometry for each system is detailed in the sixth column of Table~\ref{tab:accel_all}.

By combining relative astrometry with RV and absolute astrometry, we can derive a 3D acceleration that constrains the orbit. The absence of any one of these data types leads to a loss of information: RV data alone can only provide the minimum mass of the companion ($\mathrm{M_2\sin{i}}$), while relative astrometry alone lacks information regarding the companion’s mass. 

We require RV datasets with a minimum number of epochs, a sufficient time baseline, and a significant astrometric acceleration signal from the HGCA (see Section \ref{sec:TargetSelection}). These criteria—as well as the heterogeneity of the archival RV and relative‑astrometry sources—introduce systematic biases inherited from their original target‑selection processes. Robust occurrence‑rate studies demand uniform target selection and well‑controlled observing strategies (e.g., \cite{Bowler2016,Howard2012_occ,Petigura2013_occ,Rosenthal2021_CLS}). Consequently, the sample presented here is not suitable for occurrence‑rate analyses.

\subsection{Target Selection}
\label{sec:TargetSelection}

We aim to derive companion masses and orbits for a systematic sample of main-sequence stars by cross-matching archival RV data from \HIRES \citep{Hires,HIRESdata,Rosenthal2021_CLS} and \HARPS \citep{Harps,HARPSdata} with the Hipparcos-Gaia Catalog of Accelerations \citep[HGCA;][]{HGCA}. 

To ensure that the astrometric data provide significant constraints on our orbit solutions, we select stars from the HGCA that exhibit an acceleration corresponding to $\chi^2 \geq 25$. This threshold ensures a substantial astrometric acceleration. It nearly (though not exactly) corresponds to a $5\sigma$ detection given the two degrees of freedom for the two orthogonal directions of astrometric motion. However, the $\chi^2$ cut preferentially selects systems with strong astrometric accelerations—i.e.\ those hosting stellar or high‐mass substellar companions—and therefore underrepresents lower‐mass, planet‐hosting systems; we return to this completeness bias in Section \ref{sec:data_bias}.

Among the archival RV data, we restrict our sample to stars monitored for more than 10 years, a baseline chosen to be comparable to the approximately 25-year baseline covered by HGCA data (spanning the interval between the \Hipparcos and \Gaia missions). We also require each target to have at least ten radial velocity (RV) observations. Together, these criteria were chosen to identify stars likely to yield robust orbital solutions, showing significant astrometric accelerations and sufficiently extended RV monitoring. However, the actual level of orbital constraint depends on the distribution and phase coverage of each RV dataset, so we further examine every RV time series before finalizing the target selection. To further refine our sample, we inspect the RV time series for each star and classify each of them into exactly one of three mutually exclusive categories: curved RV, straight-line RV, and packed RV. Each category requires a different treatment in terms of astrometric and RV data, leading to differences in the availability and quality of orbital solutions. 

\begin{figure}
    \begin{center}
        \includegraphics[width=7.3cm]{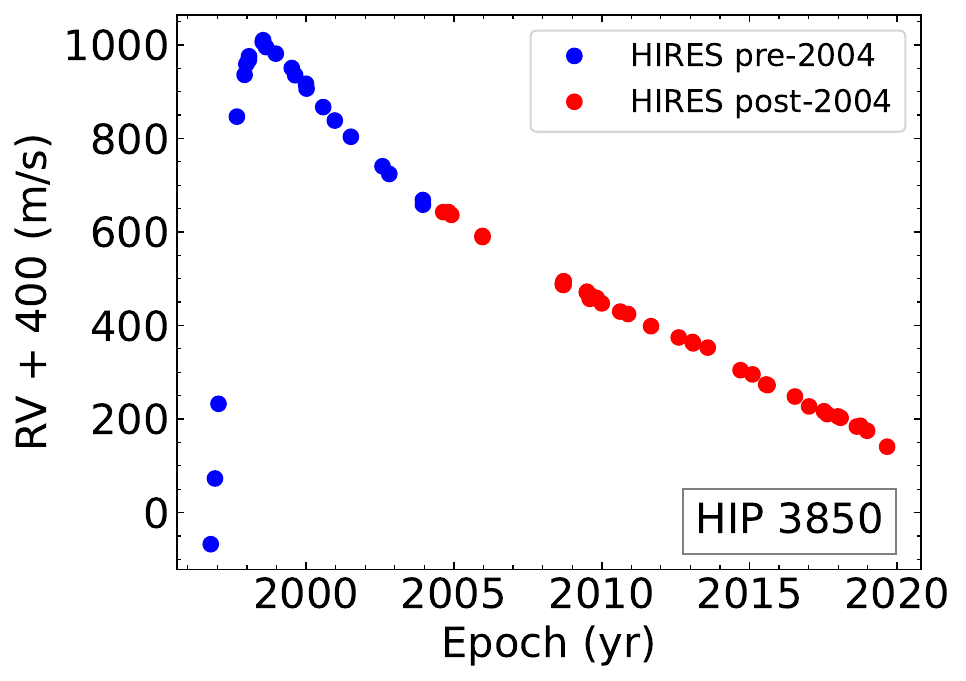}
        \hspace*{0.15cm}\includegraphics
        [width=6.7cm]{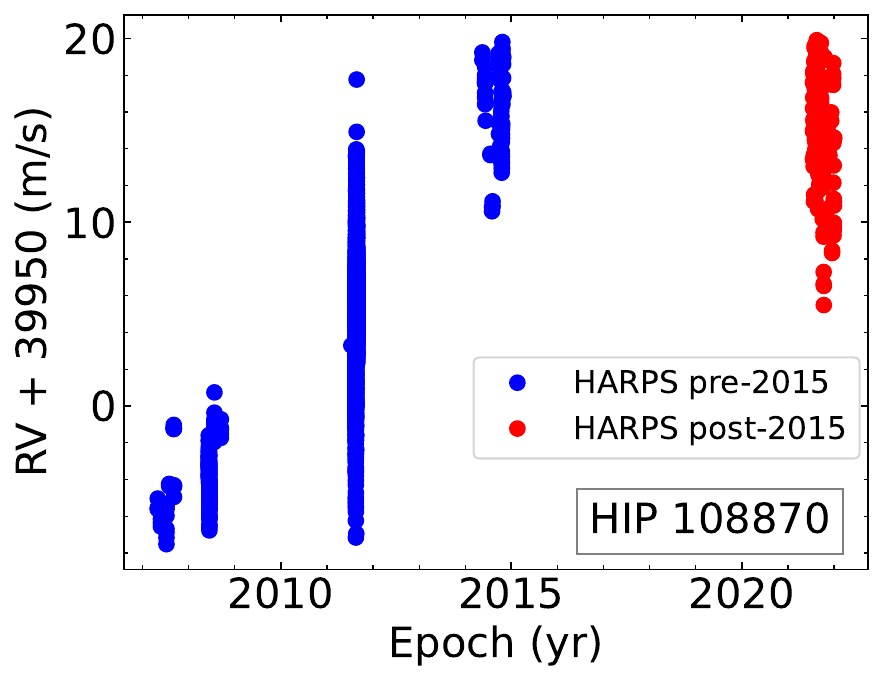}
        \includegraphics[width=7cm]{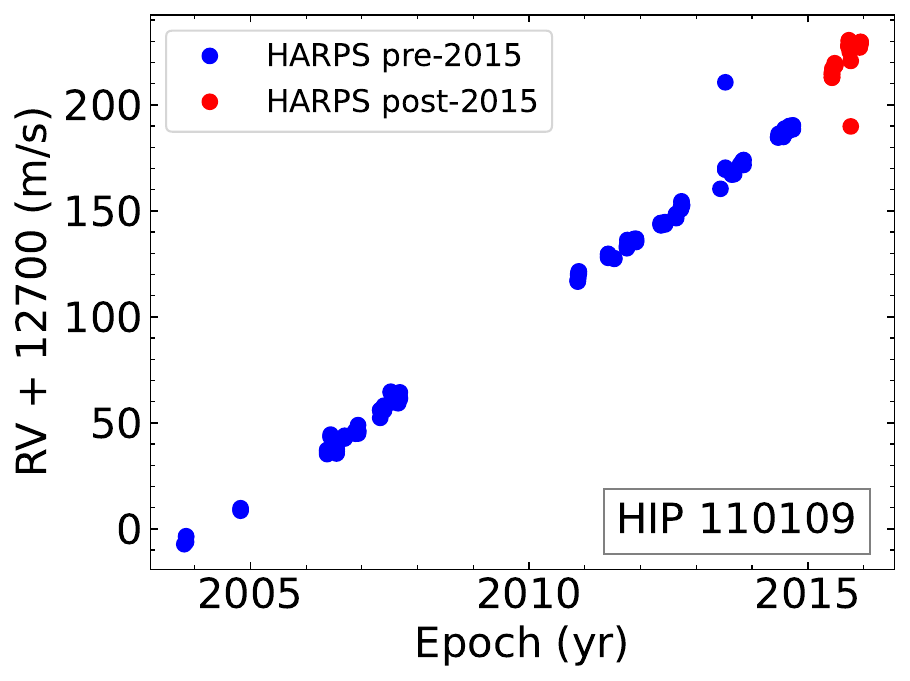}
    \end{center}
    \caption{(Top) RV time series of HIP 3850 (=HD 4747), showing substantial phase coverage of the orbit. These RVs provide enough information to derive a full orbit solution when combined with astrometric data. (Middle) RV time series of HIP 108870, illustrating packed RV observations. The measurements are closely clustered in time, reducing the number of effectively distinct observational epochs and limiting the available orbital information. (Bottom) RV time series of HIP 110109, showing a straight-line trend. This trend suggests the absence of detectable curvature, making it challenging to derive a complete orbital solution without the aid of relative astrometry.}
    \label{fig:RV_example}
\end{figure}

\textbf{Curved RV}: Stars with RV time series exhibiting clear curvature are classified under the curved RV category. These stars have significant RV orbital phase coverage, often providing enough information to solve for companion's minimum mass ($\mathrm{M_2\sin{i}}$). Substantial RV phase coverage of the orbit, when combined with absolute astrometry from the HGCA, constrains the companion's mass. An example of a star with such an RV profile---HIP 3850 (=HD 4747)---is shown in the top panel of Figure~\ref{fig:RV_example}.

\textbf{Packed RV:} We classify a star's RV data as \textit{packed} if multiple observations are tightly grouped over a short time span, yielding only a few distinct epochs ($\leq 5$ clusters). The middle panel of Figure~\ref{fig:RV_example} illustrates such a dataset with four clusters. Because fewer epochs limit our ability to track the star's reflex motion, these packed RV data offer relatively low orbital information content. To derive a full orbital solution, additional constraints (such as relative astrometry) are often required.

\textbf{Straight-Line RV}: Stars with RV measurements showing no curvature (insignificant orbital phase coverage) are classified as having straight-line RVs. They are also often referred to as ``trend'' stars \citep{2012Creep_trendStar}.  For these systems, a single acceleration measurement is insufficient to separately constrain companion mass and separation; only the combined parameter $\mathrm{M_2/R^2}$ can be determined. The bottom panel of Figure~\ref{fig:RV_example} provides an example of a star with a straight-line RV time series. In the absence of detectable RV curvature, deriving a full orbital solution requires additional relative astrometric measurements to complement the absolute astrometry from the HGCA.

Out of the 115,336 stars in the HGCA, 29,371 exhibit accelerations at or above the $\chi^2=25$ level. Within this subset, archival RV data are available for 2,465 stars from HARPS \citep{Harps, HARPSdata} and 1,548 stars from HIRES \citep{Hires, HIRESdata, Rosenthal2021_CLS}.  Cross-matching these datasets yields 106 stars from HARPS and 130 stars from HIRES that satisfy our selection criteria for RV baseline and number of observations. Of these, 14 stars are present in both the HIRES and HARPS datasets, resulting in a total of 222 unique targets.

We exclude 10 packed-RV targets and 28 straight-line-RV targets that lack archival relative astrometry, as these datasets do not provide sufficient information for a full orbital solution. An additional 40 targets are excluded for various specific reasons, summarized below. First, we remove systems in which the fitted orbital period of the longest‑period companion falls below $\approx7$ years (roughly 2.5 times the \Gaia DR3 baseline), since the long-baseline astrometry effectively averages out such short‑period signals. For example, HIP~98714 is excluded because it is a two-body system with a companion on a $\approx2$\,year orbit, too short to be constrained by our long-baseline astrometry.

We make an exception to the minimum orbital period for three-body systems where the inner companion’s short-period orbit is well-constrained by RV data alone. We retain such systems and report the inner companion’s minimum mass ($M_2 \sin i$), while employing long-term RV plus astrometry to constrain the outer companion. A single system falls outside even this exception: HIP~106006, which has a short-period inner companion well-constrained by RV but an outer companion that produces only a straight-line RV signal, rendering a two-companion orbital fit poorly constrained.

Additionally, we reject poor‐quality datasets with large uncertainties or inconsistencies between RV and astrometry (inconsistencies may arise from stellar activity–induced signals, instrument zero‑point drifts, or multiple companions; see Section~\ref{sec:data_bias}). We also omit young, highly active stars whose intrinsic variability can mask companion signals, as well as resolved binary stars with separate \Hipparcos identifications. Although we include three‐body fits for the most promising systems, we discard other multiple‐companion cases that exceed \orvara’s computational limits.

Finally, to avoid redundancy, we do not revisit several systems that have already been analyzed in detail elsewhere, referring to the pertinent publications when applicable. We do include these systems in Figure \ref{fig:general} and briefly list them here.   
HIP~26394 ($\pi$\,Men) hosts confirmed planets~b, c, and~d. There is no published semi-major axis of planet~d, therefore it is not included in Figure \ref{fig:general} \citep{26394b,HIP26394}. HIP~79248 (14\,Her) hosts a confirmed short-period planet~b and a more widely separated planet~c \citep{14Her2021,hip79248}. HIP~85647 (GJ~676) hosts confirmed short-period planets~Ad, Ae, and~Ab, as well as a more distant planet~Ac and a stellar companion~B \citep{hip85647}. HIP~29295 (GJ~229) hosts confirmed substellar companions~Ba and~Bb, and short-period candidates~c and~d. \citep{29295-1-Nature,2021AJ_GMBRANDT_sixmasses}. HIP~114189 (HR~8799) hosts confirmed planets~b, c, d, and~e \citep{HR8799/HIP114189}. HIP~14501 (HD~19467) hosts a benchmark brown dwarf HD~19467~B \citep{2024_HD19467B}.

After these exclusions, our final target list consists of \totalstar stars. A comprehensive record of all excluded targets, alongside the relevant exclusion criteria, is provided in Appendix~\ref{sec:App1}. The usage of RV and relative astrometry data for each of the remaining systems is documented in Appendix~\ref{sec:App2}, specifically in the two rightmost columns of Table~\ref{tab:accel_all}.

\subsection{Orbit Fit} \label{sec:orbit-fit}

In this work, we perform orbital analyses for \totalstar systems, which include \startwobd two-body orbit fits and \starthreebd three-body orbit fits. The data used for these analyses and the criteria applied to target selection are described in the previous section. The detailed data usage, apart from absolute astrometry from the HGCA, appears in the fifth and sixth columns of Table~\ref{tab:accel_all} in Appendix~\ref{sec:App2}.

We use the open-source Bayesian orbit-fitting package \orvara \citep{orvara} to determine the orbital parameters for each selected system. \orvara can incorporate any combination of three data types--absolute astrometry, relative astrometry, and radial velocities--to accurately fit orbits of stars and their faint companions. When the RV time series traces a substantial fraction of the orbit, even if it does not cover the full orbital period, \orvara can still constrain the companion’s mass and provide orbital solutions by integrating absolute astrometry data from HGCA. If the RV data have limited orbital phase coverage, relative astrometry may be required to break the inclination degeneracy.

\orvara uses parallel-tempered Markov Chain Monte Carlo (PT-MCMC) to sample the posterior distributions of orbital parameters. In each step of the fitting process, absolute astrometry data are processed using \htof \citep{htof}. The simulation involves running multiple MCMC chains in parallel at different temperatures. By periodically exchanging information among these parallel chains, the efficiency of the MCMC process is enhanced as it allows chains to make occasional large jumps in parameter space through position swaps, improving convergence and reducing the chance of getting stuck in local minima.

Our MCMC chains use a minimum of 20 temperature levels and employ 100 walkers per temperature level, running for at least 200,000 steps per walker to ensure thorough exploration of the parameter space. We use the coldest chain for inference and discard the initial half of each chain as a burn-in period when calculating the final results. We apply Gaussian priors on the host star masses, which are derived from \citet{mpri,mpri1, mpri2, mpri3, mpri4}, and impose flat priors on eccentricity, argument of periastron, mean longitude, and the longitude of ascending node. For companion mass, semi-major axis, and RV jitter, we used log-flat priors. Additionally, we apply a geometric prior on the inclination.\orvara produces a FITS file output for each orbit fit, containing the full set of orbital parameters sampled at each step of the MCMC chains. To assess chain convergence and the quality of the fit, we use diagnostic plots generated by \orvara, ensuring that each parameter's best-step $\chi^2$ value and the distribution of samples for each parameter indicates satisfactory convergence. We verify each orbital solution by reviewing \orvara's diagnostic plots—corner diagrams of the Keplerian parameters, astrometric prediction plots, simulated RV orbits, proper‐motion plots in right ascension and declination, and relative astrometry (separation and position angle)—to ensure that the fitted model accurately describes the data without systematic residuals.

\section{Results}\label{sec:Results}

We select \totalstar stars from the available RV and astrometry data, requiring that each star exhibit significant Hipparcos–Gaia acceleration and either sufficient RV phase coverage or less extensive RV coverage supplemented by archival relative astrometry. Following the methodology described in Section~\ref{sec: Method}, we perform orbit fitting and identify \totalcompanion companions around these \totalstar stars. Among them, \starthreebd systems host two companions, while the remaining \startwobd systems host a single companion.

For the purposes of listing and visualizing our results, we adopt hydrogen- and deuterium- burning limits of 80\,\Mjup and 13\,\Mjup as the upper mass boundaries for brown dwarf and planetary companions, respectively. Based on the median of their mass posteriors, we identify \planetcomp planetary companions, \BDcomp brown dwarf companions, and \stellarcomp stellar companions. However, many classifications remain tentative where the mass posterior overlaps the 13 or 80\,\Mjup boundaries. In this section, we present the substellar companions' results, including their MCMC-derived parameters in Table~\ref{tab:substellar}, and discuss their masses, orbits, and implications for our understanding of such systems. A full set of MCMC results for stellar companions is provided in Appendix~\ref{sec:app_stellar}, Table~\ref{tab:stellar_long}.

Astrometry from the HGCA, with an approximate 25-year baseline (the time difference between the \Hipparcos and \Gaia missions), averages out much of the signal from a companion with a shorter orbital period.  As the orbital period approaches the $\approx$3-4 year duration of either \Hipparcos or \Gaia itself, the signal can be lost entirely from a five-parameter astrometric fit.  
Consequently planets within a few AU typically have only $\mathrm{m_2 \sin i}$ (minimum mass) constraints from RV. These close-in planetary companions are shown as red upward-pointing arrows in Figure~\ref{fig:general}, and their orbital solutions are summarized in section~2 of Table~\ref{tab:substellar}.

\begin{figure*}
    \centering
    \includegraphics[width=18cm]{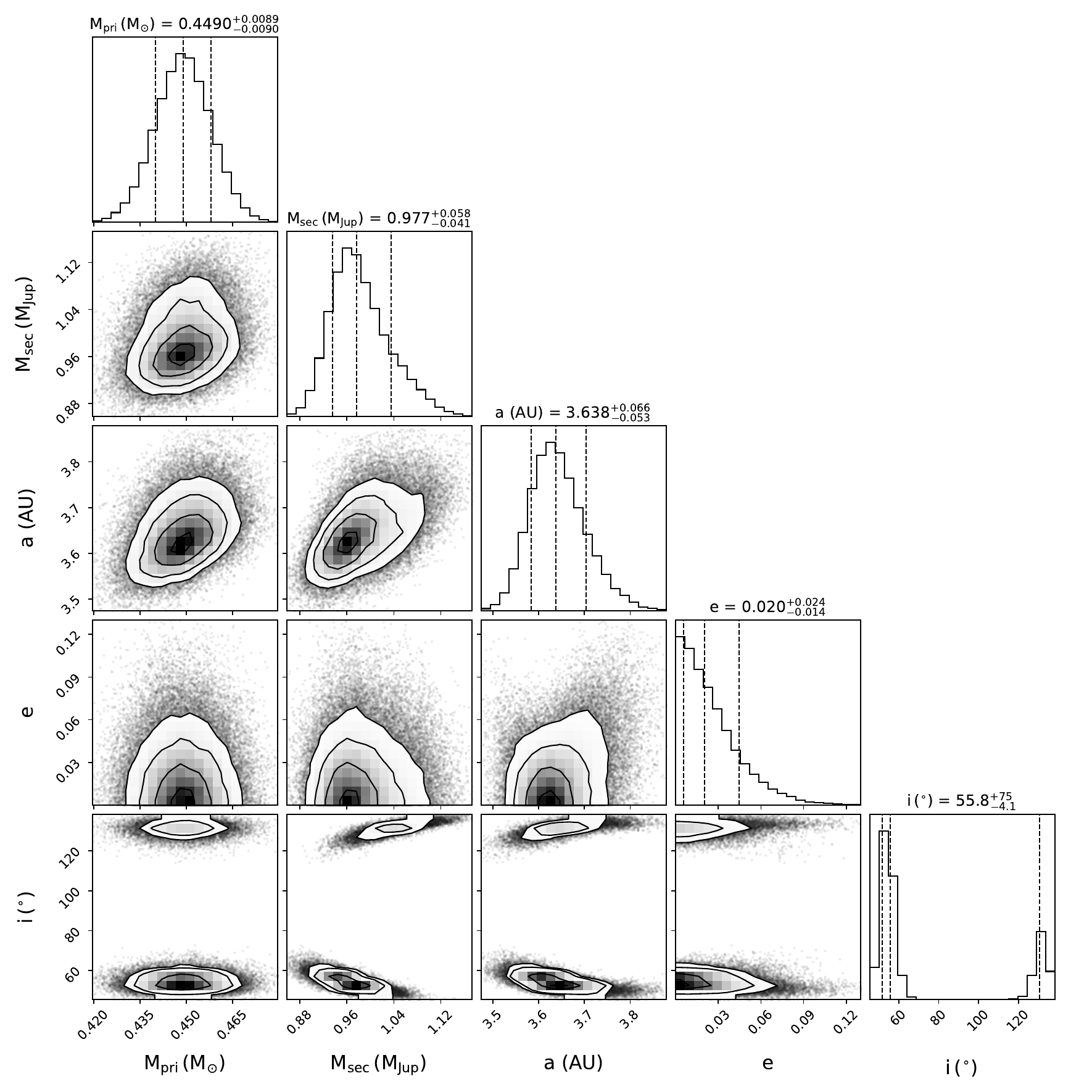}
    \caption{Corner plot of the orbital elements of HIP~106440~b, illustrating the posterior distributions and correlations among the parameters and orbital inclination. In the lower-right 1D histograms, the vertical dashed lines around the central dashed line indicate the 16\% and 84\% quantiles. In the 2D histograms, the contours correspond to the 1$\sigma$, 2$\sigma$, and 3$\sigma$-equivalent uncertainties. The inclination distribution is bimodal, with peaks near $55^\circ$ and $125^\circ$, symmetric about $90^\circ$. This reflects the degeneracy in distinguishing between prograde and retrograde orbits.}
    \label{fig:split_inc}
\end{figure*}

Astrometry and RV commonly present an inclination degeneracy, with an inability to distinguish prograde from retrograde orbits.  As an example, Figure~\ref{fig:split_inc} presents the corner plot for HIP~106440~b, where the posterior of orbital inclination has two peaks symmetric around $90^\circ$ (at approximately $65^\circ$ and $115^\circ$). These two modes are not reflected in the posterior distributions of the primary mass ($M_{\rm pri}$), secondary mass ($M_{\rm sec}$), semi-major axis ($a$), or eccentricity ($e$). Such degeneracies commonly arise in systems where the astrometric data are effectively limited to two proper motion measurements in the HGCA catalog \citep{HGCA} (when the \Hipparcos data are too imprecise to significantly impact the posterior). \citet{HD28185} offers another example of the the inclination bimodality phenomenon with detailed discussion. Consequently, the combined constraints from \Gaia and \Hipparcos--\Gaia proper motions, along with RV data, may not suffice to distinguish prograde from retrograde orbits \citep[e.g.,][]{2024An_HATP11}. Future \Gaia data releases, providing intermediate astrometric measurements, are expected to help resolve such degeneracies \citep{Gaiamission}.

For orbits where the posterior is multi-modal, i.e., where \orvara\ cannot distinguish between prograde ($i<90^\circ$) and retrograde ($i>90^\circ$) solutions, we partition the result chain into prograde and retrograde modes. We then compute two distinct sets of orbital parameters and corresponding astrometric accelerations. The retrograde-mode results for substellar companions 
are provided in section~3 of Table~\ref{tab:substellar}, and for stellar companions in section~2 of Table~\ref{tab:stellar_long}).

\subsection{Inclination‐dependent Selection Biases} \label{bias_inc_dep}

We now turn to a validation of our orbit fits based on the posterior distribution of the orbital inclination.  An unbiased sample should result in a geometric inclination posterior, matching the prior.  If there are large selection effects in our sample or systematically mis-estimated errors in one or more data sources, the posterior could depart from the geometric prior.

Each technique that contributes to our orbital fits responds to a different trigonometric projection of the orbit.  As a result, any survey that relies on a single data type is exposed to a selection bias toward extreme inclination systems. Astrometric accelerations from the HGCA \citep{HGCA} or from the Gaia DR3 scale with $\mathrm{\cos{i}}$, while RV semi-amplitude scales with $\mathrm{\sin{i}}$ \citep{RadioVelocity}. The inclination bias is mitigated when orthogonal observables are modeled together. A joint likelihood that must satisfy both $\mathrm{\cos{i}}$ and $\mathrm{\sin{i}}$ scalings constrains the inclination near its true value, reproducing the isotropic prior unless the data carry genuine information about orientation \citep{2014_McArthur_inc_bias}.

\begin{figure} 
    \centering 
    \includegraphics[width=0.5\textwidth]{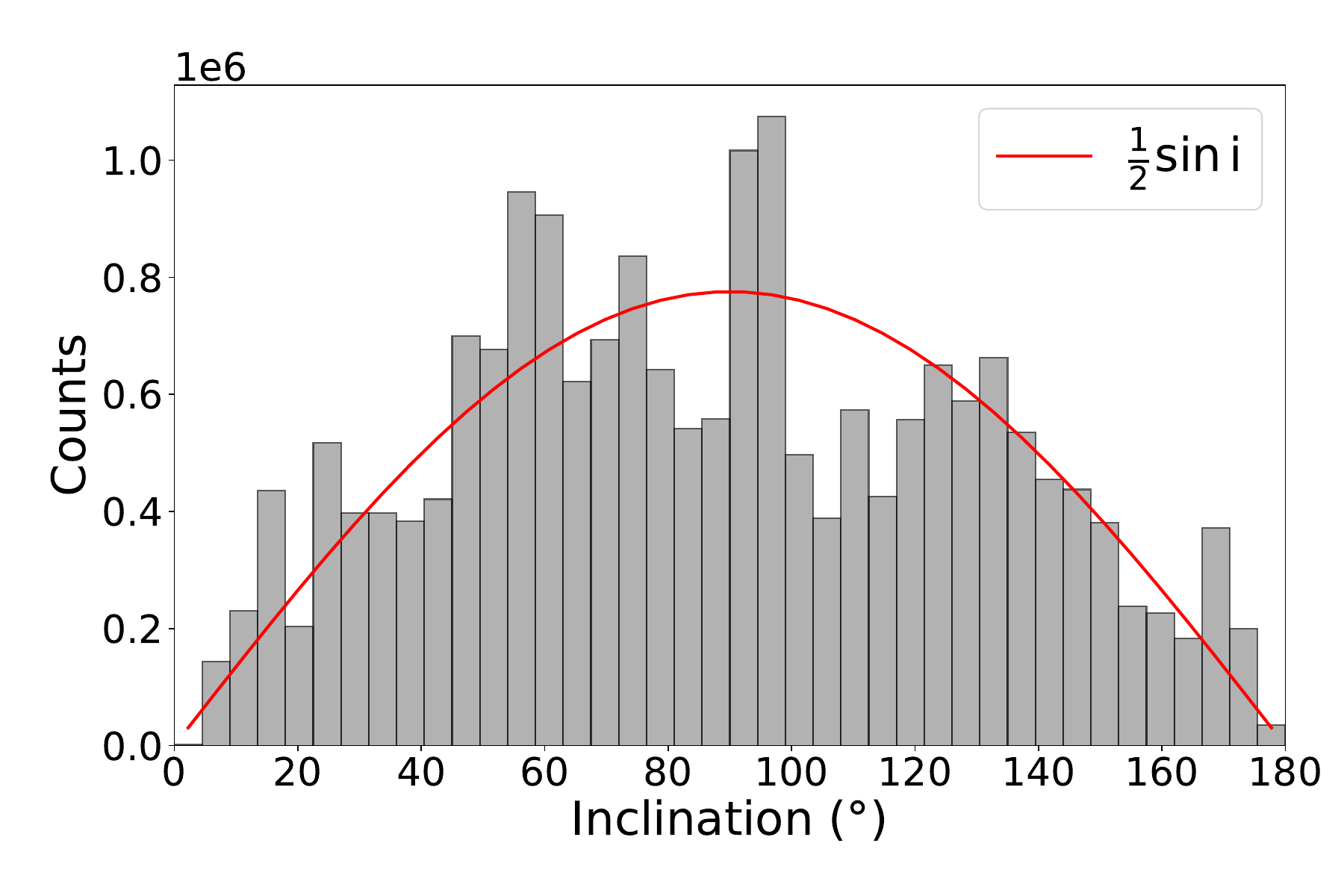}
    \caption{Combined posterior distribution of orbital inclinations from every \orvara chain in our sample. Grey bars show the histogram of $\sim1.6\times10^{6}$ MCMC samples (both \texttt{inc0} and \texttt{inc1} if there are two companion fitted), binned in $4.5^{\circ}$ intervals. The red curve is the theoretical probability density $(1/2)\sin i$—scaled to the same total number of samples and bin width—that describes an isotropic ensemble of orbital inclinations. This figure confirms that our joint RV + absolute/relative-astrometry selection and fitting introduces little systematic bias toward extreme inclination orbits.} \label{fig:inc_all}
\end{figure} 

Figure \ref{fig:inc_all} illustrates this behavior for our sample. After merging all \orvara chains we find that the combined inclination posterior closely follows the isotropic prior, $(1/2)\sin i$, demonstrating that our joint RV–astrometry approach mitigates the extreme‐inclination biases inherent to single‐data‐type surveys. Furthermore, our inclination distribution provides a better match to the prior than previous studies (e.g., \citealt{2022Feng_3Dselection_167substellar,hip79248}), underscoring the robustness of our methodology.

\input{table_substellar}
\subsection{Substellar Demographics}

\begin{figure*}     
    \centering
    \includegraphics[width=16cm]{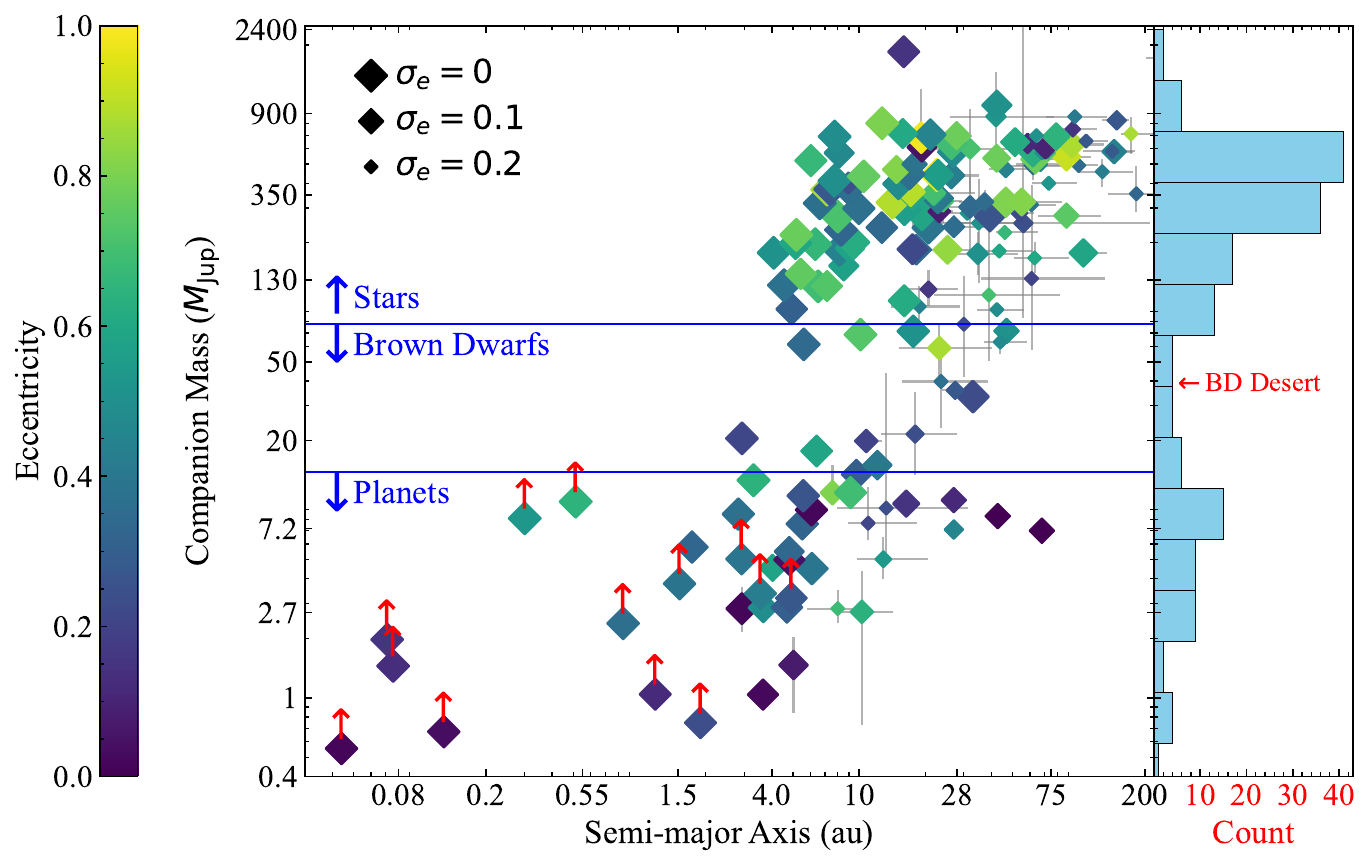}
    \includegraphics[width=16cm]{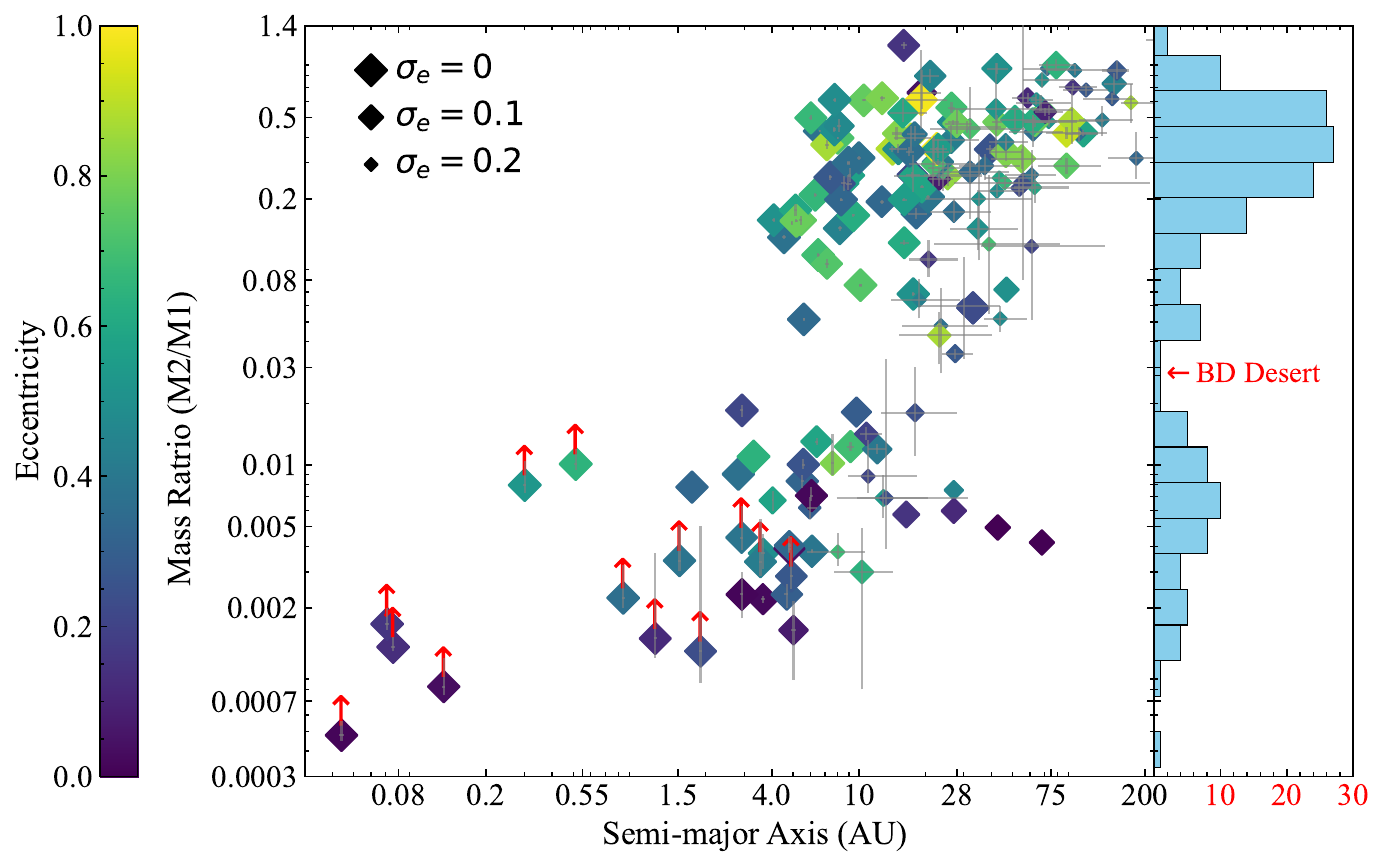}
    \caption{A comprehensive collection of orbital solutions from this work; the results from previous detailed studies that would otherwise have been in our sample are also included for completeness.  Very close‑in, low‑mass planets lie outside the axis limits. Top: companion masses and semi‑major axes, color‑coded by orbital eccentricity. Marker size indicates the uncertainty in eccentricity ($\sigma_e$), with larger markers corresponding to better‑constrained eccentricities. Blue lines separate stellar, brown‑dwarf, and planetary-mass companions. Red upward‑pointing arrows denote companions with minimum‑mass constraints ($\mathrm{M_2\sin i}$).  The histogram of companion masses is shown to the right, with the the brown dwarf desert highlighted around 25--65\,$\mathrm{M_{Jup}}$. (Bottom) Same as the top panel, but for mass ratio ($\mathrm{M_2/M_1}$). The brown dwarf desert appears to be slightly more pronounced. \label{fig:general}}
\end{figure*} 

Figure \ref{fig:general} presents the full collection of \totalcompanion orbital solutions derived in this work, together with a handful of HARPS and HIRES targets with detailed literature results that would otherwise appear in our orbit fitting sample, in two different but complementary ways. The top and bottom panels both show the inferred semi-major axis (x-axis) for each companion plotted against its absolute mass (top panel) or its companion-to-host-star mass ratio (bottom panel). Blue lines in the upper panel separate stellar, brown dwarf, and planetary companions, with adopted boundaries at $\approx$80\,\Mjup and $\approx$13\,\Mjup. In both panels, the marker color indicates the derived orbital eccentricity, while the marker size reflects the eccentricity uncertainty (larger markers correspond to smaller $\sigma_e$). Light gray error bars show the inferred uncertainties in mass and semi-major axis. The histograms on the right depict the companion distributions by mass (top) and by mass ratio (bottom).

In both panels, low-mass or low–mass-ratio companions cluster preferentially at close-in orbits and exhibit relatively small eccentricities, while more massive or higher–mass-ratio companions extend to a wider range of separations and eccentricities. The lack of planets at wide separations is partly a selection effect, as wide-separation planets induce little acceleration on their host stars. 

The lower-mass, lower-eccentricity companions in Figure~\ref{fig:general} are separated from the higher-mass, higher-eccentricity stellar companions by a sparsely populated region.  A similar scarcity has long been noted at smaller separations, where it was termed the brown dwarf desert \citep[BD Desert,][]{2006Grether_BDdesert_quantify} and identified as an artifact of their distinct formation and dynamical evolution mechanisms.  
The upper panel of Figure~\ref{fig:general} illustrates this phenomenon at wider separations, showing the companion mass as a function of separation (in AU), with points color-coded by eccentricity. The right-hand horizontal histogram presents the companion occurrence rate as a function of mass, revealing a noticeable gap in the mass distribution between $\approx$25--65 $\mathrm{M_{Jup}}$. This result aligns with the findings of \citet{BD_mass_and_ecc}, who identified a boundary separating low- and high-mass brown dwarfs at $\approx$45 $\mathrm{M_{Jup}}$. 

The lower panel of Figure~\ref{fig:general} complements this analysis by presenting the results in terms of separation (in AU) and mass ratio (companion/host star). The color-coding and marker size schemes are consistent with those in the upper panel. The scarcity of companions within the $\approx$25--65 $\mathrm{M_{Jup}}$ range is even more evident in mass-ratio space as a deficit around 0.03, reinforcing the nature of the Brown Dwarf Desert. Taken together, the two panels of Figure \ref{fig:general} confirm the existence of the Brown Dwarf Desert at wider separations up to $\approx$10\,AU and suggest that brown dwarf companions near $\approx$40\,$\mathrm{M_{Jup}}$ (or a mass ratio of $\approx$0.03) are uncommon, potentially reflecting distinct formation mechanisms and dynamical evolution compared to both planetary and stellar companions.

Although the histograms in both panels of Figure \ref{fig:general} clearly exhibit the BD desert, quantitative occurrence‐rate measurements demand corrections for detection sensitivity and survey completeness. Because BDs induce stronger RV and astrometric signals than planets at similar separations, a proper completeness correction would deepen the observed deficit—in other words, the true occurrence‐rate gap at $\sim$25–65 \Mjup is likely more pronounced than shown by this work. We defer a full completeness analysis to future study.

There are two notable companions situated in the Brown Dwarf Desert in our companion-mass presentation (upper panel). First is the previously unidentified substellar companion HIP~58289\,B. Because its parameters are poorly constrained (mass $39.8^{+20.9}_{-16.8}$\,\Mjup, semi-major axis $23.6^{+15.0}_{-7.9}$\,AU), we cannot conclusively determine whether it actually lies in the brown dwarf regime or within the Brown Dwarf Desert. Additional relative astrometry is needed to refine its orbit and mass. Second, the binary brown dwarf pair GJ~229\,Ba/Bb also appears to occupy the Brown Dwarf Desert if we consider absolute masses (upper panel), though not in the mass-ratio representation (lower panel), making it a special case. Consequently, the only unambiguous occupant of the brown dwarf desert in our sample is GJ~758\,B (HIP~95319\,B).

The higher eccentricities observed in more massive companions are likely indicative of their formation through mechanisms such as molecular cloud fragmentation and gravitational interactions in dense stellar environments \citep{2012Bates_stellar_formation, 2011BO_stellar_BD_formation}. These star-like formation processes often lead to dynamically complex systems, as they lack the dissipative forces that act within protoplanetary disks. Lower-mass companions are thought to assemble within protoplanetary disks \citep{1996Pollack_substallar_formation}, where gas drag and disk–planet interactions efficiently damp orbital eccentricity. After disk dispersal, objects that migrate inward can experience additional tidal dissipation inside the host star; the circularisation timescale scales steeply with both orbital separation and companion mass, becoming significant inside $\approx 0.1$ au \citep{2009Tidal_Evo, 2018Arzamasskiy_disk_circulation}. This tidal regime accounts for the handful of close-in, low-mass planets—drawn from earlier detailed studies—that lie beyond the bounds of Figure \ref{fig:general}.

\subsection{New Companions and Updated Results}

In this section, we summarize our findings for individual systems hosting potential substellar companions, emphasizing cases where our analysis deviates from previously published results (e.g., first-time true mass estimates or new discoveries). We also highlight one possible discoveries: HIP 58289 B, which may be a brown dwarf with its first published mass and orbit constraints in this work.

\subsubsection{HIP 9683 b and c}
HIP 9683 (HD 12661) is a G6 main-sequence star, with a fitted mass of $1.16\pm0.06$ \Msun, located at a distance of 37.14~pc. \citet{2009_HIP9683bc} identified two planets around this star: planet~b and planet~c, with minimum masses and semimajor axes of 2.3 \Mjup at 0.83 AU and 1.57 \Mjup at 2.56 AU, respectively. Our analysis updates the minimum mass of inner planet~b to $2.38\pm0.08$ \Mjup at $0.84\pm0.01$ AU, and the true mass of outer planet~c to $2.83^{+0.82}_{-0.67}$ \Mjup at $2.93^{+0.04}_{-0.05}$ AU. 

\subsubsection{HIP 17960 b}
HIP~17960 (HD~24040) is a G-type star, with a fitted mass of $1.22\pm0.06$ \Msun, located at a distance of 46.51~pc. \citet{2012_24040b} reported the discovery of planet~b, with an estimated minimum mass of $4.01\pm0.49$ \Mjup at $4.92\pm0.38$~AU, and suggested the existence of a third inner bound companion. Subsequently, \citet{Rosenthal2021_CLS} 
estimated the inner planet~c to have a minimum mass of $0.201\pm0.027$ \Mjup, with a close-in orbit of 1.3~AU and a period of 1.41~years. 

In this work, we refine the true mass of the outer planet~b to $5.00^{+0.36}_{-0.26}$ \Mjup at a semi-major axis of $4.82\pm0.08$~AU. The innermost planet~c, owing to its short period and close-in orbit, is not detectable via long-baseline absolute astrometry and is therefore excluded from our analysis.

\subsubsection{HIP 40687 b and c}
HIP 40687 (HD 68988) is a G-type star, with a fitted mass of $1.20\pm0.02$ \Msun, located at a distance of 58~pc. \citet{2002_40687b_discover} reported the discovery of planet b from the Keck precision Doppler velocity survey, with an estimated minimum mass of 1.9 \Mjup at 0.071 AU. \citet{2006Bulter_nearby_exoplanets} refined the 
minimum mass of planet~b to $1.86\pm0.16$ \Mjup at a semi-major axis of $0.070\pm0.004$ AU. \citet{2007_40687c} proposed an additional, outer 
planet c with an incomplete orbit, estimating its minimum mass at 11--20~\Mjup and a separation of 5-7 AU. Subsequently, \citet{Rosenthal2021_CLS} refined the outer planet's parameters to $15.0^{+2.8}_{-1.5}$ \Mjup at $13.2^{+5.3}_{-2.0}$ AU.

In this work, we update the minimum mass ($\mathrm{m_b\sin{i_b}}$) of the inner planet b to $1.97\pm0.07$ \Mjup at a semi-major axis of $0.071\pm0.001$ AU. We also derive a true mass for the outer planet c of $15.03_{-0.75}^{+0.90}$ \Mjup at $12.1\pm1.1$ AU.

\subsubsection{HIP 50653 b}
HIP 50653 (HD 89839) is an F-type star with a fitted mass of $1.3\pm0.06$ \Msun, located at a distance of 17.65~pc. \citet{2011_50653b_discover} and \citet{2023_50653b} derived the minimum mass of planet b to be $3.9\pm0.4$ \Mjup at $6.8_{-2.4}^{+3.3}$ AU and $3.81\pm0.05$ \Mjup at $4.76\pm0.04$ AU, respectively.

In this work, we derive the true mass of planet b to be $5.50_{-0.48}^{+0.63}$ \Mjup at a semi-major axis of $4.81\pm0.08$ AU.

\subsubsection{HIP 58289 B}
HIP 58289 (BD--06 3481) is a K-type star with a fitted mass of $0.80 \pm 0.04$ \Msun, located at a distance of 38.78~pc. We identify a potential substellar companion, HIP~58289~B, with a mass of $39.8_{-16.8}^{+20.9}$ \Mjup and a semi-major axis of  $23.6_{-7.9}^{+15.0}$ AU. Unfortunately, these constraints have large uncertainties, and we cannot definitively classify the companion as a brown dwarf. As such, it is not regarded as a secure substellar object (i.e., it is not a secure brown dwarf in the brown dwarf desert). Additional relative astrometry or RV data will be required to better characterize its nature and orbital parameters.

\subsubsection{HIP 95467 c and d}
HIP~95467 (HD~181433) is a K-type subgiant at 26.15~pc, with a fitted host-star mass of $0.87\pm0.04$\,\Msun. \citet{2009_95467bcd_discover} initially reported three planets (b, c, d) with minimum masses of 0.024, 0.64, and 0.54\,\Mjup, respectively. Later analyses refined these values \citep{2011_95467bcd_update,2019_95467bcd_latest}, culminating in estimates of $0.0223\pm0.0003$\,\Mjup\ at $0.0801\pm0.0001$\,AU for planet~b, $0.674\pm0.003$\,\Mjup\ at $1.819\pm0.001$\,AU for planet~c, and $0.612\pm0.004$\,\Mjup\ at $6.60\pm0.22$\,AU for planet~d \citep{2019_95467bcd_latest}.

We exclude the innermost planet~b because its short period renders it undetectable 
with our long-baseline astrometry. Focusing on the two outer companions, we find 
a minimum mass for planet~c of $0.75\pm0.03$\,\Mjup\ at 
$1.90\pm0.03$\,AU, which is consistent with previous studies, and a 
dynamical mass for planet~d of $2.72^{+1.68}_{-1.99}$\,\Mjup\ at 
$10.3^{+3.9}_{-2.6}$\,AU, representing an updated true mass for planet~d.

\subsubsection{HIP 97336 b and c}
HIP 97336 (HD 187123) is a G3 main-sequence star with a fitted mass of $1.19^{+0.06}_{-0.11}$ \Msun, located at a distance of 50~pc. \citet{1998_97336b_discover} reported the discovery of planet b, with a minimum mass of 0.52 \Mjup at 0.042 AU, and \citet{2009_HIP9683bc} reported the discovery of an outer planet c, with a minimum mass of $1.99\pm0.25$ \Mjup at $4.89\pm0.53$ AU. \citet{Rosenthal2021_CLS} updated the minimum masses of planet b and planet c to $0.501\pm0.016$ \Mjup and $1.71\pm0.06$ \Mjup, with semi-major axes 
of $0.041\pm0.001$ AU and $4.43\pm0.07$ AU, respectively.

Our analysis updates the minimum mass of the inner planet b to $0.56^{+0.02}_{-0.03}$ \Mjup, with a semi-major axis of $0.044\pm0.001$ AU, and derives the true mass of the outer planet c to $2.87^{+0.34}_{-0.42}$ \Mjup, with a semi-major axis of $4.69^{+0.08}_{-0.13}$ AU.

\subsection{Comparison with Previous Work}

In this section, we compare our substellar orbital fits with earlier analyses; in most cases, they agree well, partly due to the substantial overlap in datasets used. We include these objects in our sample to provide a more systematic selection of nearby stars. We also highlight instances where our results diverge from previous findings, discussing possible reasons for these discrepancies.

\subsubsection{HIP 3850 B}
HIP 3850 (HD 4747) is a G8 main sequence star located 18.8~pc away, with a fitted mass of $0.90\pm0.05$ \Msun. The benchmark brown dwarf companion, HD 4747 B, was initially reported to have a mass of $65.3^{+4.4}_{-3.3}$ \Mjup by \citet{2018Crepp_3850B}, who conducted follow-up observations with the Gemini Planet Imager to obtain its spectrum. \citet{2019Peretti_HD4747B} later performed detailed orbital and spectral analyses, refining the mass to $70.0\pm1.6$ \Mjup and the separation to $10.01\pm0.21$ AU; \citet{Brandt2019} also derived a similar mass with comparable uncertainties. In this work, using relative astrometry from imaging data \citep{Brandt2019,2019Peretti_HD4747B,2012Creep_trendStar,2018Crepp_3850B}, we determine a mass of $68.71_{-1.36}^{+1.47}$\Mjup at a separation of $10.15\pm0.17$~au. Our result agrees with previous measurements and provides slightly improved precision.

\subsubsection{HIP 8770 b and d}
HIP~8770 (HD~11506) is a G0 main-sequence star at 51.3\,pc, with a fitted mass of $1.27\pm0.06$\,\Msun. \citet{2018_HIP8770bc} reported two RV planets: HD~11506\,b, at a minimum mass of $4.83\pm0.52$\,\Mjup\ and $2.9\pm0.14$\,AU, and HD~11506\,c, at $0.41\pm0.06$\,\Mjup\ and $0.77\pm0.04$\,AU. \citet{2022Feng_3Dselection_167substellar} subsequently proposed a third companion, HD~11506\,d, with an estimated mass of $12.8^{+0.6}_{-0.4}$\,\Mjup\ at $18.2\pm0.1$\,AU, while updating planet~b to $4.88^{+1.99}_{-0.33}$\,\Mjup\ at $2.6\pm0.1$\,AU. More recently, \citet{Ruggier2024} combined RV and direct‐imaging data to derive $M_d=10.2\pm0.5\,M_{\rm Jup}$ at $17.8\pm0.2\,$AU for planet d, and found $M_b=5.02\pm0.15\,M_{\rm Jup}$ at $2.95\pm0.04\,$AU for planet b. 

Because the innermost planet~c has a period under one year and orbits within 1\,AU, its astrometric signature remains undetectable over the long \Hipparcos--\Gaia baseline. Consequently, \citet{2022Feng_3Dselection_167substellar} omitted planet~c, and we do the same here. We derive a tighter minimum mass for planet~b of $5.04\pm0.21$\,\Mjup\ at $2.92\pm0.05$\,AU, whereas our constraint on planet~d, $9.1^{+34.6}_{-4.0}$\,\Mjup\ at $13.3^{+18.0}_{-5.4}$\,AU, is much less precise than the estimate by \citet{2022Feng_3Dselection_167substellar} and \citet{Ruggier2024}. These large uncertainties likely arise from limited orbital phase coverage for the outer companion; additional long-term RV monitoring will be critical for improving planet~d’s orbital parameters and mass. 

\subsubsection{HIP 10337 c}
HIP~10337 (BD-21~397) is a K7 main-sequence star at 23.74\,pc, with a fitted mass of $0.691\pm0.013$\,\Msun. \citet{202306_10337bc} identified two RV planets (b and c), estimating minimum masses of $0.7\pm0.1$\,\Mjup\ at $2.63^{+0.06}_{-0.05}$\,AU for planet~b and $2.4^{+0.5}_{-0.2}$\,\Mjup\ at $5.9^{+3.4}_{-0.5}$\,AU for planet~c. Soon after, \citet{Philipot2023b} announced the same outer planet, reporting $4.8\pm0.6$\,\Mjup\ at $5.9\pm0.1$\,AU.

Here, we exclude the inner planet~b, as its short period and close-in orbit render it undetectable by long-baseline absolute astrometry. Focusing on the outer planet~c, we obtain a mass of $4.47^{+0.57}_{-0.54}$\,\Mjup\ at $5.96^{+0.49}_{-0.16}$\,AU, in line with previous measurements and comparable in precision.

\subsubsection{HIP 12436 Ab}
HIP~12436 (HD~16905) is a K3 main-sequence star at 39.8\,pc, with a fitted mass of $0.85\pm{0.04}$\,\Msun. \citet{2022Feng_3Dselection_167substellar} reported a substellar companion (HD~16905\,Ab) with an estimated mass of $\approx9.1$\,\Mjup\ at a separation of 6.4\,AU, which should be detectable through the astrometric and RV data considered here. More recently, \citet{Philipot2023b} refined this to $M=11.3^{+0.6}_{-0.7}$ \Mjup at $8.8^{+0.4}_{-0.3}$ AU using combined RV and direct‐imaging data.

\begin{figure}
    \centering
    \includegraphics[width=0.48\textwidth]{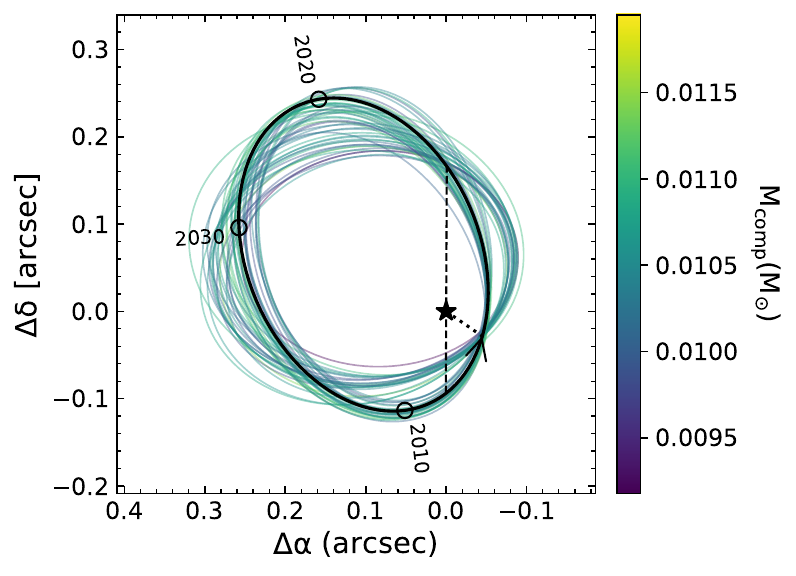}\\[1ex]
    \includegraphics[width=0.45\textwidth]{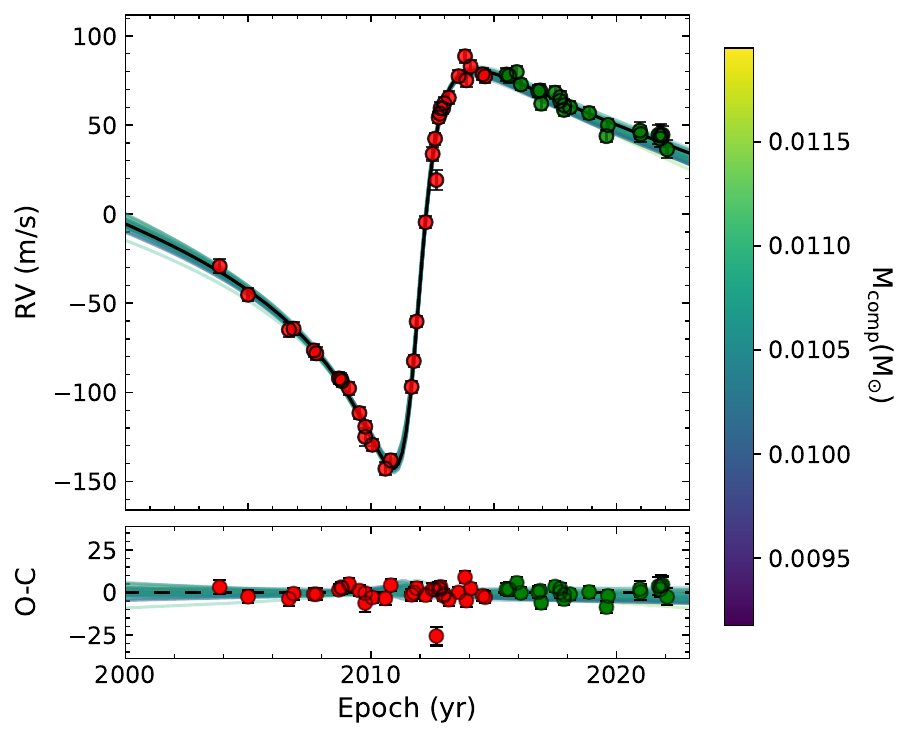}
    \caption{
    \textbf{Top:} Relative orbits of HIP~12436\,Ab (in arcseconds), fit jointly to RVs and absolute astrometry. We show 50 random draws from the MCMC orbit distribution, color-coded by companion mass, with the black curve representing the maximum-likelihood solution. The star symbol at the origin marks the host star, and selected epochs along the maximum-likelihood orbit are annotated. \textbf{Bottom:} Observed RVs of HIP~12436 overplotted with the best-fit orbital solution (black) and a random sampling of orbits from the MCMC chain. Residuals from the best-fit orbit appear in the lower subpanel.
    }
    \label{fig:12436_orbit}
\end{figure}

Our two-body orbital model fits the system well, allowing us to refine the parameters of companion~Ab to $10.2 \pm 0.5$\,\Mjup\ at a separation of $11.9 \pm 1.0$\,AU, with a period of $44.1_{-5.6}^{+7.7}$\,years and bimodal inclination distribution. This suggests that Ab is both slightly more massive and resides on a wider orbit than previously indicated.Our result lies between the values reported by \citet{2022Feng_3Dselection_167substellar} and \citet{Philipot2023b}, and is slightly closer to the latter. Figure~\ref{fig:12436_orbit} shows the orbit of HIP~12436\,Ab, with the top panel displaying the relative orbit and its predicted positions at selected epochs, and the bottom panel illustrating the observed RVs alongside our best-fit orbital solution and random MCMC draws.

\subsubsection{HIP 19428 B}
HIP~19428 (HD~26161) is a G0 star at 40.91\,pc, with a fitted mass of $1.30\pm0.07$\,\Msun. \citet{Rosenthal2021_CLS} first reported evidence for a companion from long‐term RV monitoring. \citet{Lagrange2023} emphasized the challenges in constraining its orbit due to limited phase coverage. Nonetheless, \citet{2022Feng_3Dselection_167substellar} reported the parameters of a companion at $28.46^{+20.05}_{-0.22}$\,\Mjup\ and $10.08^{+5.62}_{-1.09}$\,AU.

\begin{figure}
    \centering
    \includegraphics[width=0.4\textwidth]{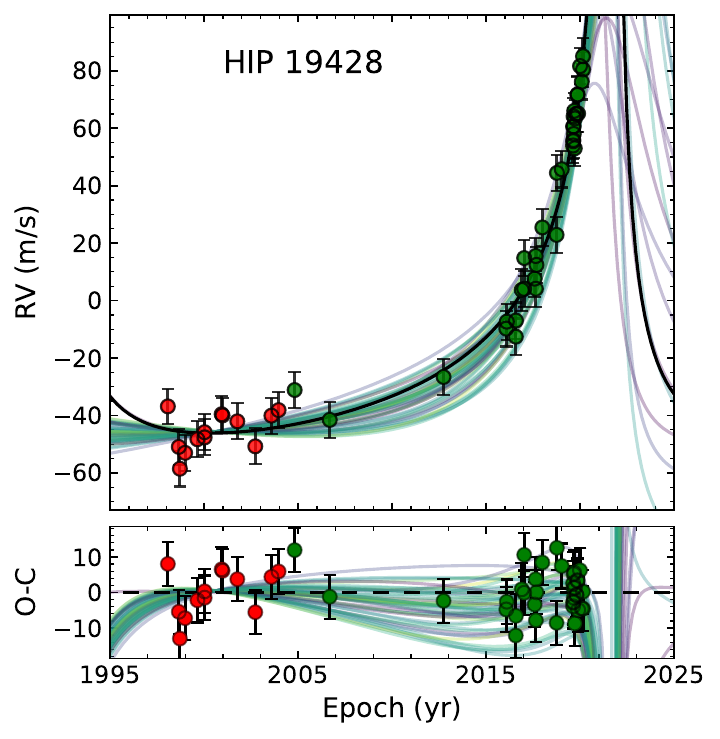}
    \caption{Observed RVs of HIP~19428, plotted alongside the best-fit orbital solution (black) and random orbits from the MCMC chain. The lower subpanel presents the residuals relative to the best-fit orbit.}
    \label{fig:19428_RV}
\end{figure}

Our updated analysis identifies a likely brown dwarf companion, HIP~19428\,B, with a mass of $58.65 \pm 17.80$\,\Mjup\ at a separation of $23.2^{+21.0}_{-8.0}$\,AU. Figure~\ref{fig:19428_RV} illustrates the observed RVs overlaid with our best-fit orbit and a random sampling of MCMC draws. The current RV phase coverage suggests that the true orbital period could be longer than our best‑fit solution but is unlikely to be shorter, corroborating the skepticism expressed by \citet{Lagrange2023} regarding earlier orbital constraints.

\subsubsection{HIP 20277 b and d}
HIP~20277 (HD~27894) is a K2 main-sequence star at 42\,pc, with a fitted mass of $0.87\pm0.04$\,\Msun. \citet{2005_20277b_discover} reported the discovery of an inner planet~b, with a minimum mass of $0.62$\,\Mjup\ at $0.12$\,AU. \citet{2017_20277bcd} then announced two additional planets, c and d, with minimum masses of $0.16$\,\Mjup\ at $0.20$\,AU and $5.45$\,\Mjup\ at $5.42$\,AU, respectively. \citet{2022Feng_3Dselection_167substellar} later updated the mass of planet~d to $6.49^{+0.99}_{-0.35}$\,\Mjup\ at $5.36^{+0.21}_{-0.22}$\,AU.

In this work, we exclude the innermost planet~c (period of $\sim18$\,days), whose short-period signal is effectively undetectable in long-baseline absolute astrometry. We then perform a three-body fit (star + planet~b + planet~d), finding a minimum mass for planet~b of $0.67\pm0.02$\,\Mjup\ at $0.129\pm0.002$\,AU, and a dynamical mass for planet~d of $7.58^{+0.60}_{-0.56}$\,\Mjup\ at $5.51\pm0.08$\,AU. These results agree with previous studies while offering slightly improved precision.

\subsubsection{HIP 21865 b}
HIP~21865 (HD~29985) is a K4 main-sequence star at 31.27\,pc, with a fitted mass of $0.72\pm0.04$\,\Msun. \citet{2022Feng_3Dselection_167substellar} reported a substellar companion (planet~b) at $5.47^{+1.94}_{-1.16}$\,\Mjup\ and $13.39^{+2.07}_{-3.68}$\,AU. Our updated analysis instead indicates that the planet is closer-in and less massive, with $2.83^{+0.67}_{-0.43}$\,\Mjup\ at $8.0^{+2.6}_{-2.2}$\,AU. 

\begin{figure}
    \centering
    \includegraphics[width=0.38\textwidth]{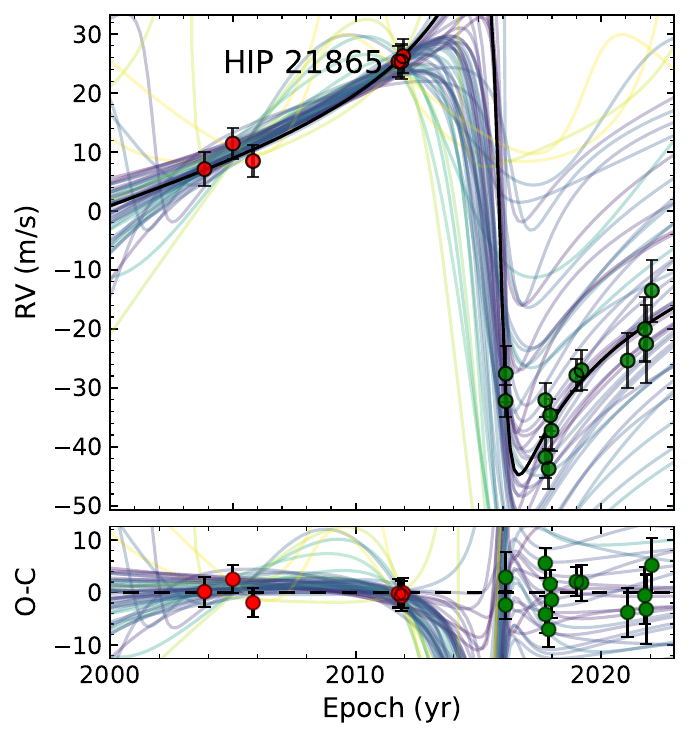}
    \caption{Observed RVs of HIP~21865, plotted with the best-fit orbital model (black) and random draws from the MCMC chain. The lower subpanel shows the residuals relative to the best-fit orbit. The HARPS fiber swap in 2015 unfortunately limits the constraining power of the data.}
    \label{fig:21865_RV}
\end{figure}

Because HARPS underwent a significant fiber swap \citep{2015HARPS_upgrade}, the combined pre- and post-upgrade RV dataset for this target has limited constraining power. As seen in Figure~\ref{fig:21865_RV}, the relatively sparse sampling between pre- and post-upgrade further reduces our ability to pin down the planet’s orbit. Additional long-term RV monitoring would help improve orbital and mass estimates for HIP~21865~b.

\subsubsection{HIP 36941 b}
HIP 36941 (HD 62364) is an F7 main-sequence star, with a fitted mass of $1.3\pm0.06$ \Msun, located at a distance of 52.96~pc. \citet{202306_10337bc} reported a companion with a minimum mass of $12.7\pm0.2$ \Mjup based on radial velocities, and a true mass of $18.77\pm0.66$ \Mjup from an astrometric solution, at a semi major axis of $6.15\pm0.04$ AU. \citet{Philipot2023b}—using a multi‑technique approach combining RV, high‑resolution imaging, and astrometry—derive $M=18.8\pm0.7$ \Mjup at $6.1\pm0.1$ AU. Our result agrees with these findings and attains a comparable level of precision.

\subsubsection{HIP 39417 b and c}
HIP~39417 (HD~66428) is a G5 main-sequence star at 53.3\,pc, with a fitted mass of $1.14\pm0.06$\,\Msun. \citet{2006Bulter_nearby_exoplanets} first reported a minimum mass of $2.82\pm0.27$\,\Mjup\ at $3.18\pm0.19$\,AU for planet~b. Later, \citet{Rosenthal2021_CLS} identified a second companion, planet~c, with a minimum mass of $27^{+22}_{-17}$\,\Mjup\ at $23.0^{+19.0}_{-7.6}$\,AU, and updated planet~b’s mass to $3.19\pm0.11$\,\Mjup\ at 3.5\,AU. Most recently, \citet{2022Feng_3Dselection_167substellar} refined both companions’ orbits, obtaining $10.95^{+2.44}_{-3.85}$\,\Mjup\ at $3.40^{+0.14}_{-0.16}$\,AU for planet~b, and $1.76^{+3.40}_{-0.041}$\,\Mjup\ at $9.41^{+1.95}_{-1.27}$\,AU for planet~c.

\begin{figure}
    \centering    \includegraphics[width=0.47\textwidth]{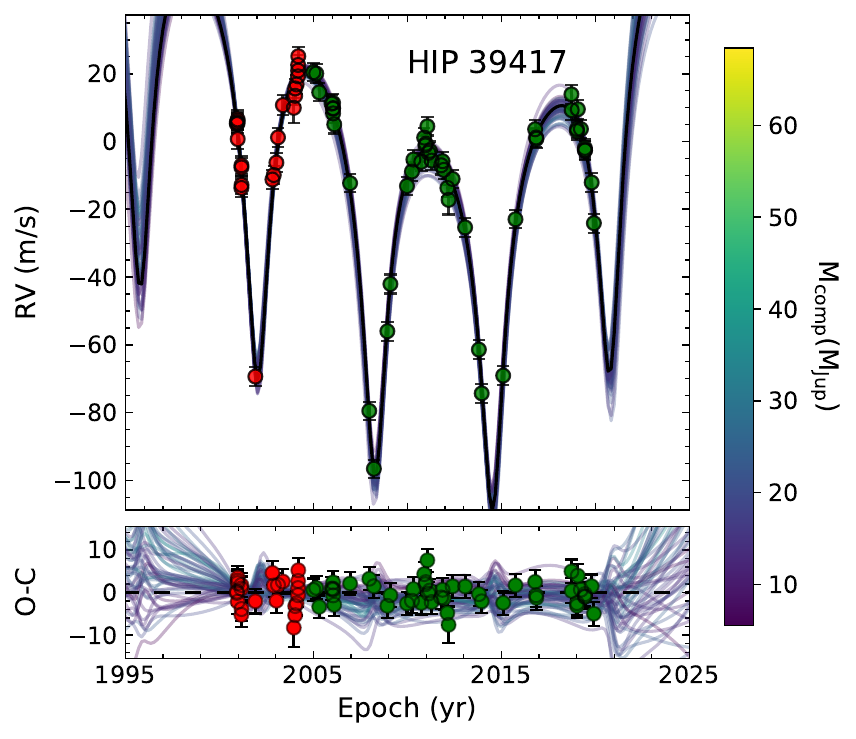}
    \caption{Observed RVs of HIP~39417, plotted with the best-fit orbital model (black) and random draws (color coded by mass of HIP~39417~c) from the MCMC chain. The short-term deviations primarily constrain the inner planet~b, while the limited long-term coverage shapes the outer planet~c.}
    \label{fig:39417_RV}
\end{figure}

In our reanalysis, we find a minimum mass for planet~b of $3.3\pm0.13$\,\Mjup\ at $3.56\pm0.06$\,AU, consistent with previous work. However, we derive a dynamical mass for planet~c of $21.6^{+13.6}_{-8.2}$\,\Mjup\ (and a minimum mass of $21.6^{+14.0}_{-8.1}$\,\Mjup) at $18.0^{+10.0}_{-5.4}$\,AU—consistent with the result of \citet{Rosenthal2021_CLS}, but differing from \citet{2022Feng_3Dselection_167substellar}. The large uncertainties likely stem from limited orbital phase coverage for the outer companion, as shown in Figure~\ref{fig:39417_RV}. Additional long-term RV monitoring will be essential to further constrain planet~c’s orbital parameters and mass.

\subsubsection{HIP 42030 b and c}
HIP~42030 (HD~72659) is a G2 main-sequence star at 49.8\,pc, with a fitted mass of $1.25\pm0.06$\,\Msun. \citet{2003_42030b_discover} initially reported a planet~b from the Keck Precision Doppler Survey, with a minimum mass of 2.55\,\Mjup\ at 3.24\,AU. More recently, \citet{2022Feng_3Dselection_167substellar} refined the planet~b mass to $3.00^{+2.59}_{-0.10}$\,\Mjup\ at $4.69^{+0.19}_{-0.20}$\,AU, and identified a second companion, planet~c, at $18.81^{+4.44}_{-4.80}$\,\Mjup\ and $13.96^{+0.88}_{-0.86}$\,AU.

\begin{figure}
    \centering
    \includegraphics[width=0.47\textwidth]{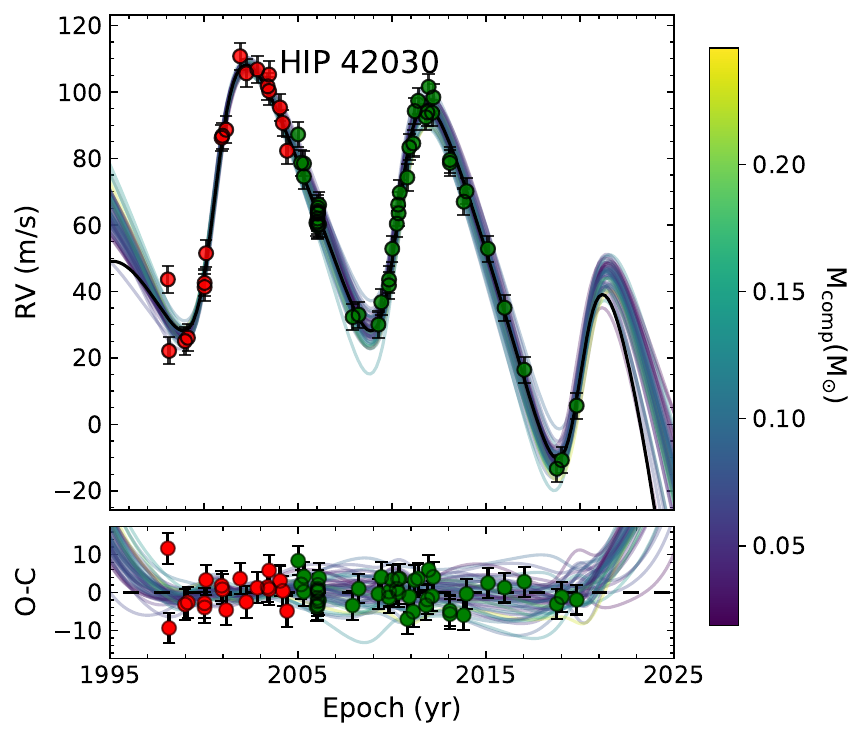}
    \caption{Observed RVs of HIP~42030, plotted alongside the best-fit orbital 
    model (black) and random draws (color coded by mass of HIP~42030~c) from the MCMC chain. The short-term 
    deviations reveal the presence of planet~b, whereas the limited 
    long-term coverage drives the orbit of planet~c.}
    \label{fig:42030_RV}
\end{figure}

In our analysis, we recover a minimum mass for planet~b of $3.19\pm0.15$\,\Mjup\ at $4.92\pm0.08$\,AU, consistent with previous estimates. However, the limited observational baseline for the long-period planet~c results in a substantially larger and more uncertain mass of $0.07^{+0.06}_{-0.03}$\,\Msun\ at $30^{+14}_{-11}$\,AU. As illustrated in Figure~\ref{fig:42030_RV}, the short-term deviations in the observed RVs primarily constrain planet~b, while the uncertain long-term trend shapes the outer planet~c. Additional long-baseline RV monitoring will be crucial for refining planet~c’s orbital parameters and mass.

\subsubsection{HIP 60648 b}
HIP~60648 (HD~108202) is a K-type main-sequence star at 38.65\,pc, with a fitted mass of $0.74\pm0.04$\,\Msun. \citet{2023_70849_mass} announced a substellar companion, HIP~60648\,b, with a true mass of $3.0^{+0.7}_{-0.5}$\,\Mjup\ and a semi-major axis of $3.7\pm0.1$\,AU. Our analysis refines the companion’s mass to $2.86^{+0.69}_{-0.45}$\,\Mjup\ at $3.66\pm0.07$\,AU, consistent with the original detection and offering slightly improved precision.

\subsubsection{HIP 64457 b and c}
HIP 64457 (HD 114783) is a K-type main-sequence star with a fitted mass of $0.90\pm0.05$ \Msun, located at a distance of 20.4~pc. \citet{2002_40687b_discover} discovered planet~b with a minimum mass of about 1 \Mjup and a period of approximately 500~days, and \citet{2009_64457b} refined its minimum mass to $1.10\pm0.06$ \Mjup at $1.16\pm0.02$~AU. Using HIRES RV measurements, \citet{Rosenthal2021_CLS} reported two planets: planet~b with a minimum mass of $1.03\pm0.03$ \Mjup at $1.16\pm0.02$~AU, and planet~c with $0.66\pm0.05$ \Mjup at $5.0\pm0.1$~AU. \citet{2023_70849_mass} subsequently derived the true mass of the outer planet~c to be $2.0\pm0.4$ \Mjup at $5.0\pm0.1$~AU.

Our analysis is consistent with previous findings. We obtain a minimum mass for the inner planet~b of $1.04\pm0.04$ \Mjup at $1.18\pm0.02$~AU and a true mass for the outer planet~c of $1.47_{-0.63}^{+0.58}$ \Mjup at $5.03\pm0.12$~AU, which aligns with earlier results at a comparable level of precision.

\subsubsection{HIP 70849 b}
HIP 70849 (GJ 9482) is a K7 main-sequence star with a fitted mass of $0.647\pm0.013$ \Msun, located at a distance of 24~pc. \citet{2011_70849b_discover} reported the discovery of planet~b, with a minimum mass of about 3--15 \Mjup at 4.5--36~AU. \citet{2024_70849_m2sini} narrowed the mass range to 4.58--4.6 \Mjup, while \citet{2023_70849_mass} confined the true mass to $4.5^{+0.4}_{-0.3}$ \Mjup and a semi-major axis of $3.99^{+0.06}_{-0.07}$~AU. Our analysis yields a true mass for planet~b of $4.55\pm0.51$ \Mjup at a semi-major axis of $4.04\pm0.03$~AU, in good agreement with previous studies.

\subsubsection{HIP 84171 b and c}
HIP 84171 (HD 156279) is a G7 main-sequence star with a fitted mass of $1.00\pm0.05$ \Msun, located at a distance of 36.6~pc. \citet{2012_84171b} reported the discovery of planet~b, with a minimum mass of $9.71\pm0.66$ \Mjup at $0.50\pm0.02$~AU. \citet{2016_84171c} subsequently announced an additional planet~c, with a mass of $8.60^{+0.55}_{-0.50}$ \Mjup, and \citet{2022Feng_3Dselection_167substellar} refined the true mass of planet~c to $9.75^{+1.32}_{-0.61}$ \Mjup and a semi-major axis of $5.49^{+0.22}_{-0.21}$~AU.

Our analysis detects both planet~b and planet~c, finding 
a minimum mass of $9.82\pm0.29$ \Mjup at $0.51\pm0.01$~AU for planet~b, and a true mass of $10.53_{-0.59}^{+0.65}$ \Mjup at $5.57\pm0.09$~AU for planet~c. These results are consistent with previous studies and achieve slightly improved precision.

\subsubsection{HIP 89620 B}
HIP 89620 (HD 167665) is an F9 main-sequence star with a fitted mass of $1.30\pm0.06$ \Msun, located at a distance of 30.86~pc. \citet{2023_89620_BD} confirmed that its companion lies in the brown dwarf mass range based on proper motion anomaly analysis.  \citet{2024_89620_imaging} subsequently imaged this companion directly and constrained its mass to be $60.3\pm0.7$ \Mjup by combining RV and absolute and relative astrometry data.

In this work, we use the relative astrometry from \citet{2024_89620_imaging} to derive a mass of $61.29^{+1.05}_{-0.96}$ \Mjup, in good agreement with prior findings, and we measure a semi-major axis of $5.60 \pm 0.05$\,au.

\subsubsection{HIP 89844 b and c}
HIP 89844 (HD 168443) is a G5 main-sequence star with a fitted mass of $1.06\pm0.05$ \Msun, located at a distance of 37.38~pc. \citet{1999_89844b_discover} first reported the discovery of the inner planet b, and  \citet{2001_89844c_discover} reported the discovery of an outer planet c. As one of the rare examples of a substellar companion with a Hipparcos‐derived orbit, \citet{Reffert2006} report a mass of $37^{+36}_{-19}\,M_{\rm Jup}$ for planet c and, assuming co‐planarity, infer $15\,M_{\rm Jup}$ for planet b. \citet{2007_89844bc_mass} reported the minimum mass and semi-major axis of planet b and 
planet c as $7.27\pm0.27$ \Mjup at $0.29\pm0.01$ AU and $16.87\pm0.64$ \Mjup at $2.84\pm0.05$ AU, respectively. \citet{Rosenthal2021_CLS} refined the minimum mass of 
the inner planet b to $7.92^{+0.15}_{-0.16}$ \Mjup, and 
\citet{2022Feng_3Dselection_167substellar} reported the true mass of the outer planet c to be $17.31^{+2.55}_{-0.91}$ \Mjup with a semi-major axis of $2.84^{+0.11}_{-0.12}$ AU.

Our analysis detects both planet b and planet c, finding the minimum mass of planet b to be $8.09^{+0.24}_{-0.25}$ \Mjup at $0.30\pm0.01$ AU, and the true mass of planet c to be $20.53\pm1.57$ \Mjup at $2.93\pm0.04$ AU, which is consistent with previous  findings with slightly improved precision.

\subsubsection{HIP 90055 b}
HIP 90055 (HD 168863) is a K1 main-sequence star with a fitted mass of $0.83\pm0.04$ \Msun, located at a distance of 39.90~pc. \citet{2022Feng_3Dselection_167substellar} reported the discovery of planet b with a mass of $6.8\pm1.1$ \Mjup and a semi-major axis of $10.96^{+3.80}_{-1.42}$ AU. Our analysis yields a bimodal inclination posterior, and a mass of $7.65_{-1.36}^{+3.98}$ \Mjup at $11.0_{-2.1}^{+7.4}$ AU, which agrees with previous findings but with slightly lower precision.

\subsubsection{HIP 95015 b}
HIP 95015 (HD 181234) is a G5 main-sequence star with a fitted mass of $1.01\pm0.05$ \Msun, located at a distance of 47.84~pc. \citet{2019_95015b_discover} reported the discovery of a planet b with a minimum mass of $8.37^{+0.34}_{-0.36}$ \Mjup at $7.52\pm0.16$ AU. 
\citet{Rosenthal2021_CLS} refined the minimum mass of planet b to $9.1^{+1.6}_{-0.8}$ \Mjup at $7.43\pm0.11$ AU. \citet{2022Feng_3Dselection_167substellar} and \citet{2023_88595b} reported the true mass of planet b to be $9.38^{+1.01}_{-0.99}$ \Mjup at $7.43^{+0.30}_{-0.34}$ AU and $10.13^{+0.74}_{-0.63}$ \Mjup at $7.52^{+0.16}_{-0.17}$ AU, respectively. Our analysis is consistent with these previous studies, yielding a true mass of $10.89_{-1.47}^{+4.19}$ \Mjup and a semi-major axis of $7.57\pm0.14$ AU.

\subsubsection{HIP 95319 B}
HIP~95319 (GJ~758) is a K0 main-sequence star at 15.5\,pc, with a fitted mass of $0.98\pm0.05$\,\Msun. \citet{2009_95319B_discover} initially reported a possible brown dwarf companion in May and August 2009, with a projected separation of about 29\,AU. Subsequently, \citet{2017_95319_imaged} obtained direct imaging, and \citet{2018_95319_relAST} as well as \citet{Brandt2019} derived a complete orbital solution, finding a true mass of $38.1^{+1.7}_{-1.5}$\,\Mjup\ and a semi-major axis of $30^{+5}_{-8}$\,AU. \citet{2021AJ_GMBRANDT_sixmasses} later updated these values to $38.04\pm0.74$\,\Mjup\ at 
$29.7^{+5.3}_{-4.2}$\,AU.

In this work, we incorporate relative astrometry from \citet{2018_95319_relAST}, arriving at a similarly precise orbit. Our results remain consistent with earlier studies, yielding a companion mass of $35.91^{+0.69}_{-0.65}$\,\Mjup\ and a semi-major axis of $27.4^{+5.6}_{-3.6}$\,AU.

\subsubsection{HIP 95740 b and c}
HIP 95740 (HD 183263) is a G-type star with a fitted mass of $1.20\pm0.06$ \Msun, located at a distance of 53~pc. \citet{2005_95740b_discover} reported the discovery of planet b, with a minimum mass of 3.69 \Mjup at 1.52 AU, and \citet{2009_HIP9683bc} reported the discovery of an outer planet c with a minimum mass of $3.57\pm0.55$ \Mjup at $4.35\pm0.28$ AU. \citet{Rosenthal2021_CLS} updated the minimum masses to 
$3.70\pm0.10$ \Mjup at $1.51\pm0.02$ AU for planet b and $7.96\pm0.22$ \Mjup at $6.04\pm0.08$ AU for planet c. \citet{2022Feng_3Dselection_167substellar} also reported the minimum mass of planet b as $3.47^{+0.29}_{-0.30}$ \Mjup at $1.46^{+0.06}_{-0.07}$ AU, and the true mass of planet c as $9.31^{+1.52}_{-1.82}$ \Mjup at $5.59^{+0.23}_{-0.25}$ AU.

Our analysis is consistent with these previous results, achieving slightly improved precision. We obtain a minimum mass for the inner planet b of $3.78^{+0.13}_{-0.14}$ \Mjup at a semi-major axis of $1.52\pm0.03$ AU, and a true mass for the outer planet c of $8.92^{+0.91}_{-0.56}$ \Mjup at a semi-major axis of $6.07\pm0.11$ AU.

\subsubsection{HIP 98819 B}
HIP~98819 (HD~190406 / HR~7672 / GJ~779) is a G0 main-sequence star at 17.71\,pc, with a fitted mass of $1.10\pm0.06$\,\Msun. \citet{2002_98819B_discover} discovered a directly imaged brown dwarf companion using Gemini-North and Keck, estimating a model-dependent mass of 55--78\,\Mjup\ (for a host-star age of 1--3\,Gyr) and finding an RV-based minimum mass of 48\,\Mjup. \citet{2012_98819_dynamic} refined this benchmark brown dwarf’s orbit to $68.7^{+2.4}_{-3.1}$\,\Mjup\ at $18.3^{+0.4}_{-0.5}$\,AU, with high eccentricity ($e=0.50^{+0.01}_{-0.01}$) and a near-edge-on inclination ($i=97.3^{+0.4}_{-0.5}\,^\circ$). \citet{2021_98819B_update} later updated these values using HGCA astrometry, obtaining $72.8\pm6.1$\,\Mjup, $16.3\pm0.7$\,AU, $i=90.7\pm5.0\,^\circ$, and $e=0.46\pm0.01$. Because HIP~98819~B lies near the stellar/substellar (hydrogen-burning) boundary, \citet{Brandt2019} discuss its bolometric luminosity, noting that it continues to cool and will eventually transition onto the main sequence over several gigayears.

Here, we incorporate relative astrometry from \citet{Brandt2019,2002_98819B_discover,2003_98819B_imag,2009_98819B_imag,2012_98819_dynamic}, deriving a mass of $71.43\pm1.05$\,\Mjup\ at a semi-major axis of $17.6\pm0.3$\,AU. Our result is consistent with previous studies and achieves slightly improved precision.

\subsubsection{HIP 106440 b}
HIP 106440 (GJ 832) is an M-type star with a fitted mass of $0.45^{+0.01}_{-0.01}$ \Msun, located at a distance of 4.94~pc. \citet{2009_106440b_discover} reported the discovery of planet b with a minimum mass of  $0.64\pm0.06$ \Mjup at $3.4\pm0.4$ AU. \citet{2023_88595b} reported the true mass of planet b to be $0.80^{+0.12}_{-0.11}$ \Mjup at $3.53^{+0.15}_{-0.16}$ AU, and \citet{2023_70849_mass} reported the true mass to be $0.99^{+0.09}_{-0.08}$ \Mjup at $3.7\pm0.1$ AU. A false positive detection of an inner planet c was reported by \citet{2014_106440_false}, which was later claimed to be non-existent by \citet{2022_106440c}.

Our analysis finds a true mass for planet b of  $1.04\pm0.04$ \Mjup with a semi-major axis of 
$3.65^{+0.07}_{-0.06}$ AU, in good agreement with previous studies and with slightly improved precision.

\subsubsection{HIP 113421 b and c}
HIP 113421 (HD 217107) is a G-type star with a fitted mass of $1.13\pm0.04$ \Msun, located at a distance of 19.72~pc. \citet{1999_113421b_discover} reported the discovery of planet b with a minimum mass of 1.27 \Mjup and a period of 7.12 days. \citet{2005_113421c_discover} reported the discovery of an outer planet c with a minimum mass of $2.1\pm1$ \Mjup at $4.3\pm2$ AU. \citet{Rosenthal2021_CLS} updated the minimum masses of planet b and planet c to  $1.39\pm0.04$ \Mjup at $0.074\pm0.001$ AU and $4.31\pm0.13$ \Mjup at $5.94^{+0.08}_{-0.09}$ AU, respectively.
Most recently, \citet{Giovinazzi2025} reported $M_b\sin i_b=1.370^{+0.016}_{-0.020}\,M_{\rm Jup}$ and a dynamical mass $M_c=4.37^{+0.13}_{-0.10}\,M_{\rm Jup}$.

In this work, we derive the minimum mass of the inner planet b to $1.45^{+0.05}_{-0.06}$ \Mjup with a semi-major axis of $0.076^{+0.001}_{-0.002}$ AU, and the true mass of the outer planet c to $4.50\pm0.12$ \Mjup with a semi-major axis of $6.10\pm0.06$ AU, , in line with previous measurements and comparable in precision.478

\subsubsection{HIP 116250 b}
HIP 116250 (HD 221420) is a G-type star with a fitted mass of $1.34\pm0.07$ \Msun, located at a distance of 31.8~pc. \citet{2019_116250b_discover} reported the discovery of planet b with a minimum mass of $9.70^{+1.10}_{-1.00}$ \Mjup at $18.5\pm2.3$ AU.  \citet{2021_116250b_truemass} and \citet{2021_116250b_Li} reported the true mass of this planet to be $22.9\pm2.2$ \Mjup at $10.15^{+0.59}_{-0.38}$ AU and $20.6^{+2.0}_{-1.6}$ \Mjup at $9.99^{+0.74}_{-0.70}$ AU, respectively. \citet{2022Feng_3Dselection_167substellar} updated the true mass of planet b to $20.35^{+1.99}_{-3.43}$ \Mjup at $8.40^{+0.41}_{-0.38}$ AU. 

Our analysis is consistent with these previous studies, yielding a true mass of $19.9^{+1.7}_{-1.5}$ \Mjup and a semi-major axis of $10.79^{+1.90}_{-0.72}$ AU.

\subsubsection{HIP 116745 b}
HIP 116745 (HD 222237) is a K3 main-sequence star with a fitted mass of $0.70\pm0.01$ \Msun, located at a distance of 11.45~pc. \citet{2024Xiao_116745b} reported the discovery of planet b, with a true mass of $5.19\pm0.58$ \Mjup and a semi-major axis of $10.8^{+1.1}_{-1.0}$ AU. Our analysis yields a true mass of $5.03^{+1.47}_{-1.05}$ \Mjup and a semi-major axis of 
$12.9^{+7.6}_{-3.1}$ AU, which is in agreement with the previous study.

\section{Comparisons with \Gaia Non-Single Star Solutions} 
\label{sec:Gaia}

In this section, we present a detailed comparison of our findings with results from the Gaia DR3 Non-Single Star (NSS) catalog, focusing on two-body orbital solutions and astrometric acceleration measurements. We process the Gaia DR3 two-body orbit solutions to compare them with those obtained in this work, and use our orbital solutions to predict the Gaia DR3 astrometric acceleration terms. By comparing our predictions to the actual DR3 values, we validate a subset of the Gaia DR3 non-single star solutions.

Analyzing key orbital parameters—including semi-major axis, eccentricity, period, and companion mass—we assess the consistency and precision of our results relative to those derived from Gaia data. This comparison provides insights into the reliability and accuracy of Gaia’s two-body solutions and astrometric acceleration solutions, highlighting the strengths and limitations of each method in characterizing stellar multiplicity and companion masses. Detailed comparisons are presented in Table \ref{Tab:gaia-2bd} and Table \ref{tab:gaia_accel}, and further discussed in the subsequent sections.

\subsection{\Gaia Two-body Orbit Data Processing} \label{sec:method-2bd}
In Gaia DR3, approximately 813,000 non-single star (NSS) sources were identified, encompassing a range of stellar multiplicity indicators, including astrometric acceleration solutions, orbital models, and eclipsing binary classifications.
The Gaia non-single stars of interest here are the astrometric non-single stars--those for which a five-parameter astrometric sky path is a poor fit.  This is captured by the renormalized unit weight error, or RUWE \citep{GaiaDR3}.  Stars with high RUWE values have astrometric residuals that indicate nonlinear motion through space; they are candidates for Keplerian orbital fits or astrometric acceleration fits.  If either of these fits is formally good, it may be presented in the \Gaia DR3 non-single-star catalog \citep{none-single-star}.

For sources classified as two-body orbits (categorized under non-single star index 3), \Gaia DR3 provides a comprehensive set of detailed orbital parameters. These parameters include orbital period (in days), eccentricity, periastron time, and the Thiele-Innes constants (A, B, F, and G) \citep{none-single-star}. The Thiele-Innes constants are a convenient representation of a Keplerian orbit for astrometry, and are related to the Campbell elements by 
\begin{align}
&A = a_G\cdot \left(\cos \Omega \cos \omega - \sin \Omega \sin \omega \cos i \right)
\label{eq:TI-A}\\
&B = a_G \cdot \left(\sin \Omega \cos \omega + \cos \Omega \sin \omega \cos i\right)
\label{eq:TI-B}\\
&F = a_G \cdot \left(-\cos \Omega \sin \omega - \sin \Omega \cos \omega \cos i\right)
\label{eq:TI-C}\\
&G = a_G \cdot \left(-\sin \Omega \sin \omega + \cos \Omega \cos \omega \cos i\right) .
\label{eq:TI-D}
\end{align}
In this context, $a_G$ denotes the \Gaia semi-major axis, defined as the angular distance from the host star to the system’s barycenter. The parameter $i$ represents the orbital inclination, $\Omega$ is the longitude of the ascending node, and $\omega$ is the argument of periastron.

To compare our results with the \Gaia DR3 data products, we must convert the \Gaia two-body orbits into a form directly comparable to our findings. 
We begin by calculating the intermediate values $u$ and $v$ using 
\begin{align}
&u = \frac{1}{2} \left(A^2 + B^2 + F^2 + G^2 \right) \label{eq:TI-Kep-u}\\
&v = BF - AG . \label{eq:TI-Kep-v}
\end{align}
We can then derive the semimajor axis of the photocenter's orbit as observed by \Gaia $a_G$, and the semimajor axis of the orbit $a$ that we derive, by
\begin{align}
&a_G = \sqrt{u + \sqrt{u^2 - v^2}}\label{eq:TI-Kep-ag}
\end{align}
\begin{equation}
a = a_G\cdot\frac{M+m}{m} \label{eq:TI-Kep-a}
\end{equation}
This assumes that all of the luminosity is from the primary star, so that the photocenter is at the same position as the primary star.  In practice, this means that the luminosity ratio of the primary to the secondary must be much larger than the mass ratio.

To obtain the companion mass, we use Equation \eqref{eq:TI-Kep-a} to convert $a_G$ to the semi-major axis $a$ referred to in \orvara outputs, which represents the distance from the companion to the barycenter. In Equation \eqref{eq:TI-Kep-a}, $M$ denotes the host star mass, and $m$ represents the companion mass.

Finally, we apply Kepler's third law 
\begin{align}
T^2 = \frac{a^3}{M+m} ,\label{eq:TI-Kep-T}
\end{align}
where $T$ denotes the orbital period, to solve for the companion mass $m$ and orbital semi-major axis $a$. Using this equation in combination with the primary mass prior from the orbit fitting process, we derive the values for $m$ and $a$. After appropriate unit conversions (to AU, years, and \Msun), these results are then compared to the \Gaia DR3 data product.  As noted above, this comparison is valid only for faint companions.

\input{table_gaia_2bd}

\subsection{Predicting Astrometric Accelerations} \label{sec:method-accel}

While some \Gaia stars had full Keplerian orbits fitted to their astrometry, others had fitted sky paths with fewer terms.  These are the seven-parameter fits (with constant acceleration in right ascension and declination) and the nine-parameter fits (with constant jerk).  Unlike a Keplerian orbit, fitting a seven- or a nine-parameter astrometric sky path to measured positions is a linear problem.  

With the orbital posterior (i.e., many thousands of likely orbits), we use \texttt{htof} and the \gaia GOST scanning law\footnote{\url{https://gaia.esac.esa.int/gost/}} to predict what \gaia DR3 should have reported for the $\alpha$ and $\delta$ (right ascension and declination) acceleration terms. In detail, we randomly sample from the many thousands of orbital draws contained in the MCMC chain. At any given orbital draw, we calculate the reflex motion induced on the primary, sample that motion exactly according to the \gaia scanning law, and then fit a 7-parameter skypath with \texttt{htof}, retaining the acceleration terms (the 6$^{\rm th}$ and 7$^{\rm th}$ parameters). For stars with a nine-parameter \Gaia solution, we perform a corresponding nine-parameter fit, again retaining the acceleration terms at the catalog epoch. 
We use 5000 samples to build a posterior of predicted acceleration terms (resulting in roughly 1\% precision on the standard errors of those acceleration terms). We then compare the predicted acceleration posteriors to those actually reported by \gaia DR3, and identify and discuss disagreements. We note that the formal errors of the predicted acceleration terms are solely, at this point, due to the uncertainties in the orbital motion of the companions and can be significantly smaller than the Gaia DR3 measurement errors.

It is worth noting one limitation: \texttt{htof} assumes the validity of the GOST scanning law, and further assumes that all \gaia measurements are of equal precision. We guard against possible systematics induced by these assumptions using a bootstrap resampling of the given \gaia GOST scanning law for every source, and add those extra systematic uncertainties into our final error budget in Appendix \ref{sec:App2} Table \ref{tab:gaia_accel}. Dead times are automatically accounted for (those published in \citealp{Gaia_Theastrometric_solution}). 

We will not have the full intermediate astrometric data until \gaia DR4, and so within our predicted acceleration analysis, we are forced to make the assumptions of (1) uniform-along scan errors and (2) that the scans reported by GOST (minus those falling within published dead time windows) reflect the true set of observations used by \gaia DPAC. These are significant assumptions, yet there is evidence that suggests the assumptions are roughly obeyed by the real data for some sources (see e.g., the discussion in Section 6.1.1 of \citealp{2021AJ_GMBRANDT_sixmasses}), especially for faint stars. One must assess this on a source-by-source basis however, using e.g., the central epoch analysis of  \citet{2021AJ_GMBRANDT_sixmasses}. This ends up being a negligible correction, but we include it for completeness.

For orbits where the posteriors are multi-modal and \orvara cannot distinguish between prograde and retrograde solutions, we modified \htof to split these modes by restricting inclination to be either greater or less than $90^\circ$. In total, 20 systems in this work exhibit such behavior (4 stellar companions and 16 substellar companions). 
Comparing the acceleration results with the 
\Gaia NSS findings can help to distinguish between the two modes. The retrograde-acceleration solutions for these bimodal systems are reported in 
Table~\ref{tab:accel_all}. 

\subsection{Comparison: Two Body Orbit}
Using the method detailed in Section \ref{sec:method-2bd}, we analyze the \Gaia\ DR3 non-single star (NSS) two-body orbit data products and compare these results with posterior parameters derived from \orvara, as summarized in Table \ref{Tab:gaia-2bd}.
For the three binaries with published two-body solutions in Gaia DR3, the orbital parameters obtained from \orvara and from the Gaia pipeline do not agree: semi-major axes, companion masses, and eccentricities differ by tens to hundreds of $\sigma$, and the periods diverge by up to three orders of magnitude (\ref{Tab:gaia-2bd}).

This discrepancy arises from limitations in the \Gaia\ DR3 binary pipeline, which uses a uniform grid constrained to period search ranges less than twice the observational time baseline—approximately 2000 days ($\approx$5.5 years) \citep{gaia_2bd}. As a result, \Gaia\ two-body orbit solutions become less reliable for systems with orbital periods approaching or exceeding this threshold. The constrained grid inherently limits the pipeline's ability to model longer-period binaries accurately. We caution researchers to critically assess \Gaia\ DR3 orbital solutions, particularly for binaries with long periods near or above this upper limit.

\begin{figure}
    \begin{center}
        \includegraphics[width=9cm]{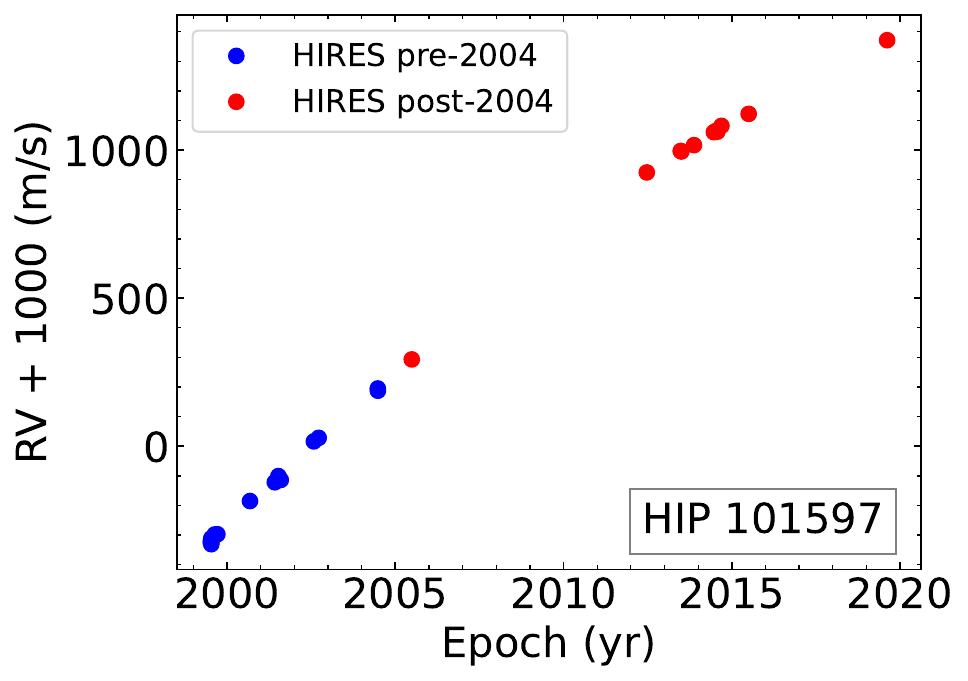}
    \end{center}
    \caption{RV time series of HIP 101597.}
    \label{fig:RV_HIP101597}
\end{figure}

The case of HIP~101597 is puzzling: the \Gaia two-body orbit suggests an RV semi-amplitude of several km/s with a 26-day period. This is clearly incompatible with the measured RVs, which show a steady acceleration (Figure \ref{fig:RV_HIP101597}), and the Washington Double Stars catalog, which astrometrically resolves the binary at a separation of $\approx$0$.\!\!''5$ and a contrast of $\approx$25 in flux.  \Gaia does detect the presence of a luminous companion, with 51\% of the transits listed as having multiple photometric peaks (parameter {\tt ipd\_frac\_multi\_peak}).  

\subsection{Comparison: Astrometric Accelerations}
\input{table_gaia_accel}
\input{table_gaia_accel_9para}

We next compare our predicted results for astrometrically accelerating stars in \Gaia.  Using each \orvara chain, we propagate the orbit with \htof through the GOST scanning law to predict \Gaia\-frame accelerations ({\tt accel\_ra} and {\tt accel\_dec}), as described in Section \ref{sec:method-accel}. A full list of acceleration predictions is attached in Appendix \ref{sec:App2}, Table \ref{tab:accel_all}.

Eleven systems in our sample have published \Gaia\ DR3 astrometric-acceleration solutions. Three of them— HIP~18320, HIP~39064, and HIP~76626—were fitted by \Gaia\ with the full 9-parameter model, which adds time derivative of the accelerations (the jerk terms); accordingly, we generate nine-parameter predictions for those stars to ensure a like-for-like comparison. Table \ref{tab:gaia_accel} compares the \Gaia\ accelerations with our predictions and lists the corresponding $\chi^{2}$ values. For the three nine-parameter cases, we also compare the jerk terms and report their $\chi^{2}$ values in Table \ref{tab:gaia_accel_9para}.

\begin{figure}
    \centering\label{fig:gaia-agreement}
    \includegraphics[width=8.8cm]{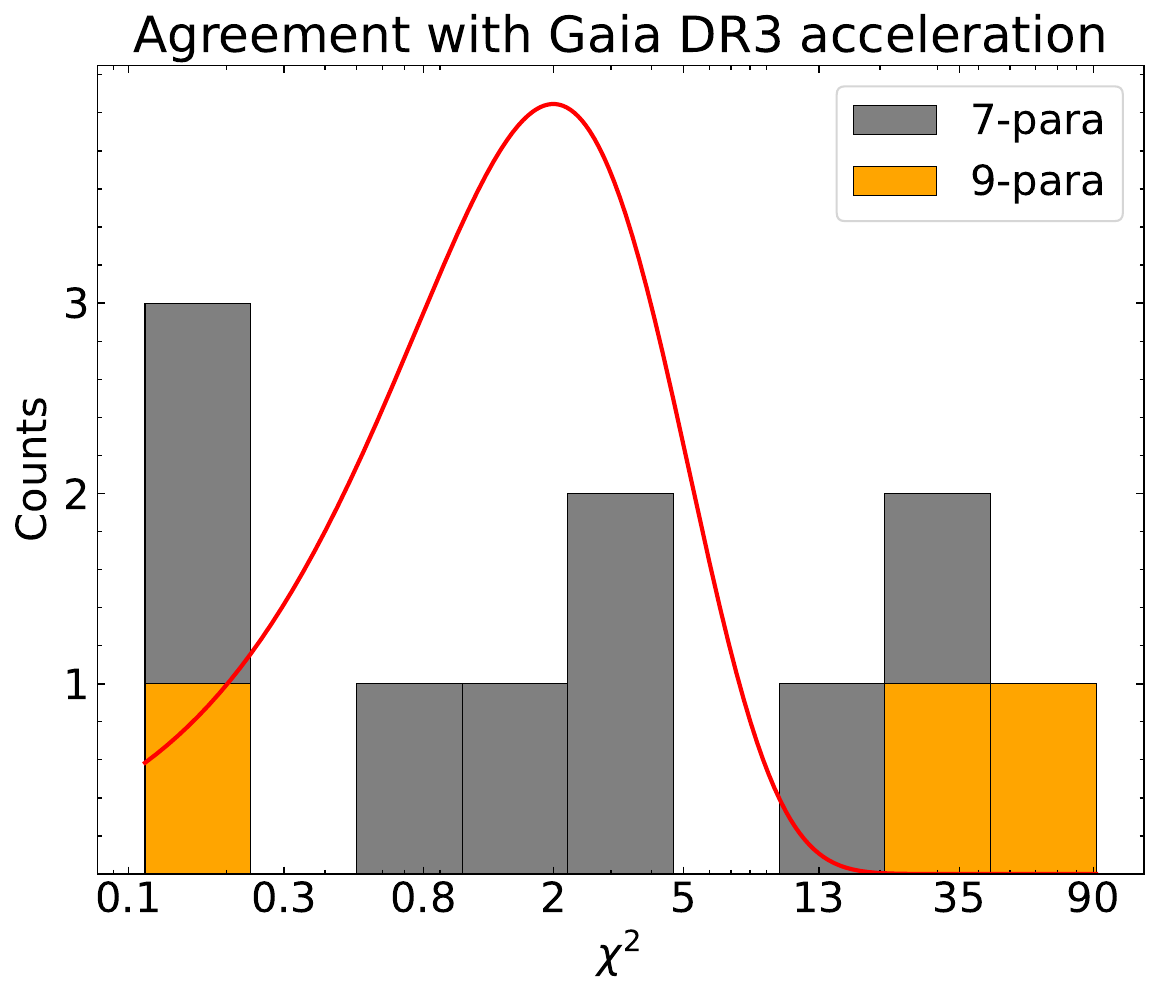}
    \caption{Agreement between measured \Gaia DR3 astrometric accelerations and results expected based on this work. The agreement is quantified by $\chi^2$ between the two sets of values, predicted and measured, for each of eleven stars. Seven-parameter fit systems are shaded in grey, and nine-parameter fit systems are shaded in orange. The probability density function for the $\chi^2$ distribution with 2 degrees of freedom (as $dp/d(\ln \chi^2)$) is plotted in red for reference. Seven of the eleven available comparisons are consistent with the red theoretical curve, while the remaining four are moderately discrepant.}
    \label{fig:gaia-accel-chi2}
\end{figure}

The agreement between the predicted astrometric accelerations from our orbital solutions and the \Gaia DR3 acceleration measurements is illustrated in Figure \ref{fig:gaia-accel-chi2}. This histogram plots the $\chi^2$ values of the eleven matched systems (seven-parameter fits in grey and nine-parameter fits in orange), with a red curve representing the expected probability density function for two degrees of freedom ($dp/d(\ln \chi^2)$). Seven out of the eleven systems fall within the predicted theoretical range, indicating a reasonable level of agreement. However, the remaining systems exhibit larger deviations, suggesting potential discrepancies that warrant further investigation.

Several factors may contribute to the discrepancies between our predicted astrometric accelerations and the \Gaia data products. Firstly, systematic errors in the \Gaia astrometric data could play a significant role. These systematics are not necessarily due to internal issues within \Gaia but may arise from external factors affecting the measurements, such as calibration uncertainties or unmodeled instrumental effects \citep{Gaia_Theastrometric_solution}. Inflating \Gaia DR3's uncertainties by a factor of $\approx$1.5-2 would bring the points into quantitative agreement.  This is similar to the error inflation in proper motion found from agreement with \Hipparcos \citep{HGCA}, in parallax from hierarchical triples \citep{Nagarajan+ElBadry_2024}, and from a full orbital fit \citep{GaiaBH_2023}.  
Secondly, \Gaia does not publish the intermediate astrometric data. Instead, users have access to the GOST scan-forcast tool \footnote{\url{https://gaia.esac.esa.int/gost/}} , which provide the nominal focal-plane crossings expected for a target. GOST omits periods of dead time, ignores focal-plane gaps, and offers an effective on-target observing probability of roughly 80\%; it also reflects \Gaia’s magnitude limits. Therefore, we cannot preform end-to-end sky path fit that uses exactly the same scans that \Gaia used. 

The most severe disagreements are for the nine-parameter fits. A seven-parameter solution fits only a constant acceleration, whereas a nine-parameter solution introduces two additional jerk terms. Over the 34-month \Gaia\ DR3 baseline, the jerk and acceleration coefficients are strongly covariant: a small change in jerk can mimic part of the quadratic curvature, forcing a compensating change in the fitted acceleration. Because our prediction is anchored by decades of RVs and by 25-year baseline HGCA astrometry, it constrains the true long-period curvature far better than the short \Gaia baseline can. We do note that, for HIP 18320 and HIP 30964, the jerk terms show better agreement with our predictions than the acceleration terms do. These terms, as the highest-order coefficients in the fitted sky path fit, do not depend on the adopted central epoch of the fit. While the \Gaia acceleration fits, and especially their uncertainties, should be treated with caution, our results overall suggest broad reliability. 

Although \Gaia\ DR3 includes linear astrometric accleration solutions for a subset of binaries, the overwhelming majority of the 1.46 billion sources are still modelled with the standard five-parameter solution (position, parallax, and proper motion), which cannot reproduce the subtle sky-plane curvature produced by long-period or low-mass companions \citep{GaiaDR3}. Future \Gaia\ data releases that provide epoch astrometry and use an upgraded calibration pipeline \citep{Gaia_Theastrometric_solution} should promote more stars to seven- or nine-parameter status, thereby enabling a more precise, one-to-one comparison with the accelerations predicted in this work. With epoch astrometry, it will also be possible to perform seven- and nine-parameter fits to any \Gaia target, even those that only have published five-parameter fits.

\section{Discussion} \label{sec:Dis}

\subsection{Unusually Massive Companions}

Our \orvara fits reveal two systems where the secondary mass is unexpectedly high—exceeding the mass of the primary star. These systems, HIP~20802 and HIP~59021, stand out as outliers in the mass ratio distribution shown in the bottom panel of Figure~\ref{fig:general}. For HIP~20802, we derive primary and secondary masses of ${1.57} \pm 0.08$~\Msun\ and ${1.77} \pm 0.06$~\Msun, respectively, while HIP~59021 exhibits a primary mass of ${1.37} \pm 0.07$~\Msun\ and a secondary mass of ${1.65} \pm 0.22$~\Msun. These yield mass ratios of $\mathrm{M_2/M_1} = 1.124$ for HIP~20802 and $1.125$ for HIP~59021, making them outliers with secondary components more massive than their primaries.

Assuming that the host stars are main-sequence stars, the secondary companions’ higher masses suggest they should be more luminous than their respective primaries. Using the mass–luminosity relation for main‐sequence stars \citep{mass-luminosity}, we estimate the absolute magnitudes of HIP 20802 B and HIP 59021 B to be approximately $2.48$ and $2.70$, respectively—brighter than their host stars, which we calculate to have $4.12$ for HIP 20802 A and $4.35$ for HIP 59021 A. However, observational data indicate that HIP~20802~A and HIP~59021~A are the brighter components in their systems \citep{Hipparcos}, contradicting this expectation.

A plausible resolution to this discrepancy is that HIP~20802~B and HIP~59021~B are not single stars but rather unresolved binary systems, each consisting of two stars with individual masses of approximately $0.8$~\Msun. This scenario would account for the high total mass inferred from our orbital fits while maintaining a brightness hierarchy consistent with observations. Such a hierarchical triple system could evade detection in photometric surveys due to the small angular separation of the secondary binary components, yet still influence the dynamical mass derived from astrometric and RV measurements \citep{multiple-stars}. Further high-resolution imaging or spectroscopic monitoring could help confirm or refute this hypothesis.

\subsection{Brown Dwarf Mass Estimation}

Estimating the masses of brown dwarfs, especially those orbiting stars as circumstellar companions, is inherently challenging due to their unique physical properties and the low occurrence rate of such systems—an observational phenomenon often termed the brown dwarf desert \citep{2006Grether_BDdesert_quantify}. Brown dwarfs occupy the mass range between the heaviest gas giant planets and the lightest stars, lacking sufficient mass to sustain stable hydrogen fusion in their cores. As a result, they cool and dim over time, leading to degeneracies in mass-luminosity relationships and making it difficult to determine their masses from photometric or spectroscopic observations alone \citep{2000Chabrier}. The scarcity of well-characterized circumstellar brown dwarfs further complicates efforts to test and calibrate theoretical models of their formation and evolution \citep{2011BO_stellar_BD_formation,2021AJ_GMBRANDT_sixmasses}. 

Traditional mass estimates for brown dwarfs often rely on theoretical cooling models, which predict how luminosity and temperature evolve over time based on assumptions about initial conditions and internal processes \citep{2003Baraffe_BDcoolingModel}. However, these models carry uncertainties due to limited empirical data and the complex physics involved, such as cloud formation and atmospheric dynamics \citep{atmo2020, 2015Marley_coolBD}. Discrepancies between model predictions and observations highlight the need for independent mass measurements to refine these models.

Our results provide precise dynamical mass estimations for multiple circumstellar brown dwarfs, derived independently of cooling models through combined astrometric and RV data. By accurately measuring their orbital motions, we obtain direct mass measurements that are not contingent on theoretical luminosity or temperature evolution \citep{Brandt2019}. These dynamical masses serve as critical benchmarks for testing and improving theoretical models of brown dwarf cooling and atmospheres. 

\subsection{Data Bias}
\label{sec:data_bias}
Our sample selection and data usage introduce a bias towards detecting massive, long-period companions. This bias arises because the combination of RV (RV) and astrometry data is inherently more sensitive to companions that exert significant gravitational influences on their host stars—typically massive companions in wider orbits \citep{2008Cumming, 2012Wright}. Such companions induce larger RV amplitudes and more substantial astrometric signals, making them easier to detect with the precision of current instruments \citep{Bowler2016}. However, this effect is mitigated by the fact that astrometric acceleration becomes more detectable for stars closer to Earth, which differs from the selection effects of the RV method \citep{2024Brandt_PASP}.

This bias towards massive, long-period companions affects our understanding of the true distribution of exoplanet masses and orbital parameters. It can skew statistical analyses and impact theoretical models of planet formation and evolution that rely on accurate population demographics \citep{Mordasini2012}. For instance, core accretion models predict a certain frequency and distribution of planet masses and orbital separations, which may not align with observations if the sample is biased \citep{2004Ida}. 

The more serious limitation, however, is the poorly documented selection function of the RV survey. Stars were added to monitoring lists over several decades according to heterogeneous and often unpublished criteria. Without a quantitative record of those choices the completeness of the parent sample cannot be reconstructed, so the present catalog should be viewed as an informative but opportunistic collection rather than a statistical survey. 

In our Keplerian fits we further exclude systems exhibiting inconsistencies between RV and astrometry. These mismatches can arise from stellar activity–induced signals (e.g.\ rotating starspots or faculae causing RV jitter and photocenter shifts; \citealt{Lagrange2010}), instrument zero‐point drifts, or multi‐body architectures (e.g.\ an inner companion dominating the RV signal while an outer object drives the long‐term astrometric acceleration; \citealt{Montet2014}). By discarding these mismatched systems, we underpopulate regions of parameter space where complex dynamics, activity‐induced noise, or instrumental systematics are important—such as compact multi‑planet systems, edge‑on outer brown‑dwarf companions, and jitter‑dominated M dwarfs.

\section{Conclusion}\label{sec:conclude}

In this paper, we present dynamical mass and orbital solutions for \totalstar systematically selected stellar systems. These solutions are derived by combining absolute astrometry from the HGCA \citep{HGCA} with RV data \citep{HIRESdata,HARPSdata,Rosenthal2021_CLS}, using the \orvara\ software package \citep{orvara}. Our analysis provides dynamical masses and orbital parameters for \totalcompanion companions—spanning \startwobd two-body fits and \starthreebd three-body fits—including \planetcomp planets, \BDcomp brown dwarfs, and \stellarcomp stellar companions. Among these, we identify one newly reported potential substellar companion (HIP 58289 B) that have not appeared in the literature before.

Our results expand the catalog of nearby planetary and substellar systems, providing orbital solutions with higher precision due to the synergistic use of astrometric and RV data. This work contributes to a deeper understanding of the architecture and mass distribution of planetary systems in our solar neighborhood, aiding in the refinement of planet formation and evolution models \citep{Mordasini2012, Bowler2016}.

Furthermore, our target systems offer a promising opportunity for future direct imaging studies. The characteristics of the companions—their relatively large masses, wide orbital separations, and favorable contrast ratios with the host star—make them suitable candidates for observation with high-contrast imaging instruments. Direct imaging can provide spatially resolved measurements that are crucial for constraining atmospheric properties, rotation rates, and potential signatures of formation mechanisms \citep{Bowler2016}. Combining astrometric data with direct imaging observations enhances our understanding of the orbital architecture and dynamical interactions within these systems \citep{Li2023_imaging}.

The synergy between astrometry and direct imaging is particularly valuable for refining orbital parameters and validating dynamical mass estimates. Future facilities equipped with advanced adaptive optics systems and/or coronagraphs, such as the Extremely Large Telescope (ELT) and the James Webb Space Telescope (JWST), will possess the sensitivity and resolution necessary to detect and characterize these companions in greater detail \citep{Quanz2015, Beichman2010}. Such observations will contribute significantly to our knowledge of planet formation theories and the diversity of planetary system architectures in our galaxy.

We also compare our derived orbital solutions with those available from the \Gaia DR3 NSS data product \citep{GaiaDR3} for the systems where overlap exists. Unfortunately, only a small number of systems in our study have corresponding orbital solutions in \Gaia DR3, which limits our ability to draw statistically significant conclusions about the performance and consistency of the \Gaia DR3 data processing in the context of dynamical mass determinations.  

Future \Gaia\ data releases will likely include extended time baselines, refined calibration processes, individual‐epoch astrometry, and NSS solutions for a larger number of sources \citep{Gaiamission,GaiaDR3}. This will not only improve the precision of the astrometric measurements but also enable the use of seven-parameter astrometric solutions that can account for non-linear proper motions due to orbital acceleration. Such improvements are crucial for accurately characterizing systems like those studied here and will enhance our ability to validate and refine our orbital models, as well as the models used in \Gaia data processing. The extended time coverage and improved data quality will allow for the detection and characterization of smaller and more distant companions, thereby expanding our understanding of stellar multiplicity and exoplanet~demographics \citep{Gaiamission}. Furthermore, access to individual epoch astrometry will facilitate more sophisticated orbital analyses, helping to resolve ambiguities such as inclination degeneracies and improve the accuracy of dynamical mass determinations \citep{orvara}.

\section*{acknowledgements}\label{sec:Acknowledge}

This work has made use of data from the European Space Agency (ESA) mission
\Gaia\ (\url{https://www.cosmos.esa.int/gaia}), processed by the Gaia
Data Processing and Analysis Consortium (DPAC,
\url{https://www.cosmos.esa.int/web/gaia/dpac/consortium}). Funding for DPAC
has been provided by national institutions, in particular those
participating in the Gaia Multilateral Agreement.

This work also makes use of the HIRES dataset provided by the NASA/IPAC Infrared Science Archive (IRSA) at Caltech (DOI:10.26131/IRSA534; \url{https://doi.org/10.26131/IRSA534}), and of the Gaia DR3 Source Catalogue hosted by IRSA (DOI:10.26131/IRSA544; \url{https://doi.org/10.26131/IRSA544}).  We further utilized the Hipparcos–Gaia Catalog of Accelerations \citep[HGCA;][]{HGCA}, the California Legacy Search \citep[CLS;][]{Rosenthal2021_CLS}, the High-Accuracy Radial velocity Planetary Searcher \citep[HARPS;][]{Harps} on the ESO 3.6 m telescope, and the Washington Double Star Catalog \citep[WDS;][]{WDS}. Original WDS sources are listed in Appendix~\ref{sec:WDS_ref}.

This work uses Astropy \citep{astropy:2013, astropy:2018}, scipy \citep{2020SciPy-NMeth}, numpy \citep{numpy1, numpy2}, \htofcodename \citep{htof_zenodo, htof}, \orvara \citep{orvara}, and Jupyter (\url{https://jupyter.org/}).

\bibliography{refs}

\appendix
\section{Removed Targets with Reasons Specified} \label{sec:App1}

\subsection{Packed RV}
The RV measurements are closely clustered in time, reducing the number of effectively distinct observational epochs and limiting the available orbital information. 

\quad --- HIP 3391, HIP 10812, HIP 14086, HIP 16537, HIP 36349, HIP 37419, HIP 71755, HIP 101382, HIP 105521, HIP 108870

\subsection{Straight Line RV}
The straight line RV
suggests the absence of detectable curvature, resulting an insufficient coverage of orbital phases.

\quad --- HIP 726, HIP 948, HIP 8674, HIP 12350, HIP 12929, HIP 18806, HIP 18944, HIP 37520, HIP 42885, HIP 45839, HIP 52720, HIP 54582, HIP 56884, HIP 60591, HIP 60973, HIP 61743, HIP 71803, HIP 84801, HIP 85533, HIP 87937, HIP 94751, HIP 108095, HIP 108158, HIP 108950, HIP 109787, HIP 110156, HIP 111148, HIP 116374

\subsection{Additional Removed Targets}

\textbf{Short period:} The long \Hipparcos--\Gaia baseline effectively averages out the reflex motion from short-period companions (under 7\,years), making their dynamical signatures difficult to detect in absolute astrometry. Consequently, any orbital solutions derived for these systems may be unreliable.

\quad --- HIP 6712,	HIP 19911,	HIP 22627,	HIP 24205,	HIP 42723,	HIP 43299,	HIP 62534,	HIP 73408,	HIP 76626,	HIP 86990,	HIP 88595,	HIP 98714,	HIP 100970,	HIP 109149

\textbf{Poor data:} (case 1) RV data of poor quality that exhibit large uncertainties; (case 2) inconsistencies between astrometric and RV measurements. Both cases would render accurate orbit fitting infeasible.

\quad --- (case 1) HIP 30034, HIP 34961, HIP 43565, HIP 54211, HIP 81662, HIP 102626

\quad --- (case 2) HIP 7244, HIP 24186, HIP 52498, HIP 56633, HIP 111449

\textbf{Resolved binary stars:}

\quad --- HIP 45343 (GJ 338 A) \& HIP 120005 (GJ 338 B); HIP 91768 (GJ 725 A) \& HIP 91772 (GJ 725 B); HIP 104214 (* 61 Cyg A) \& HIP 104217 (* 61 Cyg B)

\textbf{Multiple close-in companions that exceed \orvara's computational capabilities:} 

\quad ---  HIP 65271, HIP 77301 (HD 141399)

\textbf{Special Cases:}

\quad --- HIP~25486: This young, magnetically active star can mask the reflex motion from its companion(s). Consequently, significant 
stellar activity prevents clear detection of companion-induced signals.

\quad --- HIP~106006: Here, the inner companion is sufficiently constrained by RV data and has a period shorter than 7~years. However, the outer companion’s RV signature is essentially a straight-line trend, making it infeasible to derive a full orbital solution when combined with the short-period inner companion. We therefore exclude this system.

\section{Complete Results} \label{sec:App2}
\subsection{Predicted \Gaia Acceleration}
Table \ref{tab:accel_all} presents the complete list of predicted \Gaia accelerations ($\mathrm{mas/yr^2}$) for \totalstar selected stars. For system with bimodal inclination, we separately calculate ${\rm i}>\pi/2$(retrograde motion) and ${\rm i}<\pi/2$(prograde motion) situations. Here we report the retrograde mode for these systems in the bottom section to remain consistent with previous tables. This table also includes the sources of RV and relative astrometry data used in our analysis, in the last two colomns. Data sources are denoted as follows: CLS refers to RV data from \citet{Rosenthal2021_CLS}; HARPS refers to RV data from \citet{HARPSdata}; HIRES refers to RV data from \citet{HIRESdata}; WDS refers to relative astrometry data from Washington Double Star Catalog \citet{WDS}; Gaia2nd refers to relative astrometry data from \Gaia astrometry of the secondary mass \citet{GaiaDR3}y; BOTH refers to relative astrometry data from both WDS and secondary \Gaia; and \texttt{/} indicates that no relative astrometry data is used. The original source of data from WDS are listed in Appendix \ref{sec:WDS_ref}.

\input{table_accel_long}

\subsection{Stellar Companions}
\label{sec:app_stellar}
Table \ref{tab:stellar_long} presents the MCMC results of stellar companions. For system with bimodal inclination, we report the retrograde mode (${\rm i}>\pi/2$).
\input{table_stellar_long}

\section{Original WDS Data Souce} \label{sec:WDS_ref}

\hspace*{-1\parindent}HIP 184: \citet{ElB2021, Kpp2019j, TMA2003} \newline
HIP 1444: \citet{Llo2002} \newline
HIP 1475: \citet{ADP1998, Cll2003, CND2021, CrC2014, Dal2005a, Dal2014, ElB2018, ElB2021, FyM2013, Gel1989, Hei1992a, Izm2010, Izm2015, Izm2020, Kpp2018m, Kpp2020f, PkO2014b, PkO2021a, Sca1992, TMA2003, Tob2000, UPR2006, UPR2007, USN1974, USN1978, VVO1972b, WDK2015, WSI2004b} \newline
HIP 3821: \citet{USN1963, USN1974, USN1978, USN1984, Tor1985, TYC2002, TMA2003, WSI2004b, ElB2021} \newline
HIP 4393: \citet{ElB2018, ElB2021, TMA2003} \newline
HIP 6653: \citet{ElB2021, Llo2002, RAO2015} \newline
HIP 6712: \citet{Tok2012a, Tok2018c, Tok2024c} \newline
HIP 11433: \citet{Tok2020, Tok2021d} \newline
HIP 13642: \citet{Alz2003b, Btg2010, Bvd2012h, HIP1997, HLA2021, Hop1967, Izm2010, Izm2015, Izm2020, Kpp2019b, Kpr1992, Lmp2004, SHS2020e, TMA2003, Tor1980, Tor1983, TYC2002, Wly2012, WSI2001b, WSI2004a, WSI2007a, WSI2011, WSI2011b, WSI2013} \newline
HIP 17157: \citet{VSP2022b} \newline
HIP 18267: \citet{ElB2018, ElB2021, Hei1980a, HLA2021, SHS2022b} \newline
HIP 18317: \citet{Cou1964, HIP1997, Tok2019c, VBs1974, Wor1967b} \newline
HIP 18512: \citet{ADP1995, ADP1998, CrC2017, ElB2018, ElB2021, Hei1985a, HIP1997, Kpp2018m, Lmp2004, TMA2003, TYC2002, USN1974, USN1978, WIS2012} \newline
HIP 27207: \citet{HLA2021} \newline
HIP 28267: \citet{Egn2007, ElB2021, HLA2021, Kpp2019a, RAO2022} \newline
HIP 38931: \citet{ElB2018, ElB2021, Hei1985a, HLA2021, TMA2003} \newline
HIP 41844: \citet{2012Creep_trendStar} \newline
HIP 42220: \citet{Hor2010, Hor2011, Jod2013, MnA2019, RAO2022, Rdr2015} \newline
HIP 49769: \citet{Hln1965, Rst1955, Tok2014a, Tok2018c, Tok2019c} \newline
HIP 51525: \citet{ElB2018, ElB2021} \newline
HIP 55459: \citet{ElB2021, HrZ2020} \newline
HIP 56452: \citet{ElB2021, HrZ2020, Kpp2018m, Kpp2020f, Luy1979, TMA2003} \newline
HIP 59021: \citet{Knp1967, HIP1997, TYC2002, TMA2003, Kpp2018m, Kpp2019a, ElB2021} \newline
HIP 61901: \citet{ElB2021, RAO2022, Rdr2015, Tok2022f} \newline
HIP 62039: \citet{Llo2002} \newline
HIP 64150: \citet{CrJ2013b, VSP2021a} \newline
HIP 64797: \citet{Alz2003b, Baz1972, Cgl2023b, Cll2003, CrC2017, Ctt2021, ElB2018, ElB2021, EvC2019, Gir1992, Hei1978b, Hei1995, HIP1997, HLA2021, Hle1994, Kpp2018m, Kpr1979, LBu1993b, Los2010, Mlr1978b, Nug2012, Pop1975a, Pop1991, Ppr2020, Rdr2015, TMA2003, TYC2002, USN1974, USN1978, USN1984, WSI2006a, WSI2013} \newline
HIP 71631: \citet{Kng2005, Laf2007, Met2004a, Met2009, Rbr2012a} \newline
HIP 71898: \citet{ElB2021, Gki2004, Kpp2019b, RAO2020b, WDK2015} \newline
HIP 77052: \citet{Cou1967b, Baz1972, USN1974, Wor1989, Rdr2015, HLA2021} \newline
HIP 78395: \citet{Hor2021b} \newline
HIP 79619: \citet{Llo2002} \newline
HIP 81375: \citet{HLA2021, Hln1972, Kpp2018m, Rst1955, TMA2003} \newline
HIP 82032: \citet{ElB2021, Kpp2018m, Tok2019c, Tok2020, Tok2022f} \newline
HIP 83020: \citet{CrC2017, HIP1997, HLA2021, Izm2010, Izm2015, Izm2020, Kis2009b, Kpp2018m, Kpp2019b, Kpp2022a, Kpr1976b, Lmp2004, PkO2014b, StJ2018b, TYC2002, USN1974, WSI2004b} \newline
HIP 85533: \citet{TMA2003} \newline
HIP 85653: \citet{HLA2019, HLA2021} \newline
HIP 86974: \citet{Ken2007, Llo2002, Rbr2016a, Rbr2018, Trn2001} \newline
HIP 98677: \citet{ElB2021, HLA2021, Kpp2018m, RAO2022, Rbr2018} \newline
HIP 99385: \citet{Beu2004, Tok2016a, Tok2019c, Tok2022f} \newline
HIP 101345: \citet{Llo2002, Nbg1966, RAO2022, Tok2016d} \newline
HIP 101597: \citet{Gii2021, Hor2002a, Hor2008, Hor2010, Llo2002, Orl2021} \newline
HIP 101846: \citet{Kpp2019a} \newline
HIP 103768: \citet{ElB2021, HrZ2020} \newline
HIP 104239: \citet{Ary1983, HIP1997, OCC2017a, TYC2002, USN1974, USN1978, WSI2017a} \newline
HIP 110109: \citet{Knp1968, Kpp2019a, The1975, Tok2010, Tok2014a, Tok2015c, Tok2022f, Wor1978} \newline
HIP 111571: \citet{Jod2013, Tok2020, Tok2021d} \newline

\end{document}

%% file: table_substellar.tex
\begin{deluxetable*}{llllllll}
\tablecaption{Posterior 68\% confidence intervals for companions that are likely to be substellar. 
\label{tab:substellar}}
\tabletypesize{\scriptsize}
\tablewidth{0pt}
\tablehead{ 
\colhead{HIP ID}
&\colhead{RV jitter(m/s)}
&\colhead{${\rm M}_{\mathrm{star}}~(\mathrm{M}_\odot$)}
&\colhead{$\mathrm{M}_2~({\rm M}_{\rm Jup})$}
&\colhead{Semi-major~axis (AU)}
&\colhead{inclination($\degree$)}
&\colhead{Period (yr)}
&\colhead{Eccentricity}}
\startdata
\multicolumn{8}{c}{Full constraints listed} \\
\hline
3850 B	& $4.80^{+0.52}_{-0.45}$	& $0.867^{+0.030}_{-0.028}$	& $68.7^{+1.5}_{-1.4}$	& $10.15^{+0.17}_{-0.16}$	& $49.52^{+0.92}_{-0.92}$	& $33.48^{+0.41}_{-0.41}$	& $0.7319^{+0.0014}_{-0.0014}$	\\
8770 d	& $10.10^{+0.75}_{-0.72}$	& $1.268^{+0.065}_{-0.062}$	& $9.1^{+34.6}_{-4.0}$	& $13.3^{+18.0}_{-5.4}$	& $94^{+31}_{-36}$	& $43^{+110}_{-23}$	& $0.24^{+0.22}_{-0.16}$	\\
17960 b	& $6.78^{+0.57}_{-0.51}$	& $1.221^{+0.060}_{-0.061}$	& $5.00^{+0.36}_{-0.26}$	& $4.817^{+0.079}_{-0.083}$	& $89^{+20}_{-19}$	& $9.549^{+0.042}_{-0.042}$	& $0.046^{+0.025}_{-0.026}$	\\
19428 B	& $5.75^{+0.72}_{-0.62}$	& $1.303^{+0.066}_{-0.066}$	& $59^{+18}_{-18}$	& $23.2^{+17.0}_{-8.0}$	& $121^{+39}_{-86}$	& $96^{+124}_{-45}$	& $0.868^{+0.079}_{-0.120}$	\\
21865 b	& $2.91^{+1.10}_{-0.79}$	& $0.718^{+0.036}_{-0.035}$	& $2.83^{+0.67}_{-0.43}$	& $8.0^{+2.6}_{-2.2}$	& $99^{+49}_{-53}$	& $27^{+14}_{-10}$	& $0.68^{+0.14}_{-0.21}$	\\
36941 b	& $4.16^{+0.47}_{-0.42}$	& $1.300^{+0.064}_{-0.065}$	& $17.71^{+0.85}_{-0.83}$	& $6.41^{+0.10}_{-0.11}$	& $134.0^{+2.3}_{-2.4}$	& $14.130^{+0.070}_{-0.064}$	& $0.587^{+0.017}_{-0.015}$	\\
39417 c	& $2.65^{+0.31}_{-0.27}$	& $1.145^{+0.056}_{-0.057}$	& $21.6^{+13.6}_{-8.2}$	& $18.0^{+10.0}_{-5.4}$	& $120^{+33}_{-57}$	& $71^{+69}_{-29}$	& $0.24^{+0.17}_{-0.10}$	\\
40687 c	& $4.02^{+0.51}_{-0.46}$	& $1.201^{+0.022}_{-0.022}$	& $15.03^{+0.90}_{-0.75}$	& $12.1^{+1.1}_{-1.1}$	& $93^{+19}_{-23}$	& $38.1^{+5.3}_{-4.9}$	& $0.399^{+0.046}_{-0.049}$	\\
42030 c	& $3.69^{+0.44}_{-0.38}$	& $1.246^{+0.063}_{-0.061}$	& $78^{+58}_{-36}$	& $30^{+14}_{-11}$	& $84^{+41}_{-40}$	& $142^{+105}_{-71}$	& $0.23^{+0.27}_{-0.16}$	\\
50653 b	& $3.91^{+0.41}_{-0.37}$	& $1.300^{+0.064}_{-0.065}$	& $5.50^{+0.63}_{-0.48}$	& $4.810^{+0.081}_{-0.084}$	& $93^{+24}_{-28}$	& $9.231^{+0.068}_{-0.063}$	& $0.334^{+0.043}_{-0.039}$	\\
58289 B	& $3.60^{+0.62}_{-0.53}$	& $0.800^{+0.040}_{-0.040}$	& $40^{+21}_{-17}$	& $23.6^{+15.0}_{-7.9}$	& $98^{+45}_{-43}$	& $125^{+129}_{-57}$	& $0.37^{+0.25}_{-0.23}$	\\
60648 b	& $6.43^{+0.92}_{-0.75}$	& $0.739^{+0.036}_{-0.037}$	& $2.86^{+0.69}_{-0.45}$	& $3.663^{+0.071}_{-0.073}$	& $75^{+59}_{-29}$	& $8.14^{+0.13}_{-0.12}$	& $0.489^{+0.041}_{-0.037}$	\\
89620 B	& $7.0^{+1.3}_{-1.1}$	& $1.137^{+0.033}_{-0.031}$	& $61.29^{+1.05}_{-0.96}$	& $5.604^{+0.050}_{-0.050}$	& $58.70^{+0.72}_{-0.73}$	& $12.134^{+0.015}_{-0.015}$	& $0.3411^{+0.0035}_{-0.0035}$	\\
95015 b	& $3.48^{+0.61}_{-0.51}$	& $1.011^{+0.051}_{-0.050}$	& $10.9^{+4.2}_{-1.5}$	& $7.57^{+0.14}_{-0.14}$	& $79^{+35}_{-21}$	& $20.59^{+0.21}_{-0.20}$	& $0.808^{+0.110}_{-0.090}$	\\
95319 B	& $2.51^{+0.14}_{-0.12}$	& $0.982^{+0.049}_{-0.048}$	& $35.91^{+0.69}_{-0.65}$	& $27.4^{+5.6}_{-3.6}$	& $47.9^{+5.7}_{-8.6}$	& $143^{+47}_{-27}$	& $0.277^{+0.170}_{0.091}$	\\
98819 B	& $6.27^{+0.33}_{-0.30}$	& $0.994^{+0.029}_{-0.029}$	& $71.4^{+1.0}_{-1.0}$	& $17.64^{+0.28}_{-0.27}$	& $97.47^{+0.31}_{-0.31}$	& $71.9^{+2.4}_{-2.2}$	& $0.5081^{+0.0089}_{-0.0087}$	\\
113421 c	& $3.19^{+0.25}_{-0.21}$	& $1.132^{+0.036}_{-0.034}$	& $4.50^{+0.12}_{-0.12}$	& $6.100^{+0.060}_{-0.060}$	& $89.3^{+8.9}_{-9.0}$	& $14.140^{+0.041}_{-0.042}$	& $0.3908^{+0.0082}_{-0.0077}$	\\
116250 b	& $2.29^{+0.19}_{-0.17}$	& $1.340^{+0.070}_{-0.070}$	& $19.9^{+1.7}_{-1.5}$	& $10.79^{+1.90}_{-0.72}$	& $160.3^{+3.1}_{-5.9}$	& $30.4^{+8.6}_{-2.9}$	& $0.199^{+0.120}_{-0.061}$	\\
116745 b	& $1.90^{+0.22}_{-0.18}$	& $0.700^{+0.010}_{-0.010}$	& $5.0^{+1.5}_{-1.0}$	& $12.9^{+7.6}_{-3.1}$	& $94^{+28}_{-29}$	& $55^{+55}_{-19}$	& $0.54^{+0.18}_{-0.15}$	\\
\hline
\multicolumn{8}{c}{No inclination constraints ($\mathrm{M_2\sin{i}}$ only)}  \\
\hline
8770 b	& $10.10^{+0.75}_{-0.72}$	& $1.268^{+0.065}_{-0.062}$	& $5.04^{+0.22}_{-0.21}$\tablenotemark{\footnotesize a}	& $2.918^{+0.049}_{-0.049}$	& $87^{+42}_{-43}$	& $4.414^{+0.017}_{-0.017}$	& $0.390^{+0.016}_{-0.017}$	\\
9683 b	& $2.80^{+0.26}_{-0.23}$	& $1.158^{+0.058}_{-0.056}$	& $2.382^{+0.080}_{-0.080}$\tablenotemark{\footnotesize a}	& $0.843^{+0.014}_{-0.014}$	& $86^{+41}_{-38}$	& $0.718323(68)$\tablenotemark{\footnotesize b}	& $0.3598^{+0.0059}_{-0.0060}$	\\
20277 b	& $2.73^{+0.25}_{-0.23}$	& $0.873^{+0.039}_{-0.039}$	& $0.674^{+0.021}_{-0.021}$\tablenotemark{\footnotesize a}	& $0.1286^{+0.0019}_{-0.0020}$	& $89^{+37}_{-39}$	& $0.04933461(51)$\tablenotemark{\footnotesize b}	& $0.0334^{+0.0088}_{-0.0091}$	\\
39417 b	& $2.65^{+0.31}_{-0.27}$	& $1.145^{+0.056}_{-0.057}$	& $3.37^{+0.13}_{-0.13}$\tablenotemark{\footnotesize a}	& $3.558^{+0.058}_{-0.060}$	& $90^{+47}_{-45}$	& $6.259^{+0.015}_{-0.014}$	& $0.418^{+0.014}_{-0.013}$	\\
40687 b	& $4.47^{+0.50}_{-0.42}$	& $1.211^{+0.065}_{-0.071}$	& $1.972^{+0.070}_{-0.078}$\tablenotemark{\footnotesize a}	& $0.0710^{+0.0012}_{-0.0014}$	& $83^{+30}_{-28}$	& $0.017183893(34)$\tablenotemark{\footnotesize b}	& $0.1533^{+0.0049}_{-0.0048}$	\\
42030 b	& $3.69^{+0.44}_{-0.38}$	& $1.246^{+0.063}_{-0.061}$	& $3.19^{+0.15}_{-0.15}$\tablenotemark{\footnotesize a}	& $4.917^{+0.084}_{-0.083}$	& $90^{+44}_{-42}$	& $9.750^{+0.048}_{-0.047}$	& $0.272^{+0.020}_{-0.020}$	\\
64457 b	& $3.19^{+0.22}_{-0.21}$	& $0.900^{+0.045}_{-0.046}$	& $1.043^{+0.038}_{-0.038}$\tablenotemark{\footnotesize a}	& $1.180^{+0.019}_{-0.021}$	& $110^{+52}_{-52}$	& $1.34858(82)$\tablenotemark{\footnotesize b}	& $0.121^{+0.013}_{-0.014}$	\\
84171 b	& $2.32^{+0.25}_{-0.22}$	& $0.999^{+0.046}_{-0.045}$	& $9.82^{+0.29}_{-0.29}$\tablenotemark{\footnotesize a}	& $0.5126^{+0.0077}_{-0.0078}$	& $87^{+32}_{-28}$	& $0.365235(10)$\tablenotemark{\footnotesize b}	& $0.6479^{+0.0007}_{-0.00077}$	\\
89844 b	& $8.04^{+0.51}_{-0.47}$	& $1.064^{+0.050}_{-0.048}$	& $8.09^{+0.24}_{-0.25}$\tablenotemark{\footnotesize a}	& $0.3006^{+0.0045}_{-0.0047}$	& $92^{+34}_{-36}$	& $0.1591010(11)$\tablenotemark{\footnotesize b}	& $0.5288^{+0.0020}_{-0.0021}$	\\
95467 c	& $2.34^{+0.13}_{-0.12}$	& $0.871^{+0.043}_{-0.043}$	& $0.749^{+0.029}_{-0.029}$\tablenotemark{\footnotesize a}	& $1.897^{+0.031}_{-0.032}$	& $111^{+59}_{-62}$	& $2.798^{+0.010}_{-0.009}$	& $0.235^{+0.017}_{-0.016}$	\\
95740 b	& $5.23^{+0.42}_{-0.38}$	& $1.200^{+0.060}_{-0.059}$	& $3.78^{+0.13}_{-0.14}$\tablenotemark{\footnotesize a}	& $1.522^{+0.025}_{-0.025}$	& $90^{+39}_{-41}$	& $1.71016(82)$ \tablenotemark{\footnotesize b}	& $0.3842^{+0.0080}_{-0.0078}$	\\
97336 b	& $3.06^{+0.24}_{-0.21}$	& $1.192^{+0.057}_{-0.110}$	& $0.555^{+0.018}_{-0.032}$\tablenotemark{\footnotesize a}	& $0.0440^{+0.0010}_{-0.0010}$	& $79^{+29}_{-24}$	& $0.008478014(10)$\tablenotemark{\footnotesize b}	& $0.0139^{+0.0070}_{-0.0069}$	\\
113421 b	& $3.73^{+0.24}_{-0.22}$	& $1.135^{+0.059}_{-0.065}$	& $1.451^{+0.050}_{-0.056}$\tablenotemark{\footnotesize a}	& $0.0756^{+0.0013}_{-0.0015}$	& $91^{+25}_{-24}$	& $0.019512280(30)$ \tablenotemark{\footnotesize b}	& $0.1311^{+0.0030}_{-0.0029}$	\\
\hline
\multicolumn{8}{c}{Bimodal inclination constraints ($i>90^\circ$ mode shown)} \\
\hline
9683 c	& $2.78^{+0.26}_{-0.22}$	& $1.158^{+0.054}_{-0.055}$	& $2.83^{+0.82}_{-0.67}$	& $2.929^{+0.045}_{-0.047}$	& $136^{+11}_{-21}$	& $4.6535^{+0.0087}_{-0.0082}$	& $0.0131^{+0.0150}_{-0.0094}$	\\
10337 c	& $13.21^{+1.10}_{-0.95}$	& $0.691^{+0.013}_{-0.013}$	& $4.47^{+0.57}_{-0.54}$	& $5.96^{+0.49}_{-0.16}$	& $140.0^{+5.1}_{-7.8}$	& $17.45^{+2.20}_{-0.68}$	& $0.296^{+0.079}_{-0.130}$	\\
12436 Ac	& $3.32^{+0.51}_{-0.42}$	& $0.847^{+0.042}_{-0.042}$	& $10.90^{+0.62}_{-0.59}$	& $9.14^{+0.38}_{-0.33}$	& $133.5^{+2.7}_{-3.0}$	& $29.9^{+1.8}_{-1.4}$	& $0.692^{+0.012}_{-0.012}$	\\
20277 d	& $2.73^{+0.25}_{-0.23}$	& $0.871^{+0.040}_{-0.039}$	& $7.58^{+0.60}_{-0.56}$	& $5.514^{+0.084}_{-0.084}$	& $117.8^{+6.9}_{-11.0}$	& $13.813^{+0.051}_{-0.051}$	& $0.3239^{+0.0092}_{-0.0089}$	\\
64457 c	& $3.19^{+0.22}_{-0.21}$	& $0.900^{+0.044}_{-0.047}$	& $1.47^{+0.58}_{-0.63}$	& $5.03^{+0.12}_{-0.12}$	& $152.0^{+8.4}_{-27.0}$	& $11.89^{+0.30}_{-0.27}$	& $0.074^{+0.047}_{-0.045}$	\\
70849 b	& $5.12^{+0.65}_{-0.57}$	& $0.647^{+0.013}_{-0.013}$	& $4.55^{+0.51}_{-0.51}$	& $4.037^{+0.030}_{-0.030}$	& $130^{+14}_{-26}$	& $10.046^{+0.054}_{-0.046}$	& $0.584^{+0.070}_{-0.079}$	\\
84171 c	& $2.31^{+0.25}_{-0.22}$	& $0.997^{+0.047}_{-0.042}$	& $10.53^{+0.65}_{-0.59}$	& $5.570^{+0.086}_{-0.084}$	& $110.9^{+6.9}_{-9.4}$	& $13.093^{+0.081}_{-0.081}$	& $0.2602^{+0.0047}_{-0.0048}$	\\
89844 c	& $8.01^{+0.53}_{-0.47}$	& $1.063^{+0.045}_{-0.048}$	& $20.5^{+1.6}_{-1.6}$	& $2.929^{+0.040}_{-0.045}$	& $117.3^{+6.2}_{-10.0}$	& $4.8164^{+0.0052}_{-0.0030}$	& $0.2131^{+0.0034}_{-0.0034}$	\\
90055 b	& $6.26^{+0.60}_{-0.53}$	& $0.826^{+0.039}_{-0.040}$	& $7.6^{+4.0}_{-1.4}$	& $11.0^{+7.4}_{-2.1}$	& $129.9^{+7.1}_{-7.8}$	& $40^{+46}_{-11}$	& $0.21^{+0.24}_{-0.13}$	\\
95467 d	& $2.35^{+0.13}_{-0.12}$	& $0.869^{+0.042}_{-0.041}$	& $2.7^{+1.7}_{-2.0}$	& $10.3^{+3.9}_{-2.6}$	& $164.5^{+6.8}_{-53.0}$	& $35^{+22}_{-12}$	& $0.64^{+0.11}_{-0.12}$	\\
95740 c	& $5.23^{+0.43}_{-0.38}$	& $1.200^{+0.059}_{-0.059}$	& $8.92^{+0.91}_{-0.56}$	& $6.07^{+0.11}_{-0.11}$	& $110^{+12}_{-13}$	& $13.62^{+0.13}_{-0.13}$	& $0.019^{+0.011}_{-0.011}$	\\
97336 c	& $3.06^{+0.24}_{-0.22}$	& $1.192^{+0.056}_{-0.110}$	& $2.87^{+0.40}_{-0.42}$	& $4.690^{+0.080}_{-0.130}$	& $137.2^{+5.8}_{-9.1}$	& $9.297^{+0.071}_{-0.072}$	& $0.298^{+0.020}_{-0.020}$	\\
106440 b	& $2.03^{+0.13}_{-0.12}$	& $0.4491^{+0.0087}_{-0.0091}$	& $1.039^{+0.044}_{-0.043}$	& $3.651^{+0.071}_{-0.059}$	& $131.6^{+2.7}_{-3.4}$	& $10.39^{+0.29}_{-0.22}$	& $0.021^{+0.027}_{-0.015}$	\\
\hline \hline 
\enddata
\tablenotetext{a}{M$_2\sin i$ reported}
\tablenotetext{b}{A value written as $x(y)$ means $x \pm 0.y$ in the last digit(s); e.g.\ 0.718323(68) represents $0.718323 \pm 0.000068$.}
\end{deluxetable*}

%% file: table_gaia_2bd.tex
\begin{deluxetable*}{c|cccc|cccc}
\tablecaption{Comparison with Gaia DR3 two-body orbital fits.
\label{Tab:gaia-2bd}}
\tablehead{\colhead{ } &
\multicolumn{4}{c}{Gaia DR3 results\tablenotemark{\footnotesize a}} &
\multicolumn{4}{c}{MCMC results}}
\startdata
HIP ID & $\mathrm{M}_2~(\mathrm{M}_\odot)$\tablenotemark{\footnotesize b} & Semi-major axis (AU) \tablenotemark{\footnotesize c} & Period(yr) & Eccentricity & $\mathrm{M}_2~(\mathrm{M}_\odot)$ & Semi-major axis (AU)  & Period(yr)& Eccentricity\\
\hline
28902  & ${0.255 \pm 0.044}$  & ${3.17 \pm 0.18}$  & ${5.28 \pm 0.45}$  & ${0.134 \pm 0.011}$  & ${0.393 \pm 0.013}$  & ${7.72_{-0.12}^{+0.11}}$ & ${18.9 \pm 0.1}$  & ${0.500 \pm 0.036}$ \\
101597 & ${0.292 \pm 0.046}$  & ${0.190 \pm 0.004}$  & ${0.0730 \pm 0.0003}$  & ${0.611 \pm 0.116}$  & ${0.658 \pm 0.029}$  & $79^{+18}_{-13}$  & ${591 \pm 188}$  & ${0.646 \pm 0.069}$ \\
107133 & ${0.229 \pm 0.030}$  & ${2.83 \pm 0.18}$  & ${5.00 \pm 0.53}$  & ${0.407 \pm 0.010}$  & ${0.414 \pm 0.005}$  & ${10.50_{-0.20}^{+0.23}}$ & ${31.3 \pm 0.1}$  & ${0.761 \pm 0.005}$ \\
\enddata
\tablenotetext{a}{Assume the companion’s luminosity is negligible.}
\tablenotetext{b}{Adopt the primary mass used in the \orvara\ orbit fit \citep{mpri}.}
\tablenotetext{c}{Convert the \textit{Gaia} semi‑major axis using the
  procedure in Section~\ref{sec:method-2bd} to enable a direct comparison
  with the MCMC results.}
\end{deluxetable*}

%% file: table_gaia_accel.tex
\setlength{\tabcolsep}{16pt}  
\begin{deluxetable*}{c c c c c c}
  \tablewidth{\linewidth}
  \tablecaption{Comparison of acceleration predictions with Gaia DR3 non‐single star solutions.}
  \label{tab:gaia_accel}
  \tabletypesize{\scriptsize}
  \tablehead{
    \colhead{HIP ID} &
    \colhead{$\mathrm{Gaia}\;a_\alpha\;(\mathrm{mas\,yr^{-2}})$} &
    \colhead{$\mathrm{Gaia}\;a_\delta\;(\mathrm{mas\,yr^{-2}})$} &
    \colhead{Predict\;$a_\alpha\;(\mathrm{mas\,yr^{-2}})$} &
    \colhead{Predict\;$a_\delta\;(\mathrm{mas\,yr^{-2}})$} &
    \colhead{$\chi^2$}
  }
\startdata
13081  & $-2.435 \pm 0.093$  & $-0.431 \pm 0.071$  & $-2.427 \pm 0.053$  & $-0.499 \pm 0.170$  & $0.143$            \\
18320\tablenotemark{\footnotesize a}  & $17.842 \pm 0.333$  & $-3.128 \pm 0.155$  & $14.294 \pm 0.245$  & $-4.580 \pm 0.279$  & $83.4$              \\
24205  & $-1.017 \pm 0.106$  & $1.495 \pm 0.069$   & $-0.010 \pm 0.540$  & $1.106 \pm 0.239$   & $3.43$ \\
39064\tablenotemark{\footnotesize a}  & $-4.526 \pm 0.094$ & $7.625 \pm 0.073$ & $-7.108 \pm 0.428$  & $7.548 \pm 0.121$   & $35.2$ \\
57271  & $2.777  \pm 0.098$  & $0.197  \pm 0.056$   & $3.214  \pm 0.139$  & $0.147  \pm 0.058$   & $11.9$              \\
75676  & $-2.925 \pm 0.048$  & $5.016  \pm 0.045$   & $-3.293 \pm 0.106$  & $4.644  \pm 0.066$   & $22.2$             \\
76626\tablenotemark{\footnotesize a}  & $-6.760 \pm 0.034$  & $-0.125 \pm 0.042$   & $-6.678 \pm 0.236$  & $-0.164 \pm 0.218$   & $0.125$             \\
96395  & $-0.811 \pm 0.051$  & $0.667  \pm 0.052$   & $-0.773 \pm 0.015$  & $0.651  \pm 0.067$   & $0.639$             \\
99695  & $-0.779 \pm 0.041$  & $0.832  \pm 0.038$   & $-0.781 \pm 0.190$  & $0.786  \pm 0.179$   & $0.207$ \\
110440 & $-0.761 \pm 0.059$  & $3.668  \pm 0.069$   & $-0.451 \pm 0.223$  & $3.720  \pm 0.054$   & $3.42$             \\
117902 & $1.707  \pm 0.097$  & $1.315  \pm 0.043$   & $1.672  \pm 0.035$  & $1.249  \pm 0.044$   & $1.47$             \\
\hline
\enddata
\tablenotetext{\footnotesize a}{9 parameter fits}
\end{deluxetable*}

%% file: table_gaia_accel_9para.tex
\setlength{\tabcolsep}{16pt}  
\begin{deluxetable*}{c c c c c c}
  \tablewidth{\linewidth}
  \tablecaption{Comparison of derivative of acceleration terms with Gaia DR3 non‐single star 9-parameter solutions.}
  \label{tab:gaia_accel_9para}
  \tabletypesize{\scriptsize}
  \tablehead{
    \colhead{HIP ID} &\colhead{$\mathrm{Gaia}\;\dot{a}_\alpha\;(\mathrm{mas\,yr^{-3}})$} &
   \colhead{$\mathrm{Gaia}\;\dot{a}_\delta\;(\mathrm{mas\,yr^{-3}})$} &
    \colhead{Predict\;$\dot{a}_\alpha\;(\mathrm{mas\,yr^{-3}})$} &
    \colhead{Predict\;$\dot{a}_\delta\;(\mathrm{mas\,yr^{-3}})$} &
    \colhead{$\chi^{2}_{\dot{a}}$}
  }
\startdata
18320	&$-21.708\pm1.163$	&$-7.730\pm0.735$	&$-19.272\pm0.235$	&$-6.985\pm0.363$	&$4.881$	\\
39064	&$8.434\pm0.460$	&$8.865\pm0.350$	&$9.936\pm0.285$	&$10.279\pm0.247$	&$21.183$	\\
76626	&$-1.017\pm0.152$	&$4.724\pm0.184$	&$-1.473\pm0.269$	&$4.616\pm0.297$	&$3.319$	\\
\hline
\enddata

\end{deluxetable*}

%% file: table_accel_long.tex
\begin{longtable*}{lllll}
\caption{Predicted \textit{Gaia} acceleration. 
}  \label{tab:accel_all}\\
\hline\hline
\rm{HIP ID} & \rm{RA~accel $\mathrm{(mas/yr^2)}$}& \rm{Dec~accel $\mathrm{(mas/yr^2)}$} & \rm{RV} & \rm{RelAST}\\
\hline
\multicolumn{4}{c}{Full constraints listed} \\
\hline
184	& $0.016\pm0.003$	& $-0.0040\pm0.0009$	& CLS	& BOTH	\\
1444	& $0.23\pm0.02$	& $0.016\pm0.005$	& CLS	& WDS	\\
1475	& $0.108\pm0.002$	& $0.048\pm0.001$	& CLS	& WDS	\\
1893	& $0.030\pm0.006$	& $-0.052\pm0.004$	& HARPS	& Gaia2nd	\\
3185	& $1.74\pm0.06$	& $2.07\pm0.07$	& CLS	& /	\\
3579	& $-0.5\pm0.2$	& $-2.15\pm0.04$	& HARPS	& /	\\
3821	& $-0.28\pm0.02$	& $0.39\pm0.03$	& CLS	& BOTH	\\
3850	& $-0.021\pm0.001$	& $-0.30\pm0.01$	& CLS	& imag\tablenotemark{a}	\\
4393	& $0.0040\pm0.0007$	& $-0.011\pm0.002$	& CLS	& BOTH	\\
4423	& $-0.94\pm0.01$	& $-0.62\pm0.02$	& CLS	& /	\\
5189	& $0.133\pm0.009$	& $0.03\pm0.03$	& CLS	& /	\\
5315	& $0.05\pm0.04$	& $-0.29\pm0.02$	& CLS	& /	\\
5476	& $2\pm1$	& $-3\pm1$	& HIRES	& /	\\
5806	& $-0.07\pm0.05$	& $0.21\pm0.04$	& HARPS	& /	\\
6653	& $-0.020\pm0.002$	& $-0.041\pm0.004$	& CLS	& BOTH	\\
6712	& $-29.0\pm1$	& $-10.4\pm0.5$	& CLS	& WDS	\\
8770	& $-0.01\pm0.03$	& $0.00\pm0.03$	& HIRES	& /	\\
9603	& $-0.2\pm0.3$	& $-2.8\pm0.2$	& HARPS	& /	\\
11433	& $-0.048\pm0.003$	& $-0.035\pm0.002$	& HARPS	& WDS	\\
12114	& $2.52\pm0.09$	& $5.8\pm0.2$	& CLS	& /	\\
13081	& $-2.43\pm0.05$	& $-0.5\pm0.2$	& CLS	& /	\\
13642	& $-0.035\pm0.003$	& $-0.040\pm0.003$	& CLS	& BOTH	\\
14265	& $-0.3\pm0.1$	& $1.0\pm0.1$	& HARPS	& /	\\
14596	& $0.066\pm0.007$	& $-0.081\pm0.009$	& HARPS	& WDS	\\
17157	& $-0.16\pm0.01$	& $0.16\pm0.01$	& HARPS	& BOTH	\\
17666	& $0.045\pm0.009$	& $0.039\pm0.006$	& HIRES	& Gaia2nd	\\
17960	& $-0.00\pm0.09$	& $-0.11\pm0.04$	& CLS	& /	\\
18267	& $-0.083\pm0.007$	& $-0.051\pm0.004$	& CLS	& BOTH	\\
18317	& $-0.118\pm0.008$	& $-0.048\pm0.004$	& HARPS	& WDS	\\
18320	& $12.7\pm0.2$	& $-4.5\pm0.3$	& HIRES	& /	\\
18512	& $0.008\pm0.001$	& $0.031\pm0.004$	& CLS	& BOTH	\\
19428	& $-0.05\pm0.04$	& $-0.16\pm0.02$	& CLS	& /	\\
19911	& $5.3\pm0.7$	& $-13.9\pm0.8$	& HIRES	& /	\\
20752	& $-0.251\pm0.009$	& $0.32\pm0.01$	& HARPS	& Gaia2nd	\\
20802	& $1.34\pm0.1$	& $0.67\pm0.1$	& HIRES	& /	\\
21865	& $0.1\pm0.2$	& $0.1\pm0.2$	& HARPS	& /	\\
22596	& $-0.53\pm0.01$	& $-0.03\pm0.02$	& CLS	& /	\\
22919	& $0.116\pm0.008$	& $0.415\pm0.004$	& CLS	& /	\\
23926	& $-0.442\pm0.007$	& $-0.79\pm0.02$	& HARPS	& /	\\
24205	& $-0.0\pm0.5$	& $1.1\pm0.2$	& CLS	& /	\\
27207	& $-0.017\pm0.002$	& $-0.022\pm0.002$	& CLS	& WDS	\\
28267	& $0.116\pm0.004$	& $0.166\pm0.006$	& CLS	& BOTH	\\
28902	& $-12.4\pm0.3$	& $18.6\pm0.3$	& CLS	& /	\\
33109	& $0.29\pm0.02$	& $0.12\pm0.02$	& CLS	& /	\\
36941	& $-0.68\pm0.06$	& $0.37\pm0.07$	& HARPS	& /	\\
37233	& $-0.44\pm0.03$	& $-0.46\pm0.02$	& HARPS	& /	\\
38216	& $-0.08\pm0.04$	& $0.09\pm0.05$	& HIRES	& /	\\
38931	& $0.071\pm0.003$	& $0.051\pm0.002$	& HARPS	& BOTH	\\
39064	& $-8.0\pm0.5$	& $7.6\pm0.1$	& CLS	& /	\\
39417	& $0.1\pm0.1$	& $-0.0\pm0.1$	& CLS	& /	\\
39470	& $-0.98\pm0.09$	& $-0.36\pm0.09$	& HARPS	& /	\\
40118	& $-0.70\pm0.04$	& $-2.0\pm0.1$	& CLS	& /	\\
40687	& $0.000\pm0.001$	& $0.000\pm0.001$	& CLS	& /	\\
41844	& $-0.039\pm0.004$	& $-0.021\pm0.002$	& CLS	& WDS	\\
42030	& $0.00\pm0.05$	& $-0.00\pm0.05$	& CLS	& /	\\
42220	& $-9.4\pm0.4$	& $0.40\pm0.06$	& CLS	& WDS	\\
43299	& $2.2\pm0.5$	& $-23.2\pm0.7$	& HIRES	& /	\\
49350	& $-0.37\pm0.07$	& $0.47\pm0.06$	& CLS	& /	\\
49756	& $0.015\pm0.004$	& $0.032\pm0.005$	& HIRES	& Gaia2nd	\\
49769	& $-0.030\pm0.01$	& $0.24\pm0.01$	& CLS	& WDS	\\
50316	& $-0.09\pm0.03$	& $-0.18\pm0.02$	& CLS	& /	\\
50653	& $-0.01\pm0.06$	& $-0.02\pm0.02$	& HARPS	& /	\\
51525	& $0.035\pm0.004$	& $0.0000\pm0.0001$	& CLS	& WDS	\\
51579	& $0.54\pm0.06$	& $0.48\pm0.06$	& CLS	& /	\\
54102	& $-0.28\pm0.03$	& $0.08\pm0.03$	& HARPS	& /	\\
55459	& $0.021\pm0.002$	& $-0.0050\pm0.0004$	& CLS	& BOTH	\\
56452	& $0.051\pm0.004$	& $-0.041\pm0.003$	& CLS	& WDS	\\
57271	& $3.2\pm0.1$	& $0.15\pm0.06$	& HIRES	& /	\\
58289	& $0.04\pm0.03$	& $0.04\pm0.02$	& HARPS	& /	\\
58576	& $0.40\pm0.07$	& $2.23\pm0.07$	& CLS	& /	\\
59021	& $-0.009\pm0.002$	& $0.010\pm0.002$	& CLS	& BOTH	\\
60051	& $-0.51\pm0.03$	& $0.16\pm0.05$	& HARPS	& /	\\
60074	& $-0.008\pm0.001$	& $-0.012\pm0.002$	& HIRES	& WDS	\\
60648	& $0.2\pm0.2$	& $-0.1\pm0.2$	& HARPS	& /	\\
61901	& $-0.50\pm0.02$	& $-0.125\pm0.003$	& CLS	& BOTH	\\
62039	& $0.0010\pm0.0002$	& $0.033\pm0.004$	& CLS	& WDS	\\
62345	& $0.16\pm0.02$	& $-0.09\pm0.01$	& CLS	& /	\\
63510	& $7.0\pm0.3$	& $4.3\pm0.2$	& CLS	& /	\\
64150	& $0.0250\pm0.0007$	& $-0.178\pm0.004$	& CLS	& WDS	\\
64797	& $0.199\pm0.006$	& $-0.052\pm0.002$	& CLS	& BOTH	\\
70623	& $0.16\pm0.02$	& $0.09\pm0.04$	& CLS	& /	\\
71001	& $1.5\pm0.1$	& $-0.3\pm0.3$	& HARPS	& /	\\
71631	& $0.11\pm0.01$	& $-0.77\pm0.05$	& HIRES	& WDS	\\
71898	& $0.18\pm0.03$	& $-0.011\pm0.003$	& CLS	& WDS	\\
72043	& $0.26\pm0.03$	& $-0.72\pm0.02$	& CLS	& /	\\
72659	& $-1.5\pm0.1$	& $0.77\pm0.05$	& CLS	& Gaia2nd	\\
73408	& $0\pm1$	& $-5.1\pm0.5$	& HARPS	& /	\\
75676	& $-3.3\pm0.1$	& $4.64\pm0.07$	& HIRES	& /	\\
76543	& $-0.34\pm0.03$	& $-0.08\pm0.05$	& CLS	& /	\\
76626	& $-6.7\pm0.3$	& $-0.1\pm0.2$	& HARPS	& /	\\
77052	& $0.088\pm0.002$	& $0.314\pm0.005$	& HARPS	& WDS	\\
77810	& $0.2\pm0.1$	& $-0.58\pm0.04$	& CLS	& /	\\
78395	& $0.16\pm0.02$	& $0.053\pm0.008$	& HARPS	& WDS	\\
78709	& $-13.1\pm0.4$	& $-11.4\pm0.5$	& CLS	& /	\\
79619	& $-0.09\pm0.02$	& $-0.10\pm0.01$	& CLS	& WDS	\\
81375	& $0.0270\pm0.0009$	& $0.086\pm0.003$	& CLS	& WDS	\\
82032	& $-0.010\pm0.002$	& $0.0000\pm0.0001$	& HARPS	& BOTH	\\
83020	& $0.097\pm0.004$	& $0.049\pm0.002$	& CLS	& BOTH	\\
84684	& $1.02\pm0.08$	& $1.08\pm0.02$	& HARPS	& /	\\
85158	& $0.7\pm0.1$	& $0.3\pm0.1$	& CLS	& /	\\
85653	& $-0.224\pm0.005$	& $0.127\pm0.003$	& CLS	& WDS	\\
86974	& $-4.14\pm0.03$	& $-0.94\pm0.03$	& CLS	& WDS	\\
88595	& $0.3\pm0.2$	& $-0.4\pm0.2$	& HARPS	& /	\\
88622	& $0.57\pm0.02$	& $0.178\pm0.008$	& CLS	& Gaia2nd	\\
89270	& $-0.76\pm0.09$	& $-1.21\pm0.06$	& CLS	& /	\\
89620	& $-0.43\pm0.05$	& $2.24\pm0.04$	& CLS	& WDS	\\
92262	& $1.08\pm0.05$	& $-2.93\pm0.05$	& HARPS	& /	\\
94370	& $-1.15\pm0.08$	& $2.94\pm0.04$	& HARPS	& /	\\
95015	& $-0.1\pm0.5$	& $-0.1\pm0.4$	& CLS	& /	\\
95319	& $-0.02\pm0.02$	& $-0.08\pm0.01$	& CLS	& imag\tablenotemark{b}	\\
95401	& $0.14\pm0.02$	& $-0.034\pm0.009$	& HIRES	& WDS	\\
96395	& $-0.77\pm0.02$	& $0.65\pm0.07$	& CLS	& /	\\
98677	& $0.060\pm0.004$	& $0.013\pm0.001$	& CLS	& BOTH	\\
98714	& $-0.4\pm0.1$	& $0.8\pm0.2$	& HIRES	& /	\\
98819	& $0.03\pm0.01$	& $0.50\pm0.01$	& CLS	& imag\tablenotemark{c}	\\
99385	& $0.44\pm0.01$	& $0.183\pm0.006$	& HARPS	& WDS	\\
99695	& $-0.8\pm0.2$	& $0.8\pm0.2$	& HARPS	& /	\\
101345	& $0.159\pm0.01$	& $0.145\pm0.009$	& CLS	& WDS	\\
101597	& $-0.56\pm0.02$	& $0.060\pm0.003$	& CLS	& WDS	\\
101846	& $-0.053\pm0.004$	& $-0.052\pm0.004$	& HARPS	& BOTH	\\
103768	& $0.085\pm0.005$	& $-0.0120\pm0.0007$	& HARPS	& BOTH	\\
103983	& $0.0\pm0.3$	& $3.2\pm0.3$	& CLS	& /	\\
104239	& $-0.013\pm0.002$	& $-0.172\pm0.005$	& CLS	& BOTH	\\
106006	& $-0.3\pm0.2$	& $0.5\pm0.2$	& HARPS	& /	\\
107133	& $23.6\pm0.7$	& $14.9\pm0.4$	& HARPS	& /	\\
109169	& $0.67\pm0.03$	& $0.25\pm0.02$	& CLS	& /	\\
110109	& $0.092\pm0.005$	& $0.080\pm0.005$	& HARPS	& WDS	\\
110440	& $-0.5\pm0.2$	& $3.72\pm0.05$	& HARPS	& /	\\
111571	& $-1.8\pm0.2$	& $0.00\pm0.03$	& HARPS	& WDS	\\
113421	& $0.000\pm0.005$	& $0.000\pm0.002$	& CLS	& /	\\
113438	& $0.85\pm0.02$	& $-0.20\pm0.03$	& CLS	& /	\\
116250	& $-0.04\pm0.04$	& $0.3\pm0.1$	& HARPS	& /	\\
116350	& $0.54\pm0.01$	& $-0.18\pm0.03$	& HARPS	& /	\\
116745	& $0.05\pm0.07$	& $-0.0\pm0.1$	& HARPS	& /	\\
117902	& $1.67\pm0.03$	& $1.25\pm0.04$	& HARPS	& /	\\
\hline
\multicolumn{4}{c}{Bimodal inclination-- prograde ($i<90^\circ$)} \\
\hline
9683	& $0.09\pm0.08$	& $0.22\pm0.08$	& CLS	& /	\\
10337	& $0.08\pm0.03$	& $0.09\pm0.02$	& HARPS	& /	\\
12436	& $0.027\pm0.008$	& $0.118\pm0.008$	& HARPS	& /	\\
20277	& $0.000\pm0.002$	& $0.0\pm0.0$	& HARPS	& /	\\
21654	& $-0.0\pm0.6$	& $6.8\pm0.7$	& CLS	& /	\\
22627	& $-0.4\pm0.2$	& $0.25\pm0.08$	& CLS	& /	\\
42723	& $0.12\pm0.08$	& $-0.14\pm0.04$	& CLS	& /	\\
64457	& $-0.0\pm0.1$	& $0.02\pm0.07$	& CLS	& /	\\
65312	& $0.1\pm0.6$	& $-1.3\pm0.2$	& CLS	& /	\\
70849	& $0.23\pm0.06$	& $0.08\pm0.05$	& HARPS	& /	\\
84171	& $0.01\pm0.04$	& $-0.01\pm0.05$	& CLS	& /	\\
86990	& $0.20\pm0.06$	& $-0.8\pm0.4$	& HARPS	& /	\\
89844	& $0.02\pm0.07$	& $0.02\pm0.07$	& CLS	&  imag\tablenotemark{d}	\\
90055	& $-0.06\pm0.02$	& $-0.02\pm0.03$	& HARPS	& /	\\
95467	& $0.1\pm0.2$	& $-0.0\pm0.3$	& HARPS	& /	\\
95740	& $-0.00\pm0.06$	& $0.0\pm0.1$	& CLS	& /	\\
97336	& $0.0\pm0.0$	& $0.0\pm0.0$	& CLS	& /	\\
106440	& $-0.42\pm0.07$	& $-0.28\pm0.09$	& HARPS	& /	\\
109918	& $0.54\pm0.02$	& $-0.45\pm0.02$	& HARPS	& /	\\
115013	& $-5.3\pm0.4$	& $7.5\pm0.3$	& HARPS	& /	\\
\hline
\multicolumn{4}{c}{Bimodal inclination -- retrograde ($i>90^\circ$)} \\
\hline
9683	& $-0.06\pm0.03$	& $-0.10\pm0.09$	& CLS	& /	\\
10337	& $-0.12\pm0.02$	& $0.00\pm0.03$	& HARPS	& /	\\
12436	& $0.115\pm0.007$	& $-0.038\pm0.009$	& HARPS	& /	\\
20277	& $0.000\pm0.002$	& $0.0000\pm0.0005$	& HARPS	& /	\\
21654	& $-0.4\pm0.4$	& $-14\pm1$	& CLS	& /	\\
22627	& $0.2\pm0.1$	& $0.1\pm0.2$	& CLS	& /	\\
42723	& $-0.18\pm0.07$	& $-0.07\pm0.06$	& CLS	& /	\\
64457	& $0.0\pm0.3$	& $0.04\pm0.07$	& CLS	& /	\\
65312	& $-1.1\pm0.2$	& $1.2\pm0.4$	& CLS	& /	\\
70849	& $-0.1\pm0.1$	& $0.27\pm0.06$	& HARPS	& /	\\
84171	& $-0.01\pm0.03$	& $-0.00\pm0.02$	& CLS	& /	\\
86990	& $-0.7\pm0.3$	& $0.2\pm0.1$	& HARPS	& /	\\
89844	& $-0.01\pm0.09$	& $-0.01\pm0.07$	& CLS	&  imag\tablenotemark{d}	\\
90055	& $0.07\pm0.02$	& $0.01\pm0.03$	& HARPS	& /	\\
95467	& $-0.1\pm0.3$	& $0.2\pm0.3$	& HARPS	& /	\\
95740	& $-0.1\pm0.2$	& $0.01\pm0.05$	& CLS	& /	\\
97336	& $0.0\pm0.0$	& $0.0000\pm0.0002$	& CLS	& /	\\
106440	& $0.57\pm0.03$	& $0.03\pm0.08$	& HARPS	& /	\\
109918	& $0.69\pm0.03$	& $-0.109\pm0.009$	& HARPS	& /	\\
115013	& $4.2\pm0.5$	& $9.2\pm0.2$	& HARPS	& /	\\
\hline\hline
\end{longtable*}
\tablenotetext{a}{Relative astrometry data are from \citet{Brandt2019,2019Peretti_HD4747B,2012Creep_trendStar,2018Crepp_3850B}.}
\tablenotetext{b}{Relative astrometry data from \citet{2018_95319_relAST}.}
\tablenotetext{c}{Relative astrometry data from \citet{Brandt2019},\citet{2002_98819B_discover},\citet{2003_98819B_imag},\citet{2009_98819B_imag},\citet{2012_98819_dynamic}.}
\tablenotetext{d}{Relative astrometry data from \citet{2024_89620_imaging}.}

%% file: table_stellar_long.tex
\setlength{\tabcolsep}{6pt}  
\begin{longtable*}{llllllll}
\caption{MCMC result of stellar companions. 
}\label{tab:stellar_long}\\
\hline\hline
\textbf{HIP} & \textbf{RV~jit(m/s)} & \textbf{$\mathrm{M_1~(M_{\odot})}$} & \textbf{$\mathrm{M_2~(M_{\odot})}$} & \textbf{Semi-a (AU)} & \textbf{inclination($\degree$)} & \textbf{Period(yr)} &\textbf{Eccentricity}\\
\hline
184	& $5.65^{+0.93}_{-0.71}$	& $0.900^{+0.044}_{-0.045}$	& $0.437^{+0.068}_{-0.070}$	& $128^{+48}_{-28}$	& $113.1^{+23.0}_{-5.6}$	& $1250^{+770}_{-380}$	& $0.42^{+0.42}_{-0.32}$	\\
1444	& $4.57^{+0.79}_{-0.69}$	& $1.186^{+0.058}_{-0.059}$	& $0.449^{+0.026}_{-0.025}$	& $46.5^{+12.0}_{-7.5}$	& $92.5^{+3.9}_{-3.4}$	& $248^{+102}_{-57}$	& $0.44^{+0.18}_{-0.11}$	\\
1475	& $2.70^{+0.11}_{-0.11}$	& $0.4042^{+0.0079}_{-0.0079}$	& $0.1697^{+0.0043}_{-0.0041}$	& $105^{+30}_{-12}$	& $25^{+15}_{-11}$	& $1420^{+650}_{-230}$	& $0.589^{+0.062}_{-0.038}$	\\
1893	& $2.77^{+0.72}_{-0.61}$	& $1.002^{+0.050}_{-0.051}$	& $0.24^{+2.80}_{-0.16}$	& $56^{+155}_{-29}$	& $25^{+14}_{-14}$	& $380^{+1010}_{-240}$	& $0.24^{+0.16}_{-0.12}$	\\
3185	& $3.50^{+0.52}_{-0.44}$	& $1.095^{+0.055}_{-0.055}$	& $0.948^{+0.061}_{-0.057}$	& $42.4^{+5.0}_{-4.1}$	& $116.8^{+2.4}_{-2.5}$	& $193^{+33}_{-25}$	& $0.501^{+0.039}_{-0.036}$	\\
3579	& $4.24^{+0.47}_{-0.41}$	& $0.847^{+0.042}_{-0.042}$	& $0.673^{+0.054}_{-0.054}$	& $21.1^{+1.9}_{-1.8}$	& $73.4^{+6.5}_{-5.5}$	& $78.7^{+9.3}_{-8.4}$	& $0.450^{+0.022}_{-0.021}$	\\
3821	& $14.5^{+1.3}_{-1.2}$	& $1.089^{+0.050}_{-0.051}$	& $0.665^{+0.031}_{-0.031}$	& $64.2^{+12.0}_{-3.5}$	& $36.4^{+11.0}_{-5.2}$	& $389^{+115}_{-33}$	& $0.501^{+0.047}_{-0.260}$	\\
4393	& $8.9^{+1.3}_{-1.1}$	& $1.159^{+0.057}_{-0.056}$	& $0.68^{+0.15}_{-0.14}$	& $173^{+65}_{-18}$	& $77.4^{+6.1}_{-22.0}$	& $1690^{+1000}_{-260}$	& $0.87^{+0.11}_{-0.29}$	\\
4423	& $5.68^{+1.10}_{-0.88}$	& $1.200^{+0.060}_{-0.060}$	& $0.360^{+0.012}_{-0.012}$	& $8.95^{+0.13}_{-0.14}$	& $138.33^{+0.48}_{-0.48}$	& $21.451^{+0.034}_{-0.033}$	& $0.3575^{+0.0012}_{-0.0012}$	\\
5189	& $6.58^{+1.10}_{-0.90}$	& $1.329^{+0.066}_{-0.067}$	& $0.230^{+0.035}_{-0.026}$	& $27.0^{+13.0}_{-5.9}$	& $82^{+13}_{-14}$	& $113^{+87}_{-34}$	& $0.354^{+0.140}_{-0.046}$	\\
5315	& $4.81^{+0.37}_{-0.32}$	& $1.520^{+0.077}_{-0.075}$	& $0.83^{+0.56}_{-0.37}$	& $42^{+35}_{-16}$	& $64^{+41}_{-28}$	& $179^{+219}_{-82}$	& $0.53^{+0.14}_{-0.15}$	\\
5476	& $19.6^{+3.6}_{-2.8}$	& $1.299^{+0.065}_{-0.065}$	& $0.682^{+0.120}_{-0.064}$	& $15.8^{+4.5}_{-2.0}$	& $87^{+16}_{-14}$	& $44.8^{+19.0}_{-7.7}$	& $0.605^{+0.043}_{-0.030}$	\\
5806	& $10.08^{+0.70}_{-0.62}$	& $1.101^{+0.054}_{-0.055}$	& $0.111^{+0.028}_{-0.020}$	& $20.7^{+7.5}_{-3.8}$	& $73^{+20}_{-18}$	& $86^{+49}_{-22}$	& $0.16^{+0.18}_{-0.12}$	\\
6653	& $2.89^{+0.35}_{-0.32}$	& $1.140^{+0.057}_{-0.057}$	& $0.543^{+0.026}_{-0.023}$	& $61.8^{+18.0}_{-5.2}$	& $47.8^{+8.9}_{-5.6}$	& $374^{+176}_{-46}$	& $0.626^{+0.050}_{-0.072}$	\\
9603	& $10.8^{+1.5}_{-1.2}$	& $1.28^{+0.59}_{-0.55}$	& $0.33^{+0.10}_{-0.10}$	& $17.6^{+4.4}_{-3.2}$	& $46.4^{+1.6}_{-1.7}$	& $57.4^{+20.0}_{-8.5}$	& $0.591^{+0.065}_{-0.039}$	\\
11433	& $4.15^{+0.30}_{-0.26}$	& $0.905^{+0.045}_{-0.045}$	& $0.262^{+0.019}_{-0.018}$	& $88^{+38}_{-23}$	& $116^{+13}_{-13}$	& $760^{+550}_{-280}$	& $0.746^{+0.082}_{-0.100}$	\\
12114	& $2.33^{+0.18}_{-0.17}$	& $0.800^{+0.040}_{-0.040}$	& $0.0973^{+0.0037}_{-0.0036}$	& $16.06^{+0.81}_{-0.72}$	& $47.17^{+0.84}_{-0.80}$	& $68.0^{+4.8}_{-4.2}$	& $0.617^{+0.014}_{-0.014}$	\\
13081	& $13.7^{+2.0}_{-1.7}$	& $0.900^{+0.045}_{-0.045}$	& $0.1881^{+0.0059}_{-0.0060}$	& $6.32^{+0.10}_{-0.10}$	& $95.65^{+0.40}_{-0.40}$	& $15.215^{+0.033}_{-0.033}$	& $0.6626^{+0.0039}_{-0.0038}$	\\
13642	& $4.52^{+0.40}_{-0.35}$	& $0.918^{+0.046}_{-0.046}$	& $0.624^{+0.023}_{-0.023}$	& $108^{+28}_{-23}$	& $111.5^{+18.0}_{-3.5}$	& $910^{+370}_{-270}$	& $0.27^{+0.51}_{-0.20}$	\\
14265	& $11.9^{+1.2}_{-1.0}$	& $0.883^{+0.044}_{-0.044}$	& $0.581^{+0.170}_{-0.093}$	& $19.3^{+4.4}_{-2.3}$	& $46.5^{+11.0}_{-7.9}$	& $70^{+20}_{-10}$	& $0.050^{+0.046}_{-0.030}$	\\
14596	& $3.05^{+0.34}_{-0.30}$	& $1.105^{+0.054}_{-0.055}$	& $0.469^{+0.093}_{-0.071}$	& $62^{+23}_{-14}$	& $122.7^{+3.9}_{-3.2}$	& $380^{+230}_{-120}$	& $0.35^{+0.17}_{-0.14}$	\\
17157	& $7.93^{+1.10}_{-0.94}$	& $0.601^{+0.030}_{-0.031}$	& $0.269^{+0.012}_{-0.010}$	& $31.1^{+13.0}_{-5.6}$	& $95.2^{+7.2}_{-1.4}$	& $187^{+134}_{-48}$	& $0.51^{+0.40}_{-0.26}$	\\
17666	& $3.04^{+0.15}_{-0.14}$	& $0.866^{+0.043}_{-0.043}$	& $0.104^{+0.120}_{-0.056}$	& $39^{+43}_{-17}$	& $70.9^{+7.5}_{-12.0}$	& $250^{+460}_{-140}$	& $0.70^{+0.18}_{-0.20}$	\\
18267	& $4.20^{+0.22}_{-0.20}$	& $1.001^{+0.049}_{-0.050}$	& $0.5430^{+0.0100}_{-0.0092}$	& $73^{+21}_{-12}$	& $130.6^{+9.4}_{-14.0}$	& $500^{+230}_{-120}$	& $0.730^{+0.076}_{-0.610}$	\\
18317	& $24.9^{+2.7}_{-2.3}$	& $0.912^{+0.044}_{-0.044}$	& $0.306^{+0.035}_{-0.024}$	& $46.6^{+8.8}_{-6.7}$	& $75.1^{+1.5}_{-1.9}$	& $288^{+88}_{-62}$	& $0.807^{+0.034}_{-0.037}$	\\
18320	& $3.42^{+1.00}_{-0.76}$	& $1.238^{+0.062}_{-0.062}$	& $0.768^{+0.026}_{-0.026}$	& $12.74^{+0.32}_{-0.29}$	& $25.90^{+0.45}_{-0.44}$	& $32.07^{+1.00}_{-0.87}$	& $0.8010^{+0.0029}_{-0.0025}$	\\
18512	& $4.77^{+0.80}_{-0.64}$	& $0.756^{+0.037}_{-0.039}$	& $0.552^{+0.064}_{-0.059}$	& $145^{+32}_{-14}$	& $140.8^{+7.3}_{-8.2}$	& $1530^{+530}_{-220}$	& $0.435^{+0.062}_{-0.110}$	\\
20752	& $18.2^{+5.3}_{-3.6}$	& $1.097^{+0.056}_{-0.054}$	& $0.275^{+0.038}_{-0.023}$	& $23.2^{+3.7}_{-2.3}$	& $91.88^{+0.16}_{-0.15}$	& $95^{+22}_{-13}$	& $0.077^{+0.130}_{-0.056}$	\\
20802	& $95.1^{+3.6}_{-6.7}$	& $1.570^{+0.080}_{-0.080}$	& $1.767^{+0.064}_{-0.061}$	& $15.97^{+0.51}_{-0.42}$	& $73.9^{+1.7}_{-1.7}$	& $34.9^{+1.5}_{-1.1}$	& $0.151^{+0.014}_{-0.014}$	\\
21654\tablenotemark{$_{*}$}	& $8.9^{+1.8}_{-1.3}$	& $1.081^{+0.054}_{-0.054}$	& $0.190^{+0.014}_{-0.014}$	& $5.134^{+0.082}_{-0.085}$	& $150.9^{+1.9}_{-2.1}$	& $10.317^{+0.045}_{-0.042}$	& $0.5948^{+0.0094}_{-0.0068}$	\\
22596	& $4.07^{+0.71}_{-0.58}$	& $1.158^{+0.058}_{-0.058}$	& $0.2658^{+0.0082}_{-0.0085}$	& $19.34^{+0.39}_{-0.38}$	& $109.78^{+0.79}_{-0.84}$	& $71.3^{+1.4}_{-1.3}$	& $0.5898^{+0.0050}_{-0.0049}$	\\
22919	& $5.54^{+0.92}_{-0.73}$	& $1.200^{+0.059}_{-0.059}$	& $0.420^{+0.012}_{-0.013}$	& $22.57^{+0.37}_{-0.37}$	& $61.94^{+0.18}_{-0.18}$	& $84.27^{+0.89}_{-0.87}$	& $0.97792^{+0.00040}_{-0.00040}$	\\
23926	& $2.80^{+0.40}_{-0.34}$	& $1.420^{+0.070}_{-0.070}$	& $0.2815^{+0.0092}_{-0.0093}$	& $16.07^{+0.50}_{-0.46}$	& $170.84^{+0.15}_{-0.15}$	& $49.4^{+2.0}_{-1.8}$	& $0.555^{+0.011}_{-0.010}$	\\
27207	& $2.03^{+0.19}_{-0.17}$	& $0.943^{+0.049}_{-0.049}$	& $0.795^{+0.090}_{-0.065}$	& $149^{+21}_{-21}$	& $121.7^{+3.1}_{-3.4}$	& $1380^{+290}_{-270}$	& $0.27^{+0.16}_{-0.13}$	\\
28267	& $2.55^{+0.25}_{-0.22}$	& $1.062^{+0.052}_{-0.052}$	& $0.5050^{+0.0086}_{-0.0085}$	& $63.4^{+14.0}_{-6.3}$	& $30.1^{+5.4}_{-1.9}$	& $404^{+144}_{-61}$	& $0.776^{+0.026}_{-0.053}$	\\
28902	& $14.0^{+2.7}_{-2.0}$	& $0.900^{+0.044}_{-0.045}$	& $0.393^{+0.012}_{-0.013}$	& $7.72^{+0.11}_{-0.12}$	& $31.84^{+0.41}_{-0.41}$	& $18.86^{+0.10}_{-0.10}$	& $0.4998^{+0.0034}_{-0.0037}$	\\
33109	& $3.73^{+0.85}_{-0.69}$	& $1.200^{+0.059}_{-0.060}$	& $0.421^{+0.047}_{-0.042}$	& $22.9^{+2.3}_{-1.9}$	& $56.36^{+0.42}_{-0.42}$	& $86^{+12}_{-10}$	& $0.554^{+0.030}_{-0.026}$	\\
37233	& $7.75^{+0.83}_{-0.70}$	& $1.179^{+0.059}_{-0.059}$	& $0.2281^{+0.0075}_{-0.0076}$	& $12.69^{+0.26}_{-0.25}$	& $86.8^{+3.3}_{-3.2}$	& $38.09^{+0.80}_{-0.67}$	& $0.334^{+0.017}_{-0.019}$	\\
38216	& $4.38^{+1.00}_{-0.74}$	& $1.432^{+0.071}_{-0.071}$	& $0.091^{+0.024}_{-0.018}$	& $18.8^{+10.0}_{-4.8}$	& $173.2^{+2.3}_{-4.4}$	& $66^{+56}_{-23}$	& $0.43^{+0.21}_{-0.18}$	\\
38931	& $4.59^{+0.44}_{-0.38}$	& $0.801^{+0.039}_{-0.039}$	& $0.382^{+0.016}_{-0.016}$	& $73^{+25}_{-12}$	& $124.3^{+21.0}_{-8.7}$	& $580^{+320}_{-130}$	& $0.52^{+0.25}_{-0.27}$	\\
39064	& $2.66^{+0.34}_{-0.30}$	& $0.900^{+0.044}_{-0.046}$	& $0.1165^{+0.0041}_{-0.0041}$	& $4.554^{+0.071}_{-0.076}$	& $57.2^{+1.2}_{-1.1}$	& $9.64118^{+0.00085}_{-0.00085}$	& $0.37776^{+0.00053}_{-0.00053}$	\\
39470	& $8.8^{+1.5}_{-1.2}$	& $0.689^{+0.013}_{-0.013}$	& $0.179^{+0.016}_{-0.013}$	& $9.08^{+0.14}_{-0.15}$	& $40.1^{+5.6}_{-5.2}$	& $29.31^{+0.64}_{-0.62}$	& $0.576^{+0.048}_{-0.047}$	\\
40118	& $3.29^{+0.21}_{-0.20}$	& $1.043^{+0.052}_{-0.051}$	& $0.176^{+0.013}_{-0.011}$	& $18.2^{+2.1}_{-1.5}$	& $169.30^{+0.41}_{-0.49}$	& $70.2^{+12.0}_{-8.0}$	& $0.334^{+0.035}_{-0.025}$	\\
41844	& $3.10^{+0.62}_{-0.49}$	& $1.197^{+0.061}_{-0.059}$	& $0.270^{+0.027}_{-0.021}$	& $53.7^{+9.2}_{-8.9}$	& $122.1^{+3.3}_{-4.5}$	& $325^{+87}_{-77}$	& $0.22^{+0.18}_{-0.15}$	\\
42220	& $20.4^{+3.0}_{-2.5}$	& $0.653^{+0.012}_{-0.012}$	& $0.2588^{+0.0038}_{-0.0038}$	& $7.993^{+0.047}_{-0.049}$	& $121.83^{+0.39}_{-0.41}$	& $23.665^{+0.093}_{-0.100}$	& $0.7132^{+0.0062}_{-0.0061}$	\\
49350	& $3.94^{+0.44}_{-0.38}$	& $1.061^{+0.052}_{-0.053}$	& $0.308^{+0.044}_{-0.048}$	& $37.3^{+17.0}_{-8.8}$	& $77^{+19}_{-15}$	& $195^{+144}_{-62}$	& $0.35^{+0.17}_{-0.12}$	\\
49756	& $2.41^{+0.59}_{-0.48}$	& $1.080^{+0.053}_{-0.053}$	& $0.126^{+0.250}_{-0.071}$	& $61^{+71}_{-25}$	& $31^{+12}_{-15}$	& $440^{+840}_{-230}$	& $0.20^{+0.20}_{-0.14}$	\\
49769	& $6.71^{+0.81}_{-0.68}$	& $1.268^{+0.064}_{-0.063}$	& $0.617^{+0.064}_{-0.057}$	& $51.3^{+12.0}_{-7.7}$	& $57.21^{+0.65}_{-0.60}$	& $268^{+92}_{-56}$	& $0.611^{+0.055}_{-0.058}$	\\
50316	& $4.62^{+0.48}_{-0.41}$	& $1.198^{+0.060}_{-0.060}$	& $0.24^{+0.16}_{-0.12}$	& $35^{+26}_{-15}$	& $119^{+30}_{-45}$	& $175^{+204}_{-93}$	& $0.51^{+0.21}_{-0.24}$	\\
51525	& $7.96^{+1.00}_{-0.82}$	& $0.706^{+0.012}_{-0.013}$	& $0.160^{+0.032}_{-0.025}$	& $63^{+27}_{-12}$	& $85.1^{+1.8}_{-6.8}$	& $530^{+380}_{-140}$	& $0.60^{+0.33}_{-0.37}$	\\
51579	& $3.69^{+0.78}_{-0.62}$	& $1.251^{+0.063}_{-0.063}$	& $0.425^{+0.027}_{-0.025}$	& $16.95^{+0.79}_{-0.67}$	& $101.9^{+3.3}_{-3.8}$	& $53.9^{+3.2}_{-2.6}$	& $0.359^{+0.021}_{-0.023}$	\\
54102	& $5.47^{+0.80}_{-0.64}$	& $1.083^{+0.053}_{-0.054}$	& $0.418^{+0.080}_{-0.052}$	& $26.3^{+8.7}_{-4.8}$	& $55^{+10}_{-11}$	& $110^{+55}_{-27}$	& $0.424^{+0.062}_{-0.084}$	\\
55459	& $2.89^{+0.61}_{-0.49}$	& $1.138^{+0.057}_{-0.056}$	& $0.470^{+0.030}_{-0.029}$	& $100^{+22}_{-13}$	& $55.3^{+5.4}_{-13.0}$	& $790^{+280}_{-150}$	& $0.34^{+0.30}_{-0.23}$	\\
56452	& $2.55^{+0.25}_{-0.23}$	& $0.900^{+0.044}_{-0.045}$	& $0.554^{+0.022}_{-0.022}$	& $142^{+38}_{-37}$	& $83.4^{+2.1}_{-7.1}$	& $1400^{+600}_{-510}$	& $0.28^{+0.57}_{-0.22}$	\\
57271	& $4.7^{+1.6}_{-1.3}$	& $0.960^{+0.049}_{-0.048}$	& $0.354^{+0.019}_{-0.017}$	& $7.14^{+0.14}_{-0.13}$	& $133.02^{+0.44}_{-0.46}$	& $16.64^{+0.22}_{-0.19}$	& $0.8634^{+0.0076}_{-0.0064}$	\\
58576	& $2.65^{+0.31}_{-0.27}$	& $1.061^{+0.053}_{-0.053}$	& $0.321^{+0.014}_{-0.013}$	& $22.85^{+0.87}_{-0.79}$	& $25.93^{+0.68}_{-0.58}$	& $92.9^{+4.8}_{-4.2}$	& $0.400^{+0.025}_{-0.025}$	\\
59021	& $4.97^{+0.81}_{-0.65}$	& $1.370^{+0.070}_{-0.070}$	& $1.65^{+0.22}_{-0.22}$	& $237^{+106}_{-35}$	& $74.5^{+5.0}_{-19.0}$	& $2120^{+1560}_{-470}$	& $0.63^{+0.29}_{-0.26}$	\\
60051	& $11.18^{+0.94}_{-0.82}$	& $0.979^{+0.049}_{-0.049}$	& $0.307^{+0.036}_{-0.033}$	& $55^{+29}_{-15}$	& $94.47^{+0.60}_{-0.54}$	& $360^{+320}_{-130}$	& $0.813^{+0.064}_{-0.069}$	\\
60074	& $18.6^{+1.0}_{-1.3}$	& $1.083^{+0.061}_{-0.060}$	& $0.338^{+0.130}_{-0.066}$	& $183^{+104}_{-47}$	& $122^{+36}_{-68}$	& $2060^{+1960}_{-730}$	& $0.33^{+0.35}_{-0.25}$	\\
61901	& $3.37^{+0.32}_{-0.29}$	& $0.743^{+0.035}_{-0.035}$	& $0.2597^{+0.0069}_{-0.0068}$	& $39.4^{+2.0}_{-1.6}$	& $149.17^{+0.92}_{-1.20}$	& $247^{+23}_{-18}$	& $0.258^{+0.028}_{-0.024}$	\\
62039	& $2.49^{+0.28}_{-0.25}$	& $1.134^{+0.057}_{-0.055}$	& $0.297^{+0.026}_{-0.022}$	& $61^{+18}_{-15}$	& $81.1^{+1.8}_{-9.3}$	& $390^{+190}_{-140}$	& $0.37^{+0.49}_{-0.28}$	\\
62345	& $2.63^{+0.24}_{-0.22}$	& $1.169^{+0.057}_{-0.058}$	& $0.168^{+0.048}_{-0.037}$	& $34.7^{+18.0}_{-8.8}$	& $155.8^{+4.9}_{-6.7}$	& $177^{+150}_{-61}$	& $0.50^{+0.14}_{-0.13}$	\\
63510	& $98.2^{+1.3}_{-2.5}$	& $0.579^{+0.011}_{-0.011}$	& $0.0883^{+0.0019}_{-0.0018}$	& $4.935^{+0.035}_{-0.034}$	& $127.7^{+1.3}_{-1.3}$	& $13.419^{+0.076}_{-0.070}$	& $0.301^{+0.020}_{-0.019}$	\\
64150	& $3.57^{+0.40}_{-0.34}$	& $1.080^{+0.052}_{-0.052}$	& $0.603^{+0.016}_{-0.015}$	& $26.34^{+1.00}_{-0.94}$	& $88.90^{+0.46}_{-0.46}$	& $104.2^{+6.6}_{-5.9}$	& $0.694^{+0.025}_{-0.025}$	\\
64797	& $9.5^{+1.4}_{-1.1}$	& $0.641^{+0.036}_{-0.036}$	& $0.554^{+0.030}_{-0.028}$	& $72^{+20}_{-16}$	& $93.90^{+1.80}_{-0.59}$	& $560^{+250}_{-180}$	& $0.36^{+0.33}_{-0.11}$	\\
65312\tablenotemark{$_{*}$}	& $32.1^{+5.3}_{-4.1}$	& $1.536^{+0.077}_{-0.077}$	& $0.365^{+0.039}_{-0.028}$	& $8.94^{+0.35}_{-0.26}$	& $166.9^{+4.1}_{-15.0}$	& $19.37^{+1.10}_{-0.75}$	& $0.24^{+0.25}_{-0.11}$	\\
70623	& $5.55^{+0.43}_{-0.39}$	& $1.080^{+0.053}_{-0.055}$	& $0.29^{+0.12}_{-0.12}$	& $32^{+16}_{-10}$	& $79^{+48}_{-42}$	& $152^{+122}_{-61}$	& $0.34^{+0.12}_{-0.13}$	\\
71001	& $5.10^{+1.00}_{-0.78}$	& $0.866^{+0.044}_{-0.043}$	& $0.345^{+0.050}_{-0.047}$	& $17.2^{+4.1}_{-2.8}$	& $61^{+44}_{-21}$	& $64^{+23}_{-14}$	& $0.892^{+0.036}_{-0.055}$	\\
71631	& $49.78^{+0.16}_{-0.35}$	& $1.076^{+0.053}_{-0.053}$	& $0.448^{+0.017}_{-0.017}$	& $14.81^{+0.82}_{-0.69}$	& $93.2^{+1.4}_{-1.1}$	& $46.2^{+3.9}_{-3.2}$	& $0.813^{+0.063}_{-0.068}$	\\
71898	& $5.23^{+0.76}_{-0.62}$	& $0.4033^{+0.0082}_{-0.0080}$	& $0.0877^{+0.0074}_{-0.0059}$	& $42.4^{+14.0}_{-7.8}$	& $126.1^{+10.0}_{-6.5}$	& $390^{+200}_{-100}$	& $0.49^{+0.20}_{-0.20}$	\\
72043	& $3.93^{+0.86}_{-0.66}$	& $1.158^{+0.058}_{-0.057}$	& $0.477^{+0.048}_{-0.036}$	& $17.9^{+2.0}_{-1.3}$	& $60.8^{+4.3}_{-4.2}$	& $59.4^{+9.0}_{-5.9}$	& $0.382^{+0.075}_{-0.100}$	\\
72659	& $25.5^{+2.8}_{-2.9}$	& $0.975^{+0.049}_{-0.049}$	& $0.332^{+0.066}_{-0.047}$	& $22.0^{+3.6}_{-2.3}$	& $141.5^{+6.0}_{-7.3}$	& $90^{+21}_{-12}$	& $0.557^{+0.071}_{-0.140}$	\\
75676	& $9.9^{+3.2}_{-2.3}$	& $1.081^{+0.054}_{-0.054}$	& $0.658^{+0.018}_{-0.018}$	& $7.74^{+0.11}_{-0.11}$	& $97.89^{+0.55}_{-0.56}$	& $16.323^{+0.034}_{-0.037}$	& $0.4865^{+0.0010}_{-0.0011}$	\\
76543	& $4.70^{+1.10}_{-0.83}$	& $1.298^{+0.065}_{-0.066}$	& $0.57^{+0.34}_{-0.23}$	& $32^{+25}_{-11}$	& $77^{+47}_{-38}$	& $132^{+162}_{-59}$	& $0.71^{+0.12}_{-0.15}$	\\
77052	& $6.49^{+0.40}_{-0.36}$	& $1.089^{+0.049}_{-0.049}$	& $0.5775^{+0.0087}_{-0.0086}$	& $70.4^{+14.0}_{-7.0}$	& $144.8^{+5.7}_{-5.5}$	& $458^{+146}_{-67}$	& $0.095^{+0.130}_{-0.042}$	\\
77810	& $21.7^{+4.6}_{-3.5}$	& $1.081^{+0.054}_{-0.053}$	& $0.379^{+0.021}_{-0.017}$	& $15.1^{+1.4}_{-1.2}$	& $136.1^{+8.6}_{-8.8}$	& $48.6^{+6.8}_{-5.6}$	& $0.483^{+0.022}_{-0.017}$	\\
78395	& $4.10^{+0.69}_{-0.55}$	& $0.677^{+0.013}_{-0.013}$	& $0.176^{+0.022}_{-0.019}$	& $25.3^{+11.0}_{-5.5}$	& $37.8^{+4.1}_{-2.4}$	& $138^{+96}_{-42}$	& $0.838^{+0.041}_{-0.057}$	\\
78709	& $6.17^{+0.65}_{-0.56}$	& $1.000^{+0.050}_{-0.050}$	& $0.497^{+0.017}_{-0.017}$	& $6.033^{+0.087}_{-0.089}$	& $50.5^{+2.0}_{-1.8}$	& $12.114^{+0.011}_{-0.011}$	& $0.669^{+0.012}_{-0.012}$	\\
79619	& $3.03^{+0.46}_{-0.39}$	& $1.097^{+0.055}_{-0.053}$	& $0.310^{+0.090}_{-0.069}$	& $24.1^{+5.2}_{-3.4}$	& $124.2^{+4.7}_{-9.0}$	& $99^{+30}_{-18}$	& $0.70^{+0.11}_{-0.14}$	\\
81375	& $3.47^{+0.41}_{-0.35}$	& $0.976^{+0.049}_{-0.048}$	& $0.829^{+0.022}_{-0.023}$	& $96^{+38}_{-25}$	& $102.7^{+14.0}_{-5.7}$	& $700^{+450}_{-260}$	& $0.41^{+0.45}_{-0.20}$	\\
82032	& $3.62^{+0.55}_{-0.46}$	& $1.265^{+0.063}_{-0.063}$	& $0.60^{+0.25}_{-0.14}$	& $93^{+38}_{-18}$	& $69.1^{+5.7}_{-6.3}$	& $640^{+390}_{-160}$	& $0.875^{+0.031}_{-0.081}$	\\
83020	& $12.6^{+1.8}_{-1.5}$	& $0.808^{+0.040}_{-0.040}$	& $0.475^{+0.040}_{-0.033}$	& $67^{+25}_{-15}$	& $77.0^{+3.0}_{-15.0}$	& $490^{+300}_{-160}$	& $0.46^{+0.41}_{-0.24}$	\\
84684	& $1.98^{+0.51}_{-0.43}$	& $0.899^{+0.045}_{-0.045}$	& $0.285^{+0.010}_{-0.010}$	& $9.98^{+0.22}_{-0.22}$	& $16.6^{+1.4}_{-1.3}$	& $28.98^{+0.76}_{-0.74}$	& $0.3703^{+0.0092}_{-0.0082}$	\\
85158	& $23.3^{+5.2}_{-4.8}$	& $1.157^{+0.058}_{-0.058}$	& $0.341^{+0.120}_{-0.074}$	& $21.8^{+12.0}_{-5.0}$	& $68^{+26}_{-22}$	& $83^{+73}_{-26}$	& $0.70^{+0.10}_{-0.11}$	\\
85653	& $1.76^{+0.27}_{-0.24}$	& $0.976^{+0.050}_{-0.049}$	& $0.6076^{+0.0063}_{-0.0063}$	& $58.5^{+5.5}_{-8.6}$	& $72.3^{+4.5}_{-4.6}$	& $356^{+52}_{-76}$	& $0.116^{+0.150}_{-0.065}$	\\
86974	& $3.48^{+0.22}_{-0.20}$	& $1.113^{+0.036}_{-0.035}$	& $0.2274^{+0.0038}_{-0.0038}$	& $20.66^{+0.42}_{-0.40}$	& $61.95^{+0.40}_{-0.41}$	& $81.1^{+2.8}_{-2.6}$	& $0.3879^{+0.0053}_{-0.0051}$	\\
88622	& $10.4^{+1.7}_{-1.4}$	& $1.073^{+0.055}_{-0.054}$	& $0.65^{+0.49}_{-0.19}$	& $19.2^{+5.9}_{-2.8}$	& $119.3^{+8.0}_{-8.8}$	& $64^{+20}_{-10}$	& $0.9820^{+0.0063}_{-0.0170}$	\\
89270	& $7.2^{+1.0}_{-1.0}$	& $1.342^{+0.066}_{-0.066}$	& $0.2094^{+0.0065}_{-0.0067}$	& $5.182^{+0.079}_{-0.083}$	& $89.6^{+2.9}_{-2.9}$	& $9.4703^{+0.0031}_{-0.0031}$	& $0.76960^{+0.00055}_{-0.00054}$	\\
92262	& $4.09^{+0.85}_{-0.68}$	& $1.399^{+0.069}_{-0.070}$	& $0.356^{+0.012}_{-0.012}$	& $7.39^{+0.11}_{-0.11}$	& $30.04^{+0.51}_{-0.50}$	& $15.160^{+0.013}_{-0.013}$	& $0.2766^{+0.0019}_{-0.0019}$	\\
94370	& $5.83^{+0.85}_{-0.72}$	& $1.158^{+0.057}_{-0.058}$	& $0.1921^{+0.0063}_{-0.0063}$	& $9.41^{+0.15}_{-0.15}$	& $67.32^{+0.92}_{-0.92}$	& $24.84^{+0.13}_{-0.13}$	& $0.6232^{+0.0012}_{-0.0012}$	\\
95401	& $21.0^{+4.8}_{-3.6}$	& $1.431^{+0.074}_{-0.074}$	& $0.665^{+0.120}_{-0.075}$	& $27.8^{+4.9}_{-3.3}$	& $128.1^{+4.0}_{-3.6}$	& $101^{+26}_{-17}$	& $0.774^{+0.070}_{-0.062}$	\\
96395	& $3.57^{+0.32}_{-0.29}$	& $1.083^{+0.053}_{-0.053}$	& $0.1150^{+0.0036}_{-0.0037}$	& $6.51^{+0.10}_{-0.10}$	& $98.5^{+1.4}_{-1.7}$	& $15.161^{+0.017}_{-0.017}$	& $0.69321^{+0.00052}_{-0.00052}$	\\
98677	& $3.09^{+0.24}_{-0.21}$	& $0.898^{+0.046}_{-0.045}$	& $0.2154^{+0.0066}_{-0.0062}$	& $46.1^{+19.0}_{-4.0}$	& $95.8^{+10.0}_{-2.1}$	& $297^{+203}_{-38}$	& $0.67^{+0.29}_{-0.39}$	\\
99385	& $6.80^{+1.20}_{-0.94}$	& $0.677^{+0.013}_{-0.013}$	& $0.1772^{+0.0064}_{-0.0058}$	& $17.5^{+1.4}_{-1.1}$	& $93.71^{+0.18}_{-0.17}$	& $79.4^{+9.5}_{-7.2}$	& $0.219^{+0.027}_{-0.019}$	\\
99695	& $24.0^{+3.2}_{-2.6}$	& $0.847^{+0.042}_{-0.042}$	& $0.1330^{+0.0073}_{-0.0074}$	& $5.430^{+0.086}_{-0.089}$	& $90^{+12}_{-12}$	& $12.784^{+0.034}_{-0.034}$	& $0.755^{+0.019}_{-0.024}$	\\
101345	& $4.13^{+0.47}_{-0.40}$	& $1.242^{+0.062}_{-0.060}$	& $0.520^{+0.027}_{-0.026}$	& $88^{+26}_{-17}$	& $64.3^{+5.7}_{-6.0}$	& $620^{+300}_{-170}$	& $0.917^{+0.037}_{-0.064}$	\\
101597	& $27.1^{+10.0}_{-6.0}$	& $0.739^{+0.053}_{-0.048}$	& $0.658^{+0.035}_{-0.023}$	& $79^{+18}_{-13}$	& $50.2^{+1.3}_{-1.1}$	& $590^{+230}_{-150}$	& $0.646^{+0.066}_{-0.072}$	\\
101846	& $5.7^{+1.4}_{-1.0}$	& $0.716^{+0.036}_{-0.035}$	& $0.545^{+0.041}_{-0.039}$	& $68^{+27}_{-12}$	& $62.6^{+7.7}_{-21.0}$	& $500^{+330}_{-130}$	& $0.51^{+0.30}_{-0.33}$	\\
103768	& $4.27^{+0.87}_{-0.68}$	& $0.686^{+0.013}_{-0.013}$	& $0.1735^{+0.0088}_{-0.0088}$	& $43.5^{+18.0}_{-5.9}$	& $113.7^{+25.0}_{-7.5}$	& $309^{+210}_{-61}$	& $0.59^{+0.30}_{-0.35}$	\\
103983	& $49.29^{+0.53}_{-1.10}$	& $1.199^{+0.059}_{-0.059}$	& $0.544^{+0.043}_{-0.039}$	& $8.12^{+0.14}_{-0.15}$	& $162.1^{+1.0}_{-1.1}$	& $17.50^{+0.18}_{-0.19}$	& $0.470^{+0.041}_{-0.034}$	\\
104239	& $3.88^{+0.51}_{-0.42}$	& $1.022^{+0.044}_{-0.044}$	& $0.717^{+0.014}_{-0.014}$	& $94^{+27}_{-13}$	& $65.0^{+2.3}_{-2.5}$	& $690^{+320}_{-140}$	& $0.151^{+0.190}_{-0.059}$	\\
107133	& $8.7^{+1.4}_{-1.1}$	& $0.684^{+0.013}_{-0.013}$	& $0.4142^{+0.0053}_{-0.0053}$	& $10.50^{+0.23}_{-0.20}$	& $50.19^{+0.69}_{-0.64}$	& $32.46^{+1.10}_{-0.94}$	& $0.7613^{+0.0051}_{-0.0046}$	\\
109169	& $7.5^{+1.7}_{-1.3}$	& $1.226^{+0.061}_{-0.060}$	& $0.1756^{+0.0071}_{-0.0070}$	& $8.16^{+0.16}_{-0.17}$	& $153.04^{+0.48}_{-0.50}$	& $19.67^{+0.37}_{-0.33}$	& $0.4442^{+0.0051}_{-0.0050}$	\\
109918\tablenotemark{$_{*}$}	& $2.73^{+0.67}_{-0.53}$	& $0.870^{+0.043}_{-0.044}$	& $0.305^{+0.014}_{-0.013}$	& $14.19^{+0.61}_{-0.57}$	& $122.2^{+2.0}_{-2.6}$	& $49.3^{+2.8}_{-2.5}$	& $0.8851^{+0.0017}_{-0.0017}$	\\
110109	& $2.41^{+0.14}_{-0.13}$	& $1.080^{+0.054}_{-0.054}$	& $0.511^{+0.034}_{-0.030}$	& $42.3^{+2.3}_{-2.0}$	& $75.8^{+1.8}_{-2.2}$	& $218^{+17}_{-14}$	& $0.775^{+0.057}_{-0.071}$	\\
110440	& $4.10^{+0.60}_{-0.51}$	& $1.081^{+0.053}_{-0.054}$	& $0.1700^{+0.0061}_{-0.0061}$	& $4.086^{+0.063}_{-0.065}$	& $69.9^{+2.6}_{-2.2}$	& $7.3828^{+0.0011}_{-0.0011}$	& $0.5056^{+0.0019}_{-0.0018}$	\\
111571	& $97.8^{+1.6}_{-3.3}$	& $0.614^{+0.012}_{-0.012}$	& $0.146^{+0.013}_{-0.012}$	& $8.52^{+0.24}_{-0.21}$	& $32.0^{+5.2}_{-6.5}$	& $28.5^{+1.3}_{-1.2}$	& $0.569^{+0.038}_{-0.037}$	\\
113438	& $4.00^{+0.57}_{-0.47}$	& $1.158^{+0.058}_{-0.058}$	& $0.549^{+0.019}_{-0.019}$	& $26.8^{+1.6}_{-1.4}$	& $110.12^{+0.58}_{-0.61}$	& $106.1^{+9.3}_{-7.8}$	& $0.477^{+0.025}_{-0.023}$	\\
115013\tablenotemark{$_{*}$}	& $10.2^{+2.4}_{-1.7}$	& $0.707^{+0.013}_{-0.013}$	& $0.3027^{+0.0037}_{-0.0037}$	& $6.622^{+0.040}_{-0.040}$	& $101.7^{+1.1}_{-1.2}$	& $16.958^{+0.060}_{-0.060}$	& $0.3669^{+0.0015}_{-0.0015}$	\\
116350	& $5.07^{+0.77}_{-0.65}$	& $1.200^{+0.060}_{-0.059}$	& $0.1154^{+0.0077}_{-0.0064}$	& $7.15^{+0.11}_{-0.12}$	& $146.5^{+1.5}_{-2.1}$	& $16.661^{+0.028}_{-0.025}$	& $0.728^{+0.030}_{-0.025}$	\\
117902	& $4.23^{+0.89}_{-0.68}$	& $1.120^{+0.056}_{-0.056}$	& $0.2219^{+0.0070}_{-0.0071}$	& $8.28^{+0.13}_{-0.13}$	& $58.82^{+0.46}_{-0.45}$	& $20.558^{+0.063}_{-0.060}$	& $0.3291^{+0.0012}_{-0.0012}$ \\
\hline\hline
\end{longtable*}
\tablenotetext{*}{Bimodal inclination posterior, retrograde mode reported in this table.}